\documentclass{article}
\usepackage[whole]{bxcjkjatype}
\usepackage{amsmath,amssymb,latexsym,wasysym,marvosym,textcomp,stmaryrd,tikz,multirow,mathtools,tkz-euclide,url,enumitem,amsthm,titlesec,caption}
\usepackage[unicode]{hyperref} 
\usepackage[british]{babel}
\usepackage[backend=biber,style=numeric,sorting=none]{biblatex}
\DeclarePrefChars{'-冬}
\usepackage[none]{hyphenat}
\usetikzlibrary{arrows,angles,decorations}

\title{A Framework for Loop and Path Puzzle Satisfiability NP-Hardness Results}
\author{Hadyn Tang}
\newcommand{\rightang}[2]{\draw[line width=0.2pt,fill=gray] (#1-0.4,#2-0.4)--(#1-0.4,#2+0.4)--(#1+0.4,#2+0.4)--(#1+0.4,#2-0.4)--cycle;}
\newcommand{\acuteang}[2]{\draw[line width=0.2pt,fill=black] (#1-3^.5*0.3,#2-0.3)--(#1+3^.5*0.3,#2-0.3)--(#1,#2+0.6)--cycle;}
\newcommand{\obtuseang}[2]{\draw[line width=0.2pt,fill=white] (#1-0.5*2.351141/4,#2-0.5*1/4-0.5*5^.5/4)--(#1-0.5*3.804226/4,#2-0.5*1/4+0.5*5^.5/4)--(#1,#2+0.5)--(#1+0.5*3.804226/4,#2-0.5*1/4+0.5*5^.5/4)--(#1+0.5*2.351141/4,#2-0.5*1/4-0.5*5^.5/4)--cycle;}
\newcommand{\letcirc}[3]{\draw[fill=white](#1,#2) circle (0.5) node{\scalebox{0.16}{#3}};}
\begin{document}
	\maketitle
	
	\begin{abstract}		
		\noindent Building on the results published in \href{https://arxiv.org/abs/2004.12849}{arxiv:2004.12849}, we present a general framework for demonstrating the NP-hardness of satisfying many genres of loop and path puzzles using a `T-metacell' gadget. We then use this to prove the NP-completeness of a variety of such genres, and discuss some of the limitations of this gadget.
	\end{abstract}
	
	\section{Introduction}
	
	Pencil-and-paper logic puzzles are a category of problems where one is (usually) given a grid and some form of clues in the grid, and asked to mark parts of the grid in accordance with a set of rules. For example, genres of puzzles like Sudoku (Number Place) and Kakuro (Cross Sum) involve filling numbers in a grid, whereas other genres like LITS and Nurikabe involve shading a connected group of cells in a grid.
	
	In the past twenty-odd years, many works have shown that determining whether instances of particular genres of pencil-and-paper puzzles are solvable is an NP-complete problem. For example, McPhail and Fix demonstrated in \cite{mcphail2004nurikabe} that satisfying Nurikabe is an NP-complete problem. This interest possibly stems from Yato's 2000 similar result for Slitherlink in \cite{yato2000np}.
	
	Another area in which many completeness proofs have been recently developed (here, for the class PSPACE) is in video games. In 2020, Ani et al proved in \cite{ani2020walking} that a single gadget, named a `door' gadget, could be used to demonstrate the PSPACE-hardness of solving an arbitrary-size extension of various video games.
	
	Inspired by this work, and the many previous NP-completeness results for a variety of pencil-and-paper puzzles, in this paper we aim to present a similar gadget-based reduction for a certain category of pencil-and-paper puzzles -- ones where the objective is to draw a loop through the cells of the grid -- and hence show that many varieties of loop genres are NP-hard (and then NP-complete). 
	
	\section{Framework}
	
	\subsection{Revisiting Barred Simple Loop}
	
	Barred Simple Loop is a puzzle genre in which one is given a rectangular grid with some `bars' between cells, and the aim is to draw a loop passing through all of the cells exactly once and none of the bars. In \cite{tang2020np}, we showed that the question of whether a general Barred Simple Loop puzzle was satisfiable is NP-complete, by reducing from the problem of finding Hamiltonian cycles in cubic bipartite planar graphs. The reduction proceeded as follows:
	
	\begin{enumerate}[itemsep=0pt,parsep=1.5pt,topsep=0pt]
		\item Start with any cubic bipartite planar graph.
		\item Create a rectangular realisation of the graph.
		\item Realign vertices so their colours match an underlying checkerboard grid.
		\item Unfold the realisation and lengthen edges so that all vertices lie on a central row, and every cell not on the central row is used by an edge.
		\item Tile certain cells in an expanded grid with dominoes.
		\item Create an instance of Barred Simple Loop that is solvable if and only if the initial Hamiltonian path problem was.
	\end{enumerate}
	
	\subsection{Introducing Cubic Barred Simple Loop}
	
	Ideally, in accordance with the `gadget' style of reduction alluded to in the introduction, we want to reduce from instances Barred Simple Loop to instaces of some other loop genre $\mathcal G$ by replacing each cell in Barred Simple Loop with a gadget in the relevant genre. If the gadget was appropriately designed, the solution to the gadget in $\mathcal G$ would be able to exactly mimic the action of a single cell in Barred Simple Loop, and thus there would be a solution to the Barred Simple Loop puzzle if and only if there was a solution to the new $\mathcal G$ puzzle. 
	
	However, an issue with this is that cells in Barred Simple Loop can have up to four exits, and thus any gadget that we create would need to admit up to four exits. But then our gadget has two key constraints: not only must it be visited, but one must not visit it twice -- otherwise, one could connect two pairs of exits.
	
	As a way to avoid this, we create an intermediate genre named `Cubic Barred Simple Loop'. The rules are the same as standard Barred Simple Loop, except each cell in the puzzle is guaranteed to have at most three adjacent accessible cells (and thus be adjacent to at least one bar). Note that this does not impose any additional restrictions on the solution, only on the given problem. 
	
	If we can reduce from Cubic Barred Simple Loop to Barred Simple Loop to show Cubic Barred Simple Loop is NP-hard, then the gadgets in an arbitrary reduction $\mathcal G\rightarrow$ Cubic Barred Simple Loop would only need to obey the condition `visit once', rather than `visit exactly once', which is simpler.
	
	\subsection{Reduction for Cubic Barred Simple Loop}
	
	To prove Cubic Barred Simple Loop is NP-hard, we reduce to Barred Simple Loop using gadgets. Consider the following gadget, where the blue lines can become bars if need be:
	
	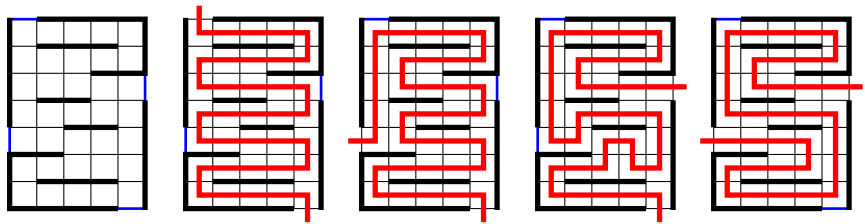
\begin{figure}[hbp]
		\centering
		\begin{tikzpicture}[scale=0.18]
			\foreach \a in {0,13,...,52}{
				\foreach \x in {0,2,...,10} \draw (\a+\x,0)--(\a+\x,14);
				\foreach \y in {0,2,...,14} \draw (\a,\y)--(\a+10,\y);
			}
			\draw [line width=1pt,color=blue] (0,0)--(10,0)--(10,14)--(0,14)--cycle;
			\draw [line width=1pt,color=blue] (13,2)--(13,12);
			\draw [line width=1pt,color=blue] (23,2)--(23,12);
			\draw [line width=1pt,color=blue] (26,12)--(26,14)--(36,14)--(36,2);
			\draw [line width=1pt,color=blue] (39,0)--(39,14)--(49,14);
			\draw [line width=1pt,color=blue] (52,0)--(62,0)--(62,2);
			\draw [line width=1pt,color=blue] (52,12)--(52,14)--(62,14);
			\foreach \a in {0,13,...,52}{
				\draw[line width=2pt] (\a+8,0)--(\a,0)--(\a,4)--(\a+4,4);
				\draw[line width=2pt] (\a+10,-0.1)--(\a+10,8);
				\draw[line width=2pt] (\a+2,2)--(\a+8,2);
				\draw[line width=2pt] (\a+4,6)--(\a+8,6);
				\draw[line width=2pt] (\a,6)--(\a,14.1);
				\draw[line width=2pt] (\a+2,8)--(\a+6,8);
				\draw[line width=2pt] (\a+6,10)--(\a+10,10)--(\a+10,14)--(\a+2,14);
				\draw[line width=2pt] (\a+2,12)--(\a+8,12);
			}
			\draw[line width=2pt,color=red] (22,-1)--(22,1)--(14,1)--(14,3)--(22,3)--(22,5)--(14,5)--(14,7)--(22,7)--(22,9)--(14,9)--(14,11)--(22,11)--(22,13)--(14,13)--(14,15);
			\draw[line width=2pt,color=red] (35,-1)--(35,1)--(27,1)--(27,3)--(35,3)--(35,5)--(29,5)--(29,7)--(35,7)--(35,9)--(29,9)--(29,11)--(35,11)--(35,13)--(27,13)--(27,5)--(25,5);
			\draw[line width=2pt,color=red] (48,-1)--(48,1)--(40,1)--(40,3)--(44,3)--(44,5)--(46,5)--(46,3)--(48,3)--(48,7)--(42,7)--(42,5)--(40,5)--(40,13)--(48,13)--(48,11)--(42,11)--(42,9)--(50,9);
			\draw[line width=2pt,color=red] (51,5)--(59,5)--(59,3)--(53,3)--(53,1)--(61,1)--(61,7)--(53,7)--(53,13)--(61,13)--(61,11)--(55,11)--(55,9)--(63,9);
		\end{tikzpicture}
		\captionof{figure}{Cubic Barred Simple Loop metacell gadget and possible solutions}
	\end{figure}
	
	Note that all cells have at most three adjacent accessible cells regardless of whether the blue bars are in place or absent. The depicted solutions (and the $180^\circ$ rotated versions of the middle two) illustrate that if the relevant bars are not in place, then for every pair of exits, there is a possible path between them visiting every cell of the gadget.
	
	Furthermore, consider colouring the gadgets like a checkerboard, where the top-left cell is black. Then all the possible entries are on black squares, and thus any path between them must alternate black-white-black-white-...-black and have one more black cell than white cell. Since there is exactly one more black cell than white cell in the whole gadget, this means that it must be visited exactly once.
	
	So, given an instance of (regular) Barred Simple Loop, we can turn it into a Cubic Barred Simple Loop puzzle by replacing each cell with one of these metacells, where every even column has its cells reflected horizontally and every even row has its cells reflected vertically (performing both reflections if both apply) so that the exits between adjacent cells line up. Then by blocking off the blue edges between gadgets corresponding to bars in the original puzzle, the new puzzle is solvable if and only if the original puzzle was, since possible solutions to the gadget exactly mimic possible solutions to a normal cell with the relevant bars around it.
	
	Since this method involves changing each cell into 35 subcells, this corresponds to a constant factor (i.e.\ polynomial) size increase, so the NP-completeness of satisfying regular Barred Simple Loop implies Cubic Barred Simple Loop is NP-hard to satisfy. Clearly satisfaction of Cubic Barred Simple Loop is also in NP because one can verify no bars are passed through in $O(n^2)$ time (where $n$ is the larger of the two side lengths of the grid), so the problem of satisfaction of an arbitrary Cubic Barred Simple Loop is NP-complete.
	
	\subsection{Creating T-Metacells}
	
	Having shown satisfaction of Cubic Barred Simple Loop is NP-complete, we now present a method to prove other loop/path genres are NP-hard to satisfy.
	
	As alluded to in subsection 2.2, suppose we had a gadget in a particular genre that functioned as a cell that must be visited and that had three exits -- let us call such gadgets `T-metacell' gadgets. Then we could attempt to reduce from Cubic Barred Simple Loop to this genre with a constant factor of dilation (i.e.\ polynomial size increase) by replacing each cell with a T-metacell facing in the appropriate orientation. However, some cells such as corner cells only have two exits, so this will not work directly. (Note that they need to have at least two exits so that the loop can visit each cell: we can easily determine that the puzzle is unsolvable in $O(n^2)$ time by looking at each cell and reduce such instances to an unsolvable puzzle like two T-metacells in a $2\times1$ grid in any orientation.)
	
	We could try placing a T-metacell in these cells anyway, so that two ends point in the correct directions and one edge is `free'. However, this might effectively remove some bars between cells, if two free edges happened to line up over an edge. So, given a Cubic Barred Simple Loop puzzle, we wish to find a method to direct all the free edges of T-metacells corresponding to cells with two adjacent bars, so that no pair of free edges line up. 
	
	Consider the graph with vertices formed on each cell of the Cubic Barred Simple Loop puzzle, as well as a vertex outside each exterior edge. Connect two edges if they represent two cells (or a cell and an edge) with a bar between them. Then, vertices corresponding to cells with only two exits must have degree 2, whereas vertices corresponding to cells with three exits and those corresponding to edges have degree 1. To select a direction for the third exit of each cell with two exits, we need to orient some of the edges so that each vertex of degree 2 has an outgoing edge, thus guaranteeing we don't introduce any additional paths between cells. However, observing that the graph is a collection of paths and cycles due to only having vertices of degree 1 and 2, this is trivial by enforcing a consistent direction in each component. This process is illustrated in the diagram overleaf. The dotted arrows do not not affect the final T-metacells, but are included to show how edges could be orientated if their relevant paths were extended. In the last grid, the thicker lines in the T-metacells are necessary, and the thinner lines the free ends. We can execute this process in $O(n^2)$ time.
	
	Thus, if we can create a constant-size T-metacell gadget in a particular genre, it would be possible to reduce from Cubic Barred Simple Loop to that genre to prove that genre's NP-hardness. The precise constraints required for this argument to work are outlined the following subsection.
	
	\newpage
	
	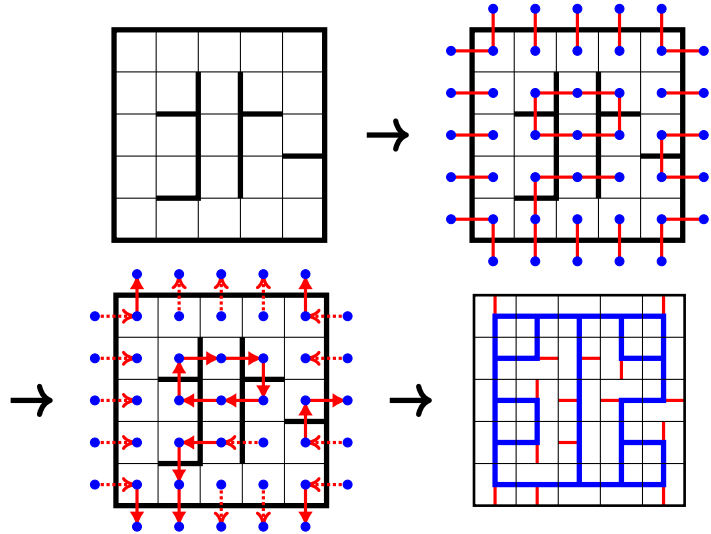
\begin{figure}
		\centering
		\begin{tikzpicture}[scale=0.28]
			\draw[line width=2pt,->,color=white] (-4,6)--(-1,6);
			\foreach \x in {4,6,8,10} \draw[line width=0.4pt] (\x,1)--(\x,11);
			\foreach \y in {3,5,7,9} \draw[line width=0.4pt] (2,\y)--(12,\y);
			\draw[line width=2pt] (2,1)--(2,11)--(12,11)--(12,1)--cycle;
			\draw[line width=2pt] (4,3)--(6,3)--(6,9);
			\draw[line width=2pt] (4,7)--(6,7);
			\draw[line width=2pt] (8,3)--(8,9);
			\draw[line width=2pt] (8,7)--(10,7);
			\draw[line width=2pt] (10,5)--(12,5);
			\draw[line width=2pt,->] (14,6)--(16,6);
			\foreach \x in {21,23,25,27} \draw[line width=0.4pt] (\x,1)--(\x,11);
			\foreach \y in {3,5,7,9} \draw[line width=0.4pt] (19,\y)--(29,\y);
			\draw[line width=2pt] (19,1)--(19,11)--(29,11)--(29,1)--cycle;
			\draw[line width=2pt] (21,3)--(23,3)--(23,9);
			\draw[line width=2pt] (21,7)--(23,7);
			\draw[line width=2pt] (25,3)--(25,9);
			\draw[line width=2pt] (25,7)--(27,7);
			\draw[line width=2pt] (27,5)--(29,5);
			\foreach \x in {18,28} \foreach \y in {2,4,6,8,10} \draw[color=red,line width=1.2pt] (\x,\y)--(\x+2,\y);
			\foreach \x in {22,24} \foreach \y in {4,6,8} \draw[color=red,line width=1.2pt] (\x,\y)--(\x+2,\y);
			\foreach \x in {20,22,24,26,28} \foreach \y in {0,10} \draw[color=red,line width=1.2pt] (\x,\y)--(\x,\y+2);
			\draw[color=red,line width=1.2pt] (22,2)--(22,4);
			\draw[color=red,line width=1.2pt] (22,6)--(22,8);
			\draw[color=red,line width=1.2pt] (26,6)--(26,8);
			\draw[color=red,line width=1.2pt] (28,4)--(28,6);
			\foreach \x in {20,22,24,26,28} \foreach \y in {0,2,...,12}\draw[color=blue,fill=blue] (\x,\y)circle(6pt);
			\foreach \x in {18,30} \foreach \y in {2,4,6,8,10}\draw[color=blue,fill=blue] (\x,\y)circle(6pt);
		\end{tikzpicture}
		
		\begin{tikzpicture}[scale=0.28]
			\draw[line width=2pt,->] (32,6)--(34,6);
			\foreach \x in {39,41,43,45} \draw[line width=0.2pt] (\x,1)--(\x,11);
			\foreach \y in {3,5,7,9} \draw[line width=0.4pt] (37,\y)--(47,\y);
			\draw[line width=2pt] (37,1)--(37,11)--(47,11)--(47,1)--cycle;
			\draw[line width=2pt] (39,3)--(41,3)--(41,9);
			\draw[line width=2pt] (39,7)--(41,7);
			\draw[line width=2pt] (43,3)--(43,9);
			\draw[line width=2pt] (43,7)--(45,7);
			\draw[line width=2pt] (45,5)--(47,5);
			\foreach \y in {2,4,6,8,10} \draw[color=red,line width=1.2pt,densely dotted,-{>[length=2mm,width=2mm]}] (36,\y)--(38,\y);
			\foreach \x in {38,40,46} \draw[color=red,line width=1.2pt,-{Latex[length=2mm,width=2mm]}] (\x,2)--(\x,0);
			\foreach \x in {38,46} \draw[color=red,line width=1.2pt,-{Latex[length=2mm,width=2mm]}] (\x,10)--(\x,12);
			\foreach \x in {42,44} \draw[color=red,line width=1.2pt,densely dotted,-{>[length=2mm,width=2mm]}] (\x,2)--(\x,0);
			\foreach \x in {40,42,44} \draw[color=red,line width=1.2pt,densely dotted,-{>[length=2mm,width=2mm]}] (\x,10)--(\x,12);
			\draw[color=red,line width=1.2pt,-{Latex[length=2mm,width=2mm]}] (46,4)--(46,6);
			\draw[color=red,line width=1.2pt,-{Latex[length=2mm,width=2mm]}] (46,6)--(48,6);
			\foreach \y in {2,4,8,10} \draw[color=red,line width=1.2pt,densely dotted,-{>[length=2mm,width=2mm]}] (48,\y)--(46,\y);
			\draw[color=red,line width=1.2pt,-{Latex[length=2mm,width=2mm]}] (40,4)--(40,2);
			\draw[color=red,line width=1.2pt,-{Latex[length=2mm,width=2mm]}] (42,4)--(40,4);
			\draw[color=red,line width=1.2pt,densely dotted,-{>[length=2mm,width=2mm]}] (44,4)--(42,4);
			\draw[color=red,line width=1.2pt,-{Latex[length=2mm,width=2mm]}] (42,6)--(40,6);
			\draw[color=red,line width=1.2pt,-{Latex[length=2mm,width=2mm]}] (40,6)--(40,8);
			\draw[color=red,line width=1.2pt,-{Latex[length=2mm,width=2mm]}] (40,8)--(42,8);
			\draw[color=red,line width=1.2pt,-{Latex[length=2mm,width=2mm]}] (42,8)--(44,8);
			\draw[color=red,line width=1.2pt,-{Latex[length=2mm,width=2mm]}] (44,8)--(44,6);
			\draw[color=red,line width=1.2pt,-{Latex[length=2mm,width=2mm]}] (44,6)--(42,6);
			\foreach \x in {38,40,42,44,46} \foreach \y in {0,2,...,12}\draw[color=blue,fill=blue] (\x,\y)circle(6pt);
			\foreach \x in {36,48} \foreach \y in {2,4,6,8,10}\draw[color=blue,fill=blue] (\x,\y)circle(6pt);
			\draw[line width=2pt,->] (50,6)--(52,6);
			\draw[line width=1.2pt,color=red] (55,2)--(55,1);
			\draw[line width=1.2pt,color=red] (55,10)--(55,11);
			\draw[line width=1.2pt,color=red] (59,4)--(58,4);
			\draw[line width=1.2pt,color=red] (57,4)--(57,3);
			\draw[line width=1.2pt,color=red] (57,2)--(57,1);
			\draw[line width=1.2pt,color=red] (57,6)--(57,7);
			\draw[line width=1.2pt,color=red] (57,8)--(58,8);
			\draw[line width=1.2pt,color=red] (59,8)--(60,8);
			\draw[line width=1.2pt,color=red] (61,8)--(61,7);
			\draw[line width=1.2pt,color=red] (61,6)--(60,6);
			\draw[line width=1.2pt,color=red] (59,6)--(58,6);
			\draw[line width=1.2pt,color=red] (63,2)--(63,1);
			\draw[line width=1.2pt,color=red] (63,10)--(63,11);
			\draw[line width=1.2pt,color=red] (63,4)--(63,5);
			\draw[line width=1.2pt,color=red] (63,6)--(64,6);
			\draw[line width=2pt,color=blue] (61,4)--(63,4)--(63,2)--(55,2)--(55,10)--(63,10)--(63,6)--(61,6)--(61,2);
			\draw[line width=2pt,color=blue] (55,4)--(57,4)--(57,6)--(55,6);
			\draw[line width=2pt,color=blue] (55,8)--(57,8)--(57,10);
			\draw[line width=2pt,color=blue] (59,2)--(59,10);
			\draw[line width=2pt,color=blue] (61,10)--(61,8)--(63,8);
			\foreach \x in {56,58,60,62} \draw[line width=0.4pt] (\x,1)--(\x,11);
			\foreach \y in {3,5,7,9} \draw[line width=0.4pt] (54,\y)--(64,\y);
			\draw[line width=1pt] (54,1)--(54,11)--(64,11)--(64,1)--cycle;
		\end{tikzpicture}
		\captionof{figure}{Orienting T-metacells appropriately}
	\end{figure}
	
	\subsection{T-Metacell Summary}
	
	To ensure the validity of the previous argument, a T-metacell must satisfy the following conditions:
	
	\begin{enumerate}[label=(\alph*),itemsep=1.5pt,topsep=-\topsep+\parskip]
		\item The gadget must be visited to solve the puzzle;
		\item The gadget must be tileable, through translations or reflections, to make an adjacency graph isomorphic to a rectangular grid;
		\item The gadget must have 3 exits on different sides of the gadget, and these must be the only places the loop can exit the gadget;
		\item The loop may only exit the gadget at a particular exit to enter the adjacent cell through the facing exit, and it cannot exit at all if the adjacent cell does not have any exit on that side or the exit faces the edge of the grid;
		\item There must be a solution to the gadget with the loop leaving the gadget at every pair of the 3 exits; and
		\item The gadget must be rotatable/reflectable so that these exits point in any 3 of the 4 directions.
	\end{enumerate}
	
	Note that (b) and (f) mean a square is the most practical gadget shape, although one could imagine a shape like a rhombus being used as well, especially for triangular grids. However, all the genres we will consider in Section 2 use square metacells. The regular square tiling also means that we can assume some properties about how adjacent gadgets or gadgets in the same row/column are arranged, which may help to restrict unwanted solutions to these gadgets.
	
	\section{Main Gallery}
	
	What follows is an assortment of T-metacells implemented most of the loop genres contained within the puzz.link puzzle repository (\cite{puzzlink}). Each section will contain the rules for the genre, followed by an outline of why the satisfiability of these puzzles belongs in NP (noting that the size of the input is quadratic in $\max(\text{grid height},\text{grid width})$). Following these comes the actual construction of the gadget, as well as any constraints on how the grid is created and tiled or how the gadget is solved. In the images of each construction, thick cyan lines represent lines forced by the gadget no matter how it is solved, and thin red lines represent a particular way the gadget could be solved to produce a solution with the desired exits. Finally, if known, the genre's origin and past satisfiability NP-completeness results are stated. Since each gadget here is a fixed size with fixed clues, the reduction can be done in polynomial time, and thus this is enough to show the the following genres are NP-complete. Example puzzles for each genre can be found in Appendix C.
	
	\subsection{Slitherlink}
	
	\textbf{Rules:} Draw a single loop passing between pairs of adjacent dots. Numbers within the grid indicate how many of the edges of the square formed by the nearest 4 dots are used by the loop. 
	
	One can clearly verify any solution of this puzzle in $O(n^2)$ time simply by checking each number is satisfied. This gadget can be rotated, reflected and translated as necessary, and works because any chain of 3s of length two or more must be traversed in a zig-zag pattern, leading to the `walls' visible in the figure (and the two 1s at the top ensure the gap in the wall cannot be used by the loop). The genre was first published by Nikoli in Puzzle Communication Nikoli \#26, and satisfiability was proved NP-complete in \cite{yato2000np}.
	
	\begin{figure}[hbp]
		\centering
		\begin{tikzpicture}[scale=0.1,font=\sffamily]
			\foreach \d in {24,48,72}{
				\draw[line width=1pt,color=cyan] (3+\d,9)--(1+\d,9)--(1+\d,7)--(3+\d,7)--(3+\d,5)--(1+\d,5)--(1+\d,3)--(3+\d,3)--(3+\d,1)--(5+\d,1)--(5+\d,3)--(7+\d,3)--(7+\d,1)--(9+\d,1)--(9+\d,3);
				\draw[line width=1pt,color=cyan] (13+\d,3)--(13+\d,1)--(15+\d,1)--(15+\d,3)--(17+\d,3)--(17+\d,1)--(19+\d,1)--(19+\d,3)--(21+\d,3)--(21+\d,5)--(19+\d,5)--(19+\d,7)--(21+\d,7)--(21+\d,9)--(19+\d,9);
				\draw[line width=1pt,color=cyan] (3+\d,13)--(1+\d,13)--(1+\d,15)--(3+\d,15)--(3+\d,17)--(1+\d,17)--(1+\d,19)--(3+\d,19)--(3+\d,21)--(5+\d,21)--(5+\d,19)--(7+\d,19)--(7+\d,21)--(9+\d,21)--(9+\d,17);
				\draw[line width=1pt,color=cyan] (13+\d,17)--(13+\d,21)--(15+\d,21)--(15+\d,19)--(17+\d,19)--(17+\d,21)--(19+\d,21)--(19+\d,19)--(21+\d,19)--(21+\d,17)--(19+\d,17)--(19+\d,15)--(21+\d,15)--(21+\d,13)--(19+\d,13);
				\draw[line width=0.6pt,color=red] (9+\d,17)--(13+\d,17);
			}
			\draw[line width=0.6pt,color=red] (24,11)--(27,11)--(27,9);
			\draw[line width=0.6pt,color=red] (33,3)--(37,3);
			\draw[line width=0.6pt,color=red] (27,13)--(35,13)--(35,9)--(43,9);
			\draw[line width=0.6pt,color=red] (43,13)--(43,11)--(46,11);
			\draw[line width=0.6pt,color=red] (48,11)--(51,11)--(51,9);
			\draw[line width=0.6pt,color=red] (51,13)--(57,13)--(57,3);
			\draw[line width=0.6pt,color=red] (59,0)--(59,3)--(61,3);
			\draw[line width=0.6pt,color=red] (67,9)--(67,13);
			\draw[line width=0.6pt,color=red] (75,9)--(75,13);
			\draw[line width=0.6pt,color=red] (81,3)--(83,3)--(83,0);
			\draw[line width=0.6pt,color=red] (85,3)--(85,13)--(91,13);
			\draw[line width=0.6pt,color=red] (91,9)--(91,11)--(94,11);
			\foreach \d in {0,24,48,72}{
				\foreach \x in {1,3,...,21} \foreach \y in {1,3,...,21} \draw[fill=black] (\x+\d,\y) circle (3pt);
				\foreach \x in {2,20} \foreach \y in {4,6,8,14,16,18} \draw (\x+\d,\y) node {\tiny 3};
				\foreach \x in {4,6,8,14,16,18} \foreach \y in {2,20} \draw (\x+\d,\y) node {\tiny 3};
				\foreach \x in {10,12} \draw (\x+\d,20) node {\tiny 1};
				\draw[dashed,line width=0.8pt,color=orange] (\d,0)--(\d,22)--(22+\d,22)--(22+\d,0)--cycle;
			}
		\end{tikzpicture}
		\captionof{figure}{Slitherlink T-metacell and possible solutions}
	\end{figure}
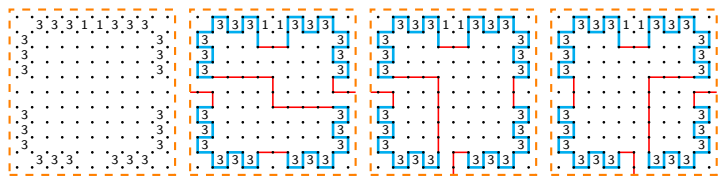

	\newpage
	
	\subsection{Masyu}
	
	\textbf{Rules:} Draw a single loop through all pearls and passing between adjacent cell centres. The loop must turn on each black pearl but go straight on the cells immediately before and after, and it must go straight through each white pearl and turn on a cell immediately before and/or after. 
	
	By checking that each pearl has been passed through and satisfied, we can verify the solution in $O(n^2)$ time. The gadget can be rotated, reflected and translated as necessary. Taking into account the tiling of the gadget, the black pearls cannot connect outside the gadget's sides because they would either hit the edge of the grid or pass straight through another black pearl, and the combination of the black cells provide the `walls' of the gadget. The genre was first published by Nikoli in Puzzle Communication Nikoli \#84 (white pearls only) or \#90 (full genre), and satisfiability was proved NP-complete in \cite{friedman2002pearl} (renamed as `Pearl').
	
	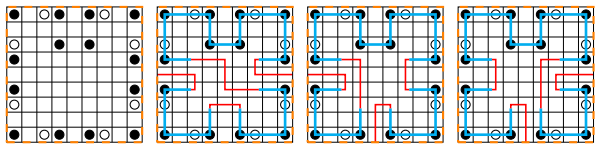
\begin{figure}[hbp]
		\centering
		\begin{tikzpicture}[scale=0.1]
			\foreach \d in {0,20,40,60}{
				\foreach \x in {2,4,...,16} \draw[line width=0.2pt] (\x+\d,0)--(\x+\d,18);
				\foreach \y in {2,4,...,16} \draw[line width=0.2pt] (\d,\y)--(18+\d,\y);
				\foreach \x in {1,17} \foreach \y in {1,7,11,17} \draw[fill=black] (\x+\d,\y) circle (0.6);
				\foreach \x in {7,11} \foreach \y in {1,13,17} \draw[fill=black] (\x+\d,\y) circle (0.6);
				\foreach \x in {5,13} \foreach \y in {1,17} \draw (\x+\d,\y) circle (0.6);
				\foreach \x in {1,17} \foreach \y in {5,13} \draw (\x+\d,\y) circle (0.6);
			}
			\foreach \d in {20,40,60}{
				\draw[line width=1pt,color=cyan] (4.5+\d,11)--(1+\d,11)--(1+\d,17)--(7+\d,17)--(7+\d,13)--(11+\d,13)--(11+\d,17)--(17+\d,17)--(17+\d,11)--(13.5+\d,11); 
				\draw[line width=1pt,color=cyan] (13.5+\d,7)--(17+\d,7)--(17+\d,1)--(11+\d,1)--(11+\d,4.5); 
				\draw[line width=1pt,color=cyan] (7+\d,4.5)--(7+\d,1)--(1+\d,1)--(1+\d,7)--(4.5+\d,7);
			}
			\draw[line width=0.6pt,color=red] (24.5,7)--(25,7)--(25,9)--(20,9);
			\draw[line width=0.6pt,color=red] (24.5,11)--(29,11)--(29,7)--(33.5,7);
			\draw[line width=0.6pt,color=red] (27,4.5)--(27,5)--(31,5)--(31,4.5);
			\draw[line width=0.6pt,color=red] (33.5,11)--(33,11)--(33,9)--(38,9);
			\draw[line width=0.6pt,color=red] (44.5,7)--(45,7)--(45,9)--(40,9);
			\draw[line width=0.6pt,color=red] (44.5,11)--(47,11)--(47,4.5);
			\draw[line width=0.6pt,color=red] (49,0)--(49,5)--(51,5)--(51,4.5);
			\draw[line width=0.6pt,color=red] (53.5,7)--(53,7)--(53,11)--(53.5,11);
			\draw[line width=0.6pt,color=red] (64.5,7)--(65,7)--(65,11)--(64.5,11);
			\draw[line width=0.6pt,color=red] (67,4.5)--(67,5)--(69,5)--(69,0);
			\draw[line width=0.6pt,color=red] (71,4.5)--(71,11)--(73.5,11);
			\draw[line width=0.6pt,color=red] (73.5,7)--(73,7)--(73,9)--(78,9);
			\foreach \d in {0,20,40,60}{
				\draw[line width=0.2pt] (\d,0)--(18+\d,0)--(18+\d,18)--(\d,18)--cycle;
				\draw[dashed,line width=0.8pt,color=orange] (\d,0)--(18+\d,0)--(18+\d,18)--(\d,18)--cycle;
			}
		\end{tikzpicture}
		\captionof{figure}{Masyu T-metacell and possible solutions}
	\end{figure}

	\subsection{Yajilin}
	
	\textbf{Rules:} Draw a single loop passing between adjacent cell centres. Some squares are marked in grey; these are not part of loop. Any other square not passed through by the loop must be shaded, and no two shaded cells may be adjacent. Grey cells may contain a number $n$ and an arrow pointing orthogonally: this means that there are $n$ shaded cells in the direction of the arrow.
	
	We can check any solution in $O(n^3)$ time, firstly by shading all unused cells and verifying no two are adjacent ($O(n^2)$), then for each clue cell checking that it is satisfied ($O(n^3)$). The gadget here can only be rotated and translated, ensuring that there is only one cell gap on each side when taking into account the grey cells. Then, the `no adjacent shaded cells' condition ensures that each gadget is visited. (We can also put $0\!\!\uparrow$ in each grey cell if we wish all grey cells to have clues.) The genre was first published by Nikoli in Puzzle Communication Nikoli \#86, and satisfiability was proved NP-complete in \cite{ishibashi2012np}.
	
	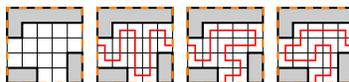
\begin{figure}[hbp]
		\centering
		\begin{tikzpicture}[scale=0.1]
			\foreach \d in {0,12,24,36} {
				\foreach \x in {2,4,6,8} \draw[line width=0.2pt] (\x+\d,0)--(\x+\d,10);
				\foreach \y in {2,4,6,8} \draw[line width=0.2pt] (\d,\y)--(10+\d,\y);
				\draw[line width=0.8pt,fill=gray!40!white] (\d,6)--(\d,10)--(10+\d,10)--(10+\d,8)--(2+\d,8)--(2+\d,6)--cycle; 
				\draw[line width=0.8pt,fill=gray!40!white] (\d,0)--(\d,2)--(4+\d,2)--(4+\d,0)--cycle; 
				\draw[line width=0.8pt,fill=gray!40!white] (8+\d,0)--(8+\d,4)--(10+\d,4)--(10+\d,0)--cycle; 
			}
			\draw[line width=0.6pt,color=red] (12,5)--(13,5)--(13,3)--(15,3)--(15,7)--(17,7)--(17,1)--(19,1)--(19,7)--(21,7)--(21,5)--(22,5);
			\draw[line width=0.6pt,color=red] (24,5)--(25,5)--(25,3)--(27,3)--(27,7)--(33,7)--(33,5)--(29,5)--(29,3)--(31,3)--(31,1)--(29,1)--(29,0);
			\draw[line width=0.6pt,color=red] (41,0)--(41,1)--(43,1)--(43,5)--(41,5)--(41,3)--(37,3)--(37,5)--(39,5)--(39,7)--(45,7)--(45,5)--(46,5);
			\foreach \d in {0,12,24,36}{
				\draw[line width=0.2pt] (\d,0)--(\d,10)--(10+\d,10)--(10+\d,0)--cycle;
				\draw[dashed,line width=0.8pt,color=orange] (\d,0)--(\d,10)--(10+\d,10)--(10+\d,0)--cycle;
			}
		\end{tikzpicture}
		\captionof{figure}{Yajilin T-metacell and possible solutions}
	\end{figure}

	\subsection{Slalom}
	
	\textbf{Rules:} Draw a single loop passing between adjacent unshaded cell centres which visits each gate (dotted line) once. When passing through a gate, the loop must go straight perpendicular to the gate. Furthermore, the loop must be assigned a direction such that, starting at the circle containing the number $n$ (the number of gates), any gate with an arrow and a number $k$ pointing to it is the $k^\text{th}$ gate the loop passes through.
	
	We can check that each gate is passed through in the right order in $O(n^2)$ time. The gadget here can be rotated, reflected and translated as necessary, with the gate ensuring it is passed through. Note that we also need to place the circle with the number of gates in it somewhere: placing it next to any gate will not affect the gadget. The genre was first published by Nikoli in Puzzle Communication Nikoli \#116, and satisfiability was proved NP-complete in \cite{kanehiro2015satogaeri}.
	
	\begin{figure}[hbp]
		\centering
		\begin{tikzpicture}[scale=0.1]
			\foreach \d in {0,12,24,36}{
				\fill[color=black] (\d,0)--(\d,4)--(2+\d,4)--(2+\d,2)--(4+\d,2)--(4+\d,0)--cycle;
				\fill[color=black] (\d+6,0)--(\d+6,2)--(\d+8,2)--(\d+8,4)--(\d+10,4)--(\d+10,0)--cycle;
				\fill[color=black] (\d+4,4)--(\d+4,6)--(\d+6,6)--(\d+6,4)--cycle;
				\fill[color=black]
				(\d,6)--(\d,10)--(\d+10,10)--(\d+10,6)--(\d+8,6)--(\d+8,8)--(\d+2,8)--(\d+2,6)--cycle;
				\foreach \x in {2,4,6,8} \draw[line width=0.2pt] (\x+\d,0)--(\x+\d,10);
				\foreach \y in {2,4,6,8} \draw[line width=0.2pt] (\d,\y)--(10+\d,\y);
			}
			\foreach \d in {12,24,36} \draw[line width=1pt,color=cyan] (\d+3,5.5)--(\d+3,7)--(\d+7,7)--(\d+7,5.5);
			\draw[line width=0.6pt,color=red] (12,5)--(15,5)--(15,5.5);
			\draw[line width=0.6pt,color=red] (19,5.5)--(19,5)--(22,5);
			\draw[line width=0.6pt,color=red] (24,5)--(27,5)--(27,5.5);
			\draw[line width=0.6pt,color=red] (31,5.5)--(31,3)--(29,3)--(29,0);
			\draw[line width=0.6pt,color=red] (41,0)--(41,3)--(39,3)--(39,5.5);
			\draw[line width=0.6pt,color=red] (43,5.5)--(43,5)--(46,5);
			\foreach \d in {0,12,24,36}{
				\draw[densely dotted, line width=1pt] (\d+5,6)--(\d+5,8);
				\draw[line width=0.2pt] (\d,0)--(\d,10)--(10+\d,10)--(10+\d,0)--cycle;
				\draw[dashed,line width=0.8pt,color=orange] (\d,0)--(\d,10)--(10+\d,10)--(10+\d,0)--cycle;
			}
		\end{tikzpicture}
		\captionof{figure}{Slalom T-metacell and possible solutions}
	\end{figure}
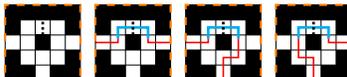
	
	\subsection{Nagareru Loop} 
	
	\textbf{Rules:} Draw a single directed loop passing between adjacent unshaded cell centres. It must pass straight through every black arrow in the direction of the arrow, and if it passes in front of a white arrow (with no shaded cells between), it must turn away from the arrow and travel that way for at least one unit.
	
	We can check that each arrow is satisfied in $O(n^3)$ time (iterating over each arrow). The gadget here can only be rotated and reflected over a side, ensuring that adjacent exits line up with each other. The black arrow ensures every gadget is visited. Note that there are six provided solutions, corresponding to each pair of exits and the direction of the loop between them. The genre was first published by Nikoli in Puzzle Communication Nikoli \#147, and satisfiability was proved NP-complete in \cite{ide2020moon}.
	
	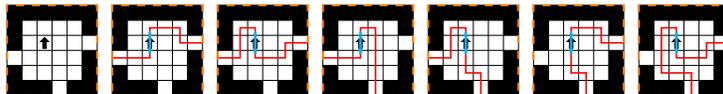
\begin{figure}
		\centering
		\begin{tikzpicture}[scale=0.1]
			\foreach \d in {0,14,...,84} {
				\fill[color=black] (\d,0)--(\d,4)--(2+\d,4)--(2+\d,2)--(6+\d,2)--(6+\d,0)--cycle;
				\fill[color=black] (\d+8,0)--(\d+8,2)--(\d+10,2)--(\d+10,6)--(\d+12,6)--(\d+12,0)--cycle;
				\fill[color=black]
				(\d,6)--(\d,12)--(\d+12,12)--(\d+12,8)--(\d+10,8)--(\d+10,10)--(\d+2,10)--(\d+2,6)--cycle;
				\foreach \x in {2,4,6,8,10} \draw[line width=0.2pt] (\x+\d,0)--(\x+\d,12);
				\foreach \y in {2,4,6,8,10} \draw[line width=0.2pt] (\d,\y)--(12+\d,\y);
				\fill[color=black] (4.7+\d,6.2)--(4.7+\d,7.1)--(4.3+\d,7.1)--(5+\d,7.8)--(5.7+\d,7.1)--(5.3+\d,7.1)--(5.3+\d,6.2)--cycle;
			}
			\foreach \d in {14,28,...,84}\draw[color=cyan,line width=1pt] (5+\d,5.5)--(5+\d,8.5);
			\draw[color=red,line width=0.6pt] (14,5)--(19,5)--(19,5.5);
			\draw[color=red,line width=0.6pt] (19,8.5)--(19,9)--(23,9)--(23,7)--(26,7);
			\draw[color=red,line width=0.6pt] (28,5)--(31,5)--(31,9)--(33,9)--(33,8.5);
			\draw[color=red,line width=0.6pt] (33,5.5)--(33,5)--(37,5)--(37,7)--(40,7);
			\draw[color=red,line width=0.6pt] (42,5)--(47,5)--(47,5.5);
			\draw[color=red,line width=0.6pt] (47,8.5)--(47,9)--(49,9)--(49,0);
			\draw[color=red,line width=0.6pt] (56,5)--(59,5)--(59,9)--(61,9)--(61,8.5);
			\draw[color=red,line width=0.6pt] (61,5.5)--(61,3)--(63,3)--(63,0);
			\draw[color=red,line width=0.6pt] (77,0)--(77,3)--(75,3)--(75,5.5);
			\draw[color=red,line width=0.6pt] (75,8.5)--(75,9)--(79,9)--(79,7)--(82,7);
			\draw[color=red,line width=0.6pt] (91,0)--(91,3)--(87,3)--(87,9)--(89,9)--(89,8.5);
			\draw[color=red,line width=0.6pt] (89,5.5)--(89,5)--(93,5)--(93,7)--(96,7);
			\foreach \d in {0,14,...,84} {
				\draw[line width=0.2pt] (\d,0)--(\d,12)--(12+\d,12)--(12+\d,0)--cycle;
				\draw[dashed,line width=0.8pt,color=orange] (\d,0)--(\d,12)--(12+\d,12)--(12+\d,0)--cycle;
			}
		\end{tikzpicture}
		\captionof{figure}{Nagareru Loop T-metacell and possible solutions}
	\end{figure}

	\subsection{Moon or Sun}
	
	\textbf{Rules:} Draw a single loop passing between adjacent cell centres. It must visit each region exactly once, either passing through all the moon cells and none of the sun cells, or vice versa. Furthermore, each time it crosses a region boundary it must alternate from visiting all suns to visiting all moons, or vice versa.
	
	We can check that in each room all cells of one type have been visited, and that the loop alternates appropriately, in $O(n^2)$ time. The gadget here can only be translated and rotated to ensure the exits line up appropriately. For this gadget, we also need to consider a checkerboard colouring of the grid of metacells, then on white cells we place copies of the gadget shown below whereas on black cells we place copies of that gadget but with suns replaced by moons and vice versa, to satisfy the last rule. The `dividing line' forces which of moon or sun can be visited in each region, and thus the marks of the other type form a wall similar to in Yajilin which cannot be passed through. The genre was first published by Nikoli in Puzzle Communication Nikoli \#154, and satisfiability was proved NP-complete in \cite{ide2020moon}.
	
	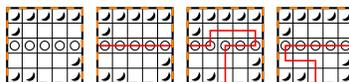
\begin{figure}[hbp]
		\centering
		\begin{tikzpicture}[scale=0.1]
			\foreach \d in {0,12,24,36} {
				\foreach \x in {2,4,6,8} \draw[line width=0.2pt] (\x+\d,0)--(\x+\d,10);
				\foreach \y in {2,4,6,8} \draw[line width=0.2pt] (\d,\y)--(\d+10,\y);
				\foreach \x in {1,3,5,7,9}{
					\draw (\x+\d,5) circle (0.6);
					\fill[color=black] (\x+\d+.7/2^.5,9+.7/2^.5) arc (45:-135:.7) arc (-90:0:.7*2^.5);
				}
				\fill[color=black] (1+\d+.7/2^.5,7+.7/2^.5) arc (45:-135:.7) arc (-90:0:.7*2^.5);
				\foreach \x in {1,3} \fill[color=black] (\x+\d+.7/2^.5,1+.7/2^.5) arc (45:-135:.7) arc (-90:0:.7*2^.5);
				\foreach \y in {1,3} \fill[color=black] (9+\d+.7/2^.5,\y+.7/2^.5) arc (45:-135:.7) arc (-90:0:.7*2^.5);
			}
			\draw[color=red,line width=0.6pt] (12,5)--(22,5);
			\draw[color=red,line width=0.6pt] (24,5)--(27,5)--(27,7)--(33,7)--(33,5)--(29,5)--(29,0);
			\draw[color=red,line width=0.6pt] (41,0)--(41,3)--(37,3)--(37,5)--(46,5);
			\foreach \d in {0,12,24,36} {
				\draw[line width=0.8pt] (\d,0)--(\d,10)--(\d+10,10)--(\d+10,0)--cycle;
				\draw[dashed,line width=0.8pt,color=orange] (\d,0)--(\d,10)--(\d+10,10)--(\d+10,0)--cycle;
			}
		\end{tikzpicture}
		\captionof{figure}{Moon or Sun T-metacell and possible solutions}
	\end{figure}

	\subsection{Country Road}
	
	\textbf{Rules:} Draw a single loop passing between adjacent cell centres and visiting each region once. A number within a region indicates the number of cells in that region through which the loop passes. Out of any pair of adjacent cells in different regions, at least one must be used by the loop.
	
	We can check that the loop passes through each region once (as well as that it passes through the appropriate number of squares), and the adjacent cell condition, both in $O(n^2)$ time. The gadget here can only be rotated and reflected over a side, ensuring that the exits line up appropriately. The numbered exterior regions force the path to traverse the entire length of the regions, hence creating the gadget's walls. The genre was first published by Nikoli in Puzzle Communication Nikoli \#65, and satisfiability was proved NP-complete in \cite{ishibashi2012np}.
	
	\begin{figure}[hbp]
		\centering
		\begin{tikzpicture}[scale=0.1,font=\sffamily]
			\foreach \d in {0,14,28,42} {
				\foreach \x in {2,4,6,8,10} \draw[line width=0.2pt] (\x+\d,0)--(\x+\d,12);
				\foreach \y in {2,4,8,10} \draw[line width=0.2pt] (\d,\y)--(12+\d,\y);
				\draw[line width=0.8pt] (\d,6)--(12+\d,6);
				\draw[line width=0.8pt] (6+\d,0)--(6+\d,10);
				\draw[line width=0.8pt] (2+\d,6)--(2+\d,10)--(10+\d,10)--(10+\d,8)--(8+\d,8)--(8+\d,6);
				\draw[line width=0.8pt] (4+\d,6)--(4+\d,4)--(2+\d,4)--(2+\d,2)--(6+\d,2);
				\draw[line width=0.8pt] (6+\d,4)--(8+\d,4)--(8+\d,2)--(10+\d,2)--(10+\d,6);
				\draw (0.5+\d,11.5) node {\scalebox{.2}{11}};
				\draw (0.4+\d,5.5) node {\scalebox{.2}{6}};
				\draw (6.4+\d,3.5) node {\scalebox{.2}{6}};
			}
			\foreach \d in {14,28,42} {
				\draw[color=cyan,line width=1pt] (\d+1,7.5)--(\d+1,11)--(\d+11,11)--(\d+11,7)--(\d+9.5,7);
				\draw[color=cyan,line width=1pt] (\d+2.5,5)--(\d+1,5)--(\d+1,1)--(\d+4.5,1);
				\draw[color=cyan,line width=1pt] (\d+7,2.5)--(\d+7,1)--(\d+11,1)--(\d+11,4.5);
			}
			\draw[color=red,line width=0.6pt] (14,7)--(15,7)--(15,7.5);
			\draw[color=red,line width=0.6pt] (16.5,5)--(17,5)--(17,9)--(21,9)--(21,7)--(23.5,7);
			\draw[color=red,line width=0.6pt] (18.5,1)--(19,1)--(19,5)--(21,5)--(21,2.5);
			\draw[color=red,line width=0.6pt] (25,4.5)--(25,5)--(26,5);
			\draw[color=red,line width=0.6pt] (28,7)--(29,7)--(29,7.5);
			\draw[color=red,line width=0.6pt] (30.5,5)--(31,5)--(31,9)--(33,9)--(33,3)--(35,3)--(35,2.5);
			\draw[color=red,line width=0.6pt] (32.5,1)--(33,1)--(33,0);
			\draw[color=red,line width=0.6pt] (39,4.5)--(39,5)--(35,5)--(35,7)--(37.5,7);
			\draw[color=red,line width=0.6pt] (43,7.5)--(43,7)--(45,7)--(45,9)--(47,9)--(47,5)--(44.5,5);
			\draw[color=red,line width=0.6pt] (46.5,1)--(47,1)--(47,0);
			\draw[color=red,line width=0.6pt] (49,2.5)--(49,7)--(51.5,7);
			\draw[color=red,line width=0.6pt] (53,4.5)--(53,5)--(54,5);
			\foreach \d in {0,14,28,42} {
				\draw[line width=0.8pt] (\d,0)--(\d,12)--(12+\d,12)--(12+\d,0)--cycle;
				\draw[dashed,line width=0.8pt,color=orange] (\d,0)--(\d,12)--(12+\d,12)--(12+\d,0)--cycle;
			}
		\end{tikzpicture}
		\captionof{figure}{Country Road T-metacell and possible solutions}
	\end{figure}
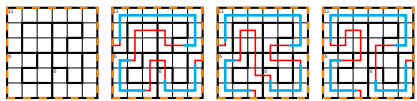

	\subsection{Onsen-meguri}
	
	\textbf{Rules:} Draw non-overlapping loops passing between adjacent cell centres. At least one loop passes through each region, but each loop visits a region at most once. Each loop must pass over exactly one circled cell, and all such cells are passed over. For each loop the number of cells the loop passes through in each region must be fixed, and if there is a number on the circled cell, equal to that.
	
	We can check these conditions by iterating over the loops and the circled cells, both in $O(n^2)$ time. The gadget here can be rotated, reflected and translated as necessary, though a circled 7 needs to be placed in the centre of exactly one gadget. Given the symmetry, the 4-loops are adjacent over a side to a different 4-loop or the edge of the puzzle, so we can show they must form small loops. Then, there aren't enough cells for the 7-loop to visit any of the 4-loops' regions and, given it must visit every other region, the gadget works as intended. The genre was first published by Nikoli in Puzzle Communication Nikoli \#155. No NP-completeness results could be found.
	
	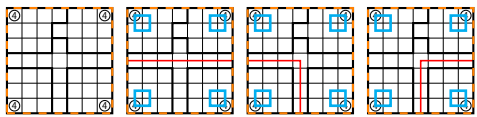
\begin{figure}[hbp]
		\centering
		\begin{tikzpicture}[scale=0.1,font=\sffamily]
			\foreach \d in {0,16,32,48} {
				\foreach \x in {2,4,...,12} \draw[line width=0.2pt] (\x+\d,0)--(\x+\d,14);
				\foreach \y in {2,4,...,12} \draw[line width=0.2pt] (\d,\y)--(14+\d,\y);
				\draw[line width=0.8pt] (\d,6)--(\d+6,6)--(\d+6,0);
				\draw[line width=0.8pt] (\d+8,0)--(\d+8,6)--(\d+14,6);
				\draw[line width=0.8pt] (\d,8)--(\d+6,8)--(\d+6,12)--(\d+8,12)--(\d+8,14);
				\draw[line width=0.8pt] (\d+6,10)--(\d+8,10)--(\d+8,8)--(\d+14,8);
				\foreach \x in {1,13} \foreach \y in {1,13}{
					\draw (\x+\d,\y) circle (0.7);
					\draw (\x+\d,\y) node {\scalebox{.4}{4}};
				}
			}
			\foreach \d in {16,32,48} {
				\foreach \x in {1,11} \foreach \y in {1,11} \draw[color=cyan,line width=1pt] (\x+\d,\y)--(\x+\d,\y+2)--(\x+\d+2,\y+2)--(\x+\d+2,\y)--cycle;
			}
			\draw[color=red,line width=0.6pt] (16,7)--(30,7);
			\draw[color=red,line width=0.6pt] (32,7)--(39,7)--(39,0);
			\draw[color=red,line width=0.6pt] (55,0)--(55,7)--(62,7);
			\foreach \d in {0,16,32,48} {
				\draw[line width=0.8pt] (\d,0)--(\d,14)--(14+\d,14)--(14+\d,0)--cycle;
				\draw[dashed,line width=0.8pt,color=orange] (\d,0)--(\d,14)--(14+\d,14)--(14+\d,0)--cycle;
			}
		\end{tikzpicture}
		\captionof{figure}{Onsen-meguri T-metacell and possible solutions}
	\end{figure}
	
	\subsection{Mejilink}
	
	\textbf{Rule:} Draw a loop along region boundaries so that the unused perimeter of each region is equal to that region's area. 
	
	We can check this condition in $O(n^2)$ time by iterating over the regions. The gadget can only be rotated and reflected over a side to align its exits. Note that every region can only have a perimeter of 10 being traversed, and if none of the inner edges of the outer regions are traversed then two of the other internal regions can't be satisfied. Thus, one 5-unit edge of each outer region has to be used, and so another one must also be used, which means the outside 2-, 8- and 10-unit edges cannot. The genre was first published by Nikoli in Puzzle Communication Nikoli \#126. No NP-completeness results could be found.
	
	\begin{figure}[hbp]
		\centering
		\begin{tikzpicture}[scale=0.1]
			\foreach \d in {0,22,44,66}{
				\draw[dash pattern=on 1pt off 0.5pt,line width=0.2pt] (\d,0)--(\d,20)--(20+\d,20)--(20+\d,0)--cycle;
				\draw[dash pattern=on 1pt off 0.5pt,line width=0.2pt] (\d,16)--(2+\d,16)--(2+\d,18)--(4+\d,18)--(4+\d,14);
				\draw[dash pattern=on 1pt off 0.5pt,line width=0.2pt] (12+\d,16)--(12+\d,14)--(14+\d,14)--(14+\d,12)--(12+\d,12)--(12+\d,8)--(14+\d,8)--(14+\d,2);
				\draw[dash pattern=on 1pt off 0.5pt,line width=0.2pt] (4+\d,0)--(4+\d,4)--(6+\d,4)--(6+\d,6)--(12+\d,6)--(12+\d,2)--(16+\d,2)--(16+\d,16)--(8+\d,16)--(8+\d,14)--(2+\d,14)--(2+\d,12)--(6+\d,12)--(6+\d,10);
				\draw[dash pattern=on 1pt off 0.5pt,line width=0.2pt] (8+\d,6)--(8+\d,8)--(4+\d,8)--(4+\d,10)--(8+\d,10)--(8+\d,12)--(10+\d,12)--(10+\d,10)--(12+\d,10);
				\draw[dash pattern=on 1pt off 0.5pt,line width=0.2pt] (16+\d,10)--(18+\d,10)--(18+\d,4)--(20+\d,4);
			}
			\draw[color=red,line width=0.6pt] (22,16)--(24,16)--(24,18)--(26,18)--(26,14)--(24,14)--(24,12)--(28,12)--(28,10)--(30,10)--(30,12)--(32,12)--(32,10)--(34,10)--(34,8)--(36,8)--(36,2)--(38,2)--(38,10)--(40,10)--(40,4)--(42,4);
			\draw[color=red,line width=0.6pt] (44,16)--(46,16)--(46,18)--(48,18)--(48,14)--(52,14)--(52,16)--(56,16)--(56,14)--(58,14)--(58,12)--(56,12)--(56,8)--(58,8)--(58,2)--(56,2)--(56,6)--(50,6)--(50,4)--(48,4)--(48,0);
			\draw[color=red,line width=0.6pt] (70,0)--(70,4)--(72,4)--(72,6)--(74,6)--(74,8)--(70,8)--(70,10)--(74,10)--(74,12)--(76,12)--(76,10)--(78,10)--(78,12)--(80,12)--(80,14)--(78,14)--(78,16)--(82,16)--(82,10)--(84,10)--(84,4)--(86,4);
			\foreach \d in {0,22,44,66}{
				\foreach \x in {0,2,...,20} \foreach \y in {0,2,...,20} \draw[fill=black] (\x+\d,\y) circle (3pt);
				\draw[dashed,line width=0.8pt,color=orange] (\d,0)--(\d,20)--(20+\d,20)--(20+\d,0)--cycle;
			}
		\end{tikzpicture}
		\captionof{figure}{Mejilink T-metacell and possible solutions}
	\end{figure}
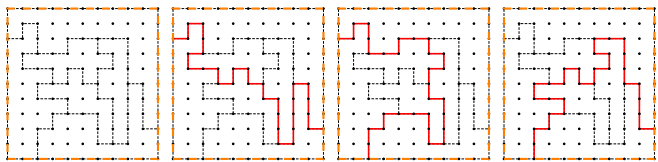
	
	\subsection{Double Back}
	
	\textbf{Rule:} Draw a loop passing between adjacent cell centres visiting every cell once and visiting each region exactly twice.
	
	We can check the region visits in $O(n^2)$ time by following the loop. The gadget here can be rotated, reflected and translated as necessary. Notice that the outer three regions each have four `dead end' cells which forces these cells to be the four entries/exits to each region, and thus the loop can only exit the gadget at the midpoints of the left, bottom and right sides. The genre was first published by Palmer Mebane at \cite{mebane2009double}. No NP-completeness results could be found.
	
	\begin{figure}[hbp]
		\centering
		\begin{tikzpicture}[scale=0.1]
			\foreach \d in {0,20,40,60}{
				\foreach \x in {2,4,...,16} \draw[line width=0.2pt] (\x+\d,0)--(\x+\d,18);
				\foreach \y in {2,4,...,16} \draw[line width=0.2pt] (\d,\y)--(18+\d,\y);
				\draw[line width=0.8pt] (\d,10)--(4+\d,10)--(4+\d,12)--(2+\d,12)--(2+\d,16)--(8+\d,16)--(8+\d,14)--(6+\d,14)--(6+\d,12)--(8+\d,12)--(8+\d,4)--(10+\d,4)--(10+\d,2);
				\draw[line width=0.8pt] (8+\d,0)--(8+\d,2)--(16+\d,2)--(16+\d,6)--(14+\d,6)--(14+\d,4)--(12+\d,4)--(12+\d,6)--(10+\d,6)--(10+\d,8)--(14+\d,8)--(14+\d,10)--(16+\d,10)--(16+\d,8)--(18+\d,8);
				\draw[line width=0.8pt] (8+\d,10)--(12+\d,10)--(12+\d,8);
				\draw[line width=0.8pt] (2+\d,10)--(2+\d,2)--(6+\d,2)--(6+\d,4)--(4+\d,4)--(4+\d,6)--(6+\d,6)--(6+\d,8)--(8+\d,8);
			}
			\foreach \d in {20,40,60}{
				\draw[line width=1pt,color=cyan] (1+\d,8.5)--(1+\d,1)--(7+\d,1)--(7+\d,3)--(8.5+\d,3);
				\draw[line width=1pt,color=cyan] (9.5+\d,1)--(17+\d,1)--(17+\d,7)--(15+\d,7)--(15+\d,8.5);
				\draw[line width=1pt,color=cyan] (3+\d,6.5)--(3+\d,3)--(5+\d,3)--(5+\d,5)--(7+\d,5)--(7+\d,6.5);
				\draw[line width=1pt,color=cyan] (11.5+\d,3)--(15+\d,3)--(15+\d,5)--(13+\d,5)--(13+\d,7)--(11.5+\d,7);
				\draw[line width=1pt,color=cyan] (3+\d,9.5)--(3+\d,11)--(1+\d,11)--(1+\d,17)--(9+\d,17)--(9+\d,15)--(11+\d,15)--(11+\d,17)--(12.5+\d,17);
				\draw[line width=1pt,color=cyan] (7+\d,9.5)--(7+\d,11)--(5+\d,11)--(5+\d,13)--(3+\d,13)--(3+\d,15)--(7+\d,15)--(7+\d,13)--(9+\d,13)--(9+\d,11)--(11+\d,11)--(11+\d,13)--(12.5+\d,13);
				\draw[line width=1pt,color=cyan] (15.5+\d,17)--(17+\d,17)--(17+\d,15.5);
				\draw[line width=1pt,color=cyan] (13+\d,9.5)--(13+\d,10.5);
				\draw[line width=1pt,color=cyan] (17+\d,9.5)--(17+\d,10.5);
				\draw[line width=1pt,color=cyan] (5+\d,7.5)--(5+\d,8.5);
				\draw[line width=1pt,color=cyan] (9.5+\d,5)--(10.5+\d,5);
				\draw[line width=0.6pt,color=red] (12.5+\d,17)--(15.5+\d,17);
				\draw[line width=0.6pt,color=red] (12.5+\d,13)--(13+\d,13)--(13+\d,15)--(15+\d,15)--(15+\d,11)--(13+\d,11)--(13+\d,10.5);
				\draw[line width=0.6pt,color=red] (17+\d,15.5)--(17+\d,10.5);
			}
			\draw[line width=0.6pt,color=red] (20,9)--(21,9)--(21,8.5);
			\draw[line width=0.6pt,color=red] (23,6.5)--(23,7)--(25,7)--(25,7.5);
			\draw[line width=0.6pt,color=red] (25,8.5)--(25,9)--(23,9)--(23,9.5);
			\draw[line width=0.6pt,color=red] (27,6.5)--(27,7)--(29,7)--(29,5)--(29.5,5);
			\draw[line width=0.6pt,color=red] (30.5,5)--(31,5)--(31,3)--(31.5,3);
			\draw[line width=0.6pt,color=red] (28.5,3)--(29,3)--(29,1)--(29.5,1);
			\draw[line width=0.6pt,color=red] (27,9.5)--(27,9)--(31,9)--(31,7)--(31.5,7);
			\draw[line width=0.6pt,color=red] (33,9.5)--(33,9)--(35,9)--(35,8.5);
			\draw[line width=0.6pt,color=red] (40,9)--(41,9)--(41,8.5);
			\draw[line width=0.6pt,color=red] (43,9.5)--(43,9)--(45,9)--(45,8.5);
			\draw[line width=0.6pt,color=red] (45,7.5)--(45,7)--(43,7)--(43,6.5);
			\draw[line width=0.6pt,color=red] (47,6.5)--(47,9.5);
			\draw[line width=0.6pt,color=red] (48.5,3)--(49,3)--(49,5)--(49.5,5);
			\draw[line width=0.6pt,color=red] (49,0)--(49,1)--(49.5,1);
			\draw[line width=0.6pt,color=red] (50.5,5)--(51,5)--(51,3)--(51.5,3);
			\draw[line width=0.6pt,color=red] (51.5,7)--(49,7)--(49,9)--(53,9)--(53,9.5);
			\draw[line width=0.6pt,color=red] (55,8.5)--(55,9)--(57,9)--(57,9.5);
			\draw[line width=0.6pt,color=red] (61,8.5)--(61,9)--(63,9)--(63,9.5);
			\draw[line width=0.6pt,color=red] (63,6.5)--(63,7)--(65,7)--(65,7.5);
			\draw[line width=0.6pt,color=red] (65,8.5)--(65,9)--(67,9)--(67,9.5);
			\draw[line width=0.6pt,color=red] (67,6.5)--(67,7)--(69,7)--(69,9)--(71,9)--(71,7)--(71.5,7);
			\draw[line width=0.6pt,color=red] (68.5,3)--(69,3)--(69,5)--(69.5,5);
			\draw[line width=0.6pt,color=red] (69,0)--(69,1)--(69.5,1);
			\draw[line width=0.6pt,color=red] (70.5,5)--(71,5)--(71,3)--(71.5,3);
			\draw[line width=0.6pt,color=red] (73,9.5)--(73,9)--(75,9)--(75,8.5);
			\draw[line width=0.6pt,color=red] (77,9.5)--(77,9)--(78,9);
			\draw[line width=0.6pt,color=red] (37,9.5)--(37,9)--(38,9);
			\foreach \d in {0,20,40,60}{
				\draw[line width=0.8pt] (\d,0)--(\d,18)--(18+\d,18)--(18+\d,0)--cycle;
				\draw[dashed,line width=0.8pt,color=orange] (\d,0)--(\d,18)--(18+\d,18)--(18+\d,0)--cycle;
			}
		\end{tikzpicture}
		\captionof{figure}{Double Back T-metacell and possible solutions}
	\end{figure}
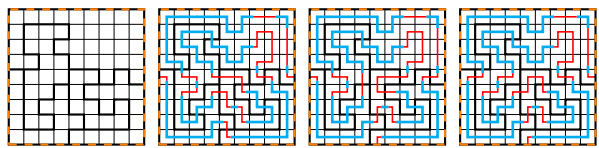
	
	\subsection{Scrin}
	
	\textbf{Rules:} Draw at least five non-overlapping grid-aligned rectangles that do not share any cell edge. Considering rectangles touching at a corner to be adjacent, the rectangles should form a single loop. All circled cells must be in a rectangle, and if they contain a number that must be the area of the rectangle.
	
	We can check that the areas of the rectangles are correct, that all circles are passed through and that valid loop is formed all in $O(n^2)$ time. The gadget can only be rotated and reflected over a side to keep the exit rectangles aligned. The 2s and the 3 must point into the gadget so that they don't hit another 2 or the edge of the grid. Then the only way for the gadgets to connect is a rectangle on one of the open 2s. Noting that no matter how this rectangle is placed it connects immediately to the next gadget, it can't connect to anywhere unintended and so this T-metacell works. The genre was first published by Nikoli in Puzzle Communication Nikoli \#165. No NP-completeness results could be found.
	
	\begin{figure}[hbp]
		\centering
		\begin{tikzpicture}[scale=0.1,font=\sffamily]
			\foreach \d in {20,40,60}{
				\fill[color=cyan!40!white] (7+\d,5)--(7+\d,3)--(9+\d,3)--(9+\d,5)--cycle;
				\fill[color=cyan!40!white] (13+\d,9)--(13+\d,7)--(15+\d,7)--(15+\d,9)--cycle;
				\fill[color=cyan!40!white] (3+\d,9)--(3+\d,11)--(5+\d,11)--(5+\d,9)--cycle;
				\fill[color=cyan!40!white] (9+\d,15)--(9+\d,13)--(11+\d,13)--(11+\d,15)--cycle;
			}
			\fill[color=red!40!white] (20,7)--(21,7)--(21,9)--(20,9)--cycle;
			\draw[color=red,line width=0.6pt] (20,7)--(21,7)--(21,9)--(20,9);
			\fill[color=red!40!white] (29,3)--(29,5)--(31,5)--(31,3)--cycle;
			\draw[color=red,line width=0.6pt] (29,3)--(31,3)--(31,5)--(27,5);
			\fill[color=red!40!white] (25,9)--(29,9)--(29,11)--(25,11)--cycle;
			\draw[color=red,line width=0.6pt] (23,9)--(29,9)--(29,11)--(25,11);
			\fill[color=red!40!white] (29,7)--(29,9)--(33,9)--(33,7)--cycle;
			\draw[color=red,line width=0.6pt] (33,7)--(29,7)--(29,9)--(35,9);
			\fill[color=red!40!white] (38,9)--(37,9)--(37,11)--(38,11)--cycle;
			\draw[color=red,line width=0.6pt] (38,9)--(37,9)--(37,11)--(38,11);
			\fill[color=red!40!white] (40,7)--(41,7)--(41,9)--(40,9)--cycle;
			\draw[color=red,line width=0.6pt] (40,7)--(41,7)--(41,9)--(40,9);
			\fill[color=red!40!white] (47,5)--(47,9)--(49,9)--(49,5)--cycle;
			\draw[color=red,line width=0.6pt] (47,5)--(47,9)--(49,9)--(49,3);
			\fill[color=red!40!white] (45,9)--(47,9)--(47,11)--(45,11)--cycle;
			\draw[color=red,line width=0.6pt] (43,9)--(47,9)--(47,11)--(45,11);
			\fill[color=red!40!white] (49,0)--(49,1)--(51,1)--(51,0)--cycle;
			\draw[color=red,line width=0.6pt] (49,0)--(49,1)--(51,1)--(51,0);
			\fill[color=red!40!white] (53,9)--(53,11)--(55,11)--(55,9)--cycle;
			\draw[color=red,line width=0.6pt] (53,7)--(53,11)--(55,11)--(55,9);
			\fill[color=red!40!white] (63,9)--(63,7)--(65,7)--(65,9)--cycle;
			\draw[color=red,line width=0.6pt] (63,9)--(63,7)--(65,7)--(65,11);
			\fill[color=red!40!white] (67,5)--(67,7)--(69,7)--(69,5)--cycle;
			\draw[color=red,line width=0.6pt] (67,5)--(67,7)--(69,7)--(69,3);
			\fill[color=red!40!white] (69,0)--(69,1)--(71,1)--(71,0)--cycle;
			\draw[color=red,line width=0.6pt] (69,0)--(69,1)--(71,1)--(71,0);
			\fill[color=red!40!white] (73,7)--(69,7)--(69,9)--(73,9)--cycle;
			\draw[color=red,line width=0.6pt] (73,7)--(69,7)--(69,9)--(75,9);
			\fill[color=red!40!white] (78,9)--(77,9)--(77,11)--(78,11)--cycle;
			\draw[color=red,line width=0.6pt] (78,9)--(77,9)--(77,11)--(78,11);
			\foreach \d in {20,40,60}{
				\draw[color=red,line width=0.6pt] (9+\d,13)--(11+\d,13);
				\foreach \x in {1,15} \foreach \y in {3,11} \draw[color=cyan,fill=cyan!40!white,line width=1pt] (\x+\d,\y)--(\x+\d+2,\y)--(\x+\d+2,\y+4)--(\x+\d,\y+4)--cycle;
				\foreach \x in {3,11} \draw[color=cyan,fill=cyan!40!white,line width=1pt] (\x+\d,1)--(\x+\d+4,1)--(\x+\d+4,3)--(\x+\d,3)--cycle;
				\draw[color=cyan,fill=cyan!40!white,line width=1pt] (3+\d,15)--(9+\d,15)--(9+\d,17)--(3+\d,17)--cycle;
				\draw[color=cyan,fill=cyan!40!white,line width=1pt] (11+\d,15)--(15+\d,15)--(15+\d,17)--(11+\d,17)--cycle;
				\draw[color=cyan,line width=1pt] (7+\d,5)--(7+\d,3)--(9+\d,3);
				\draw[color=cyan,line width=1pt] (3+\d,9)--(3+\d,11)--(5+\d,11);
				\draw[color=cyan,line width=1pt] (9+\d,13)--(9+\d,15)--(11+\d,15)--(11+\d,13);
				\draw[color=cyan,line width=1pt] (13+\d,7)--(15+\d,7)--(15+\d,9);
			}
			\foreach \d in {0,20,40,60}{
				\foreach \x in {1,3,...,17} \foreach \y in {1,3,...,17} \draw[fill=black] (\x+\d,\y) circle (3pt);
				\foreach \x in {2,16} \foreach \y in {4,14}{
					\draw[fill=white] (\x+\d,\y) circle (0.7);
					\draw (\x+\d,\y) node {\scalebox{0.4}{2}};
				}
				\foreach \y in {2,16} \foreach \x in {4,14} \draw[fill=white] (\x+\d,\y) circle (0.7);
				\foreach \y in {2,16} \draw (14+\d,\y) node {\scalebox{0.4}{2}};
				\draw[fill=white] (4+\d,10) circle (0.7);
				\draw[fill=white] (8+\d,4) circle (0.7);
				\draw[fill=white] (10+\d,14) circle (0.7);
				\draw[fill=white] (14+\d,8) circle (0.7);
				\draw (4+\d,2) node {\scalebox{0.4}{2}};
				\draw (4+\d,16) node {\scalebox{0.4}{3}};
				\draw[dashed,line width=0.8pt,color=orange] (\d,0)--(\d,18)--(18+\d,18)--(18+\d,0)--cycle;
			}
		\end{tikzpicture}
		\captionof{figure}{Scrin T-metacell and possible solutions}
	\end{figure}
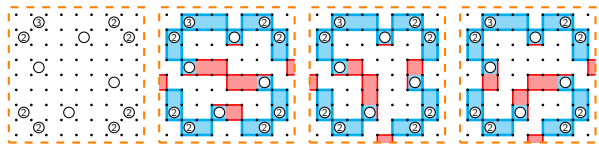
	
	\subsection{Geradeweg}
	
	\textbf{Rules:} Draw a loop passing between adjacent cell centres that visits every circled cell. Divide the loop into segments where it turns. Any numbers on circled cells indicate the lengths of all segments passing through that cell. At a circled corner, both segments must have equal length.
	
	We can check all conditions in $O(n^2)$ by going around the loop. The gadget can only be rotated and translated. Accounting for the walls of the entire grid or adjacent clues being different, the clues next to the corner cannot pass through the edge. Then, observe that in each metacell-edge row/column the corner clues alternate between 2 and 3 and there is only enough space to fit all of those segments in, they must be placed as shown. Finally, the central 3 clues must turn inwards to avoid the wall/another 3's loop edge, and the 2s turn in to avoid the wall/the adjacent 3's edge. The genre was first published by Robert Vollmert at \cite{vollmert2013geradeweg}. No NP-completeness results could be found.
	
	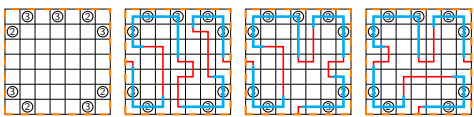
\begin{figure}[hbp]
		\centering
		\begin{tikzpicture}[scale=0.1,font=\sffamily]
			\foreach \d in {0,16,32,48}{
				\foreach \x in {2,4,...,12} \draw[line width=0.2pt] (\x+\d,0)--(\x+\d,14);
				\foreach \y in {2,4,...,12} \draw[line width=0.2pt] (\d,\y)--(14+\d,\y);
				\foreach \x in {3,11} \foreach \y in {1,13} \draw (\x+\d,\y) circle (0.7);
				\foreach \x in {1,13} \foreach \y in {3,11} \draw (\x+\d,\y) circle (0.7);
				\draw (7+\d,13) circle (0.7);
				\draw (3+\d,1) node {\scalebox{0.4}{2}};
				\draw (11+\d,1) node {\scalebox{0.4}{3}};
				\draw (3+\d,13) node {\scalebox{0.4}{3}};
				\draw (11+\d,13) node {\scalebox{0.4}{2}};
				\draw (7+\d,13) node {\scalebox{0.4}{3}};
				\draw (1+\d,3) node {\scalebox{0.4}{3}};
				\draw (1+\d,11) node {\scalebox{0.4}{2}};
				\draw (13+\d,3) node {\scalebox{0.4}{2}};
				\draw (13+\d,11) node {\scalebox{0.4}{3}};
			}
			\foreach \d in {16,32,48}{
				\draw[color=cyan,line width=1pt] (1+\d,6.5)--(1+\d,1)--(5+\d,1)--(5+\d,2.5);
				\draw[color=cyan,line width=1pt] (2.5+\d,9)--(1+\d,9)--(1+\d,13)--(7+\d,13)--(7+\d,7.5);
				\draw[color=cyan,line width=1pt] (7.5+\d,1)--(13+\d,1)--(13+\d,5)--(11.5+\d,5);
				\draw[color=cyan,line width=1pt] (9+\d,11.5)--(9+\d,13)--(13+\d,13)--(13+\d,7.5);
			}
			\draw[color=red,line width=0.6pt] (16,7)--(17,7)--(17,6.5);
			\draw[color=red,line width=0.6pt] (18.5,9)--(21,9)--(21,2.5);
			\draw[color=red,line width=0.6pt] (23,7.5)--(23,7)--(25,7)--(25,5)--(23,5)--(23,1)--(23.5,1);
			\draw[color=red,line width=0.6pt] (27.5,5)--(27,5)--(27,11)--(25,11)--(25,11.5);
			\draw[color=red,line width=0.6pt] (29,7.5)--(29,7)--(30,7);
			\draw[color=red,line width=0.6pt] (32,7)--(33,7)--(33,6.5);
			\draw[color=red,line width=0.6pt] (34.5,9)--(37,9)--(37,2.5);
			\draw[color=red,line width=0.6pt] (39,0)--(39,1)--(39.5,1);
			\draw[color=red,line width=0.6pt] (39,7.5)--(39,7)--(41,7)--(41,11.5);
			\draw[color=red,line width=0.6pt] (43.5,5)--(43,5)--(43,7)--(45,7)--(45,7.5);
			\draw[color=red,line width=0.6pt] (49,6.5)--(49,7)--(51,7)--(51,9)--(50.5,9);
			\draw[color=red,line width=0.6pt] (53,2.5)--(53,5)--(59.5,5);
			\draw[color=red,line width=0.6pt] (55,0)--(55,1)--(55.5,1);
			\draw[color=red,line width=0.6pt] (55,7.5)--(55,7)--(57,7)--(57,11.5);
			\draw[color=red,line width=0.6pt] (61,7.5)--(61,7)--(62,7);
			\foreach \d in {0,16,32,48}{
				\draw[line width=0.2pt] (\d,0)--(\d,14)--(14+\d,14)--(14+\d,0)--cycle;
				\draw[dashed,line width=0.8pt,color=orange] (\d,0)--(\d,14)--(14+\d,14)--(14+\d,0)--cycle;
			}
		\end{tikzpicture}
		\captionof{figure}{Geradeweg T-metacell and possible solutions}
	\end{figure}
	
	\subsection{Castle Wall}
	
	\textbf{Rules:} Outlined blocks of one or more cells are called walls. Draw a loop passing between adjacent non-wall cell centres. White walls and black walls can only be inside/outside the loop respectively. Walls may contain a number $n$ and an arrow pointing orthogonally: this means that the loop segments in the direction of the arrow have a total length of $n$.
	
	We can check these conditions in $O(n^3)$ time by iterating over the wall cells. The gadget here can be rotated, reflected and translated as necessary. The white and black walls force each gadget to be passed through. (We can also place a 0 and an arrow in an appropriate direction on every wall if we wish all walls to have clues.) Note that there are six provided solutions, corresponding to each pair of exits and the side which is inside the loop. The genre was first published by Palmer Mebane at \cite{mebane2009castle}. No NP-completeness results could be found.
	
	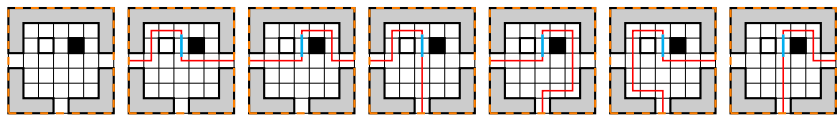
\begin{figure}[hbp]
		\centering
		\begin{tikzpicture}[scale=0.1]
			\foreach \d in {0,16,...,96}{
				\foreach \x in {2,4,...,12} \draw[line width=0.2pt] (\x+\d,0)--(\x+\d,14);
				\foreach \y in {2,4,...,12} \draw[line width=0.2pt] (\d,\y)--(14+\d,\y);
				\draw[line width=0.8pt,fill=gray!40!white] (\d,8)--(\d,14)--(\d+14,14)--(\d+14,8)--(\d+12,8)--(\d+12,12)--(\d+2,12)--(\d+2,8)--cycle;
				\draw[line width=0.8pt,fill=gray!40!white] (\d,0)--(\d,6)--(\d+2,6)--(\d+2,2)--(\d+6,2)--(\d+6,0)--cycle;
				\draw[line width=0.8pt,fill=gray!40!white] (\d+8,0)--(\d+8,2)--(\d+12,2)--(\d+12,6)--(\d+14,6)--(\d+14,0)--cycle;
				\draw[line width=0.8pt] (\d+4,8)--(\d+4,10)--(\d+6,10)--(\d+6,8)--cycle;
				\draw[line width=0.8pt,fill=black] (\d+8,8)--(\d+8,10)--(\d+10,10)--(\d+10,8)--cycle;
			}
			\foreach \d in {16,32,...,96} \draw [line width=1pt,color=cyan] (7+\d,7.5)--(7+\d,10.5);
			\draw[line width=0.6pt,color=red] (16,7)--(19,7)--(19,11)--(23,11)--(23,10.5);
			\draw[line width=0.6pt,color=red] (23,7.5)--(23,7)--(30,7);
			\draw[line width=0.6pt,color=red] (32,7)--(39,7)--(39,7.5);
			\draw[line width=0.6pt,color=red] (39,10.5)--(39,11)--(43,11)--(43,7)--(46,7);
			\draw[line width=0.6pt,color=red] (48,7)--(51,7)--(51,11)--(55,11)--(55,10.5);
			\draw[line width=0.6pt,color=red] (55,7.5)--(55,0);
			\draw[line width=0.6pt,color=red] (64,7)--(71,7)--(71,7.5);
			\draw[line width=0.6pt,color=red] (71,10.5)--(71,11)--(75,11)--(75,3)--(71,3)--(71,0);
			\draw[line width=0.6pt,color=red] (87,0)--(87,3)--(83,3)--(83,11)--(87,11)--(87,10.5);
			\draw[line width=0.6pt,color=red] (87,7.5)--(87,7)--(94,7);
			\draw[line width=0.6pt,color=red] (103,0)--(103,7.5);
			\draw[line width=0.6pt,color=red] (103,10.5)--(103,11)--(107,11)--(107,7)--(110,7);
			\foreach \d in {0,16,...,96}{
				\draw[line width=0.2pt] (\d,0)--(\d,14)--(14+\d,14)--(14+\d,0)--cycle;
				\draw[dashed,line width=0.8pt,color=orange] (\d,0)--(\d,14)--(14+\d,14)--(14+\d,0)--cycle;
			}
		\end{tikzpicture}
		\captionof{figure}{Castle Wall T-metacell and possible solutions}
	\end{figure}
	
	\subsection{Maxi Loop}
	
	\textbf{Rules:} Draw a loop passing between adjacent cell centres that visits every cell. A number within the region indicates the maximum number of cells the loop visits within the region on any one entrance to the region, across all such visits.
	
	We can check this condition in $O(n^2)$ time by following the loop. The gadget here can only be rotated and translated to ensure only one free path end lines up between adjacent cells. The clues in the outer regions force the entire length of the region to be traversed in one entry, thus creating the walls of the gadget. The genre was first published by Naoki Inaba at \cite{inaba2001max}. No NP-completeness results could be found.
	
	\begin{figure}[hbp]
		\centering
		\begin{tikzpicture}[scale=0.1,font=\sffamily]
			\foreach \d in {0,12,24,36}{
				\foreach \x in {2,4,6,8} \draw[line width=0.2pt] (\x+\d,0)--(\x+\d,10);
				\foreach \y in {2,4,6,8} \draw[line width=0.2pt] (\d,\y)--(10+\d,\y);
				\draw[line width=1pt] (\d+2,2)--(\d+2,8)--(\d+8,8)--(\d+8,2)--cycle;
				\draw[line width=1pt] (\d,4)--(\d+2,4);
				\draw[line width=1pt] (\d+8,6)--(\d+10,6);
				\draw[line width=1pt] (\d+6,0)--(\d+6,2);
				\draw (0.4+\d,9.5) node {\scalebox{.2}{8}};
				\draw (0.4+\d,3.5) node {\scalebox{.2}{4}};
				\draw (6.4+\d,1.5) node {\scalebox{.2}{4}};
			}
			\foreach \d in {12,24,36}{
				\draw[line width=1pt,color=cyan] (1+\d,5.5)--(1+\d,9)--(9+\d,9)--(9+\d,7.5);
				\draw[line width=1pt,color=cyan] (1+\d,2.5)--(1+\d,1)--(4.5+\d,1);
				\draw[line width=1pt,color=cyan] (7.5+\d,1)--(9+\d,1)--(9+\d,4.5);
				\draw[line width=1pt,color=cyan] (\d+3,5.5)--(\d+3,7)--(\d+4.5,7);
			}
			\draw[line width=0.6pt,color=red] (12,5)--(13,5)--(13,5.5);
			\draw[line width=0.6pt,color=red] (13,2.5)--(13,3)--(15,3)--(15,5.5);
			\draw[line width=0.6pt,color=red] (16.5,7)--(21,7)--(21,7.5);
			\draw[line width=0.6pt,color=red] (16.5,1)--(17,1)--(17,5)--(19,5)--(19,1)--(19.5,1);
			\draw[line width=0.6pt,color=red] (21,4.5)--(21,5)--(22,5);
			\draw[line width=0.6pt,color=red] (24,5)--(25,5)--(25,5.5);
			\draw[line width=0.6pt,color=red] (25,2.5)--(25,3)--(31,3)--(31,1)--(31.5,1);
			\draw[line width=0.6pt,color=red] (28.5,1)--(29,1)--(29,0);
			\draw[line width=0.6pt,color=red] (33,4.5)--(33,5)--(27,5)--(27,5.5);
			\draw[line width=0.6pt,color=red] (28.5,7)--(33,7)--(33,7.5);
			\draw[line width=0.6pt,color=red] (37,2.5)--(37,3)--(41,3)--(41,7)--(40.5,7);
			\draw[line width=0.6pt,color=red] (39,5.5)--(39,5)--(37,5)--(37,5.5);
			\draw[line width=0.6pt,color=red] (40.5,1)--(41,1)--(41,0);
			\draw[line width=0.6pt,color=red] (43.5,1)--(43,1)--(43,7)--(45,7)--(45,7.5);
			\draw[line width=0.6pt,color=red] (45,4.5)--(45,5)--(46,5);
			\foreach \d in {0,12,24,36}{
				\draw[line width=0.8pt] (\d,0)--(\d,10)--(10+\d,10)--(10+\d,0)--cycle;
				\draw[dashed,line width=0.8pt,color=orange] (\d,0)--(\d,10)--(10+\d,10)--(10+\d,0)--cycle;
			}
		\end{tikzpicture}
		\captionof{figure}{Maxi Loop T-metacell and possible solutions}
	\end{figure}
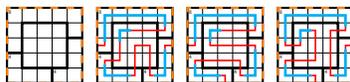
	
	\subsection{Mid-loop}
	
	\textbf{Rules:} Draw a loop passing between adjacent cell centres passing over all the black dots. Black dots indicate the midpoints of their corresponding loop segments (and cannot be on corners), though not all possible dots are necessarily given.
	
	We can check this condition in $O(n^2)$ time by following the loop. The gadget here can only be rotated and reflected over a side to ensure the exits line up. The outer clues force a unique winding wall. The genre was first published by Nikoli in Puzzle Communication Nikoli \#163. No NP-completeness results could be found.
	
	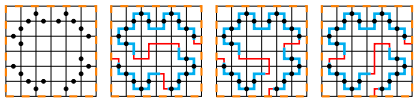
\begin{figure}[hbp]
		\centering
		\begin{tikzpicture}[scale=0.1]
			\foreach \d in {0,14,28,42}{
				\foreach \x in {2,4,6,8,10} \draw[line width=0.2pt] (\x+\d,0)--(\x+\d,12);
				\foreach \y in {2,4,6,8,10} \draw[line width=0.2pt] (\d,\y)--(12+\d,\y);
			}
			\foreach \d in {14,28,42}{
				\draw[color=cyan,line width=1pt] (\d+1,4.5)--(\d+1,3)--(\d+3,3)--(\d+3,1)--(\d+5,1)--(\d+5,3)--(\d+6.5,3);
				\draw[color=cyan,line width=1pt] (\d+7.5,1)--(\d+9,1)--(\d+9,3)--(\d+11,3)--(\d+11,5)--(\d+9,5)--(\d+9,6.5);
				\draw[color=cyan,line width=1pt] (\d+3,5.5)--(\d+3,7)--(\d+1,7)--(\d+1,9)--(\d+3,9)--(\d+3,11)--(\d+5,11)--(\d+5,9)--(\d+7,9)--(\d+7,11)--(\d+9,11)--(\d+9,9)--(\d+11,9)--(\d+11,7.5);
			}
			\draw[color=red,line width=0.6pt] (14,5)--(15,5)--(15,4.5);
			\draw[color=red,line width=0.6pt] (17,5.5)--(17,5)--(19,5)--(19,7)--(23,7)--(23,6.5);
			\draw[color=red,line width=0.6pt] (25,7.5)--(25,7)--(26,7);
			\draw[color=red,line width=0.6pt] (20.5,3)--(21,3)--(21,1)--(21.5,1);
			\draw[color=red,line width=0.6pt] (28,5)--(29,5)--(29,4.5);
			\draw[color=red,line width=0.6pt] (31,5.5)--(31,5)--(35,5)--(35,3)--(34.5,3);
			\draw[color=red,line width=0.6pt] (35,0)--(35,1)--(35.5,1);
			\draw[color=red,line width=0.6pt] (37,6.5)--(37,7)--(39,7)--(39,7.5);
			\draw[color=red,line width=0.6pt] (43,4.5)--(43,5)--(45,5)--(45,5.5);
			\draw[color=red,line width=0.6pt] (48.5,3)--(49,3)--(49,7)--(51,7)--(51,6.5);
			\draw[color=red,line width=0.6pt] (49,0)--(49,1)--(49.5,1);
			\draw[color=red,line width=0.6pt] (53,7.5)--(53,7)--(54,7);
			\foreach \d in {0,14,28,42}{
				\foreach \x in {3,5,9} \foreach \y in {2,10} \draw[fill=black] (\x+\d,\y) circle (7pt);
				\foreach \x in {4,8} \foreach \y in {1,11} \draw[fill=black] (\x+\d,\y) circle (7pt);
				\foreach \x in {2,10} \foreach \y in {3,9} \draw[fill=black] (\x+\d,\y) circle (7pt);
				\foreach \x in {1,11} \foreach \y in {4,8} \draw[fill=black] (\x+\d,\y) circle (7pt);
				\draw[fill=black] (2+\d,7) circle (7pt);
				\draw[fill=black] (10+\d,5) circle (7pt);
				\draw[fill=black] (7+\d,10) circle (7pt);
				\draw[line width=0.2pt] (\d,0)--(\d,12)--(12+\d,12)--(12+\d,0)--cycle;
				\draw[dashed,line width=0.8pt,color=orange] (\d,0)--(\d,12)--(12+\d,12)--(12+\d,0)--cycle;
			}
		\end{tikzpicture}
		\captionof{figure}{Mid-loop T-metacell and possible solutions}
	\end{figure}

	\clearpage
	
	\subsection{Balance Loop}
	
	\textbf{Rules:} Draw a loop passing between adjacent cell centres passing through all the pearls. Divide the loop into segments where it turns. A white pearl indicates the corner of two equal-length segments or the midpoint of a segment, and a black pearl indicates the corner of two unequal-length segments or a non-midpoint cell in some segment's interior. Additionally, numbers are given on some pearls, this represents the length of the segment passing straight through the pearl or the total length of the two segments touching the pearl.
	
	We can check these conditions in $O(n^2)$ time by following the loop. The gadget here can be rotated, reflected and translated as necessary. The 8s in the white pearls are adjacent to walls or more white 8s so must turn at these points, and then with the black 6s they form the walls of the gadget. (If the loop exited via the white 2, it would have too long a segment.) The genre was first published by Prasanna Seshadri at \cite{seshadri2015balance}. No NP-completeness results could be found.
	
	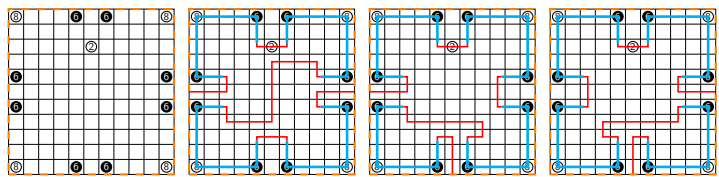
\begin{figure}[hbp]
		\centering
		\begin{tikzpicture}[scale=0.1,font=\sffamily]
			\foreach \d in {0,24,48,72}{
				\foreach \x in {2,4,...,20} \draw[line width=0.2pt] (\x+\d,0)--(\x+\d,22);
				\foreach \y in {2,4,...,20} \draw[line width=0.2pt] (\d,\y)--(22+\d,\y);
				\foreach \x in {1,21} \foreach \y in {1,21}{
					\draw (\x+\d,\y) circle (0.7);
					\draw (\x+\d,\y) node {\scalebox{0.4}{8}};
				}
				\foreach \x in {9,13} \foreach \y in {1,21}{
					\draw[fill=black] (\x+\d,\y) circle (0.7);
					\draw[color=white] (\x+\d,\y) node {\scalebox{0.4}{6}};
				}
				\foreach \y in {9,13} \foreach \x in {1,21}{
					\draw[fill=black] (\x+\d,\y) circle (0.7);
					\draw[color=white] (\x+\d,\y) node {\scalebox{0.4}{6}};
				}
				\draw (11+\d,17) circle (0.7);
				\draw (11+\d,17) node {\scalebox{0.4}{2}};
			}
			\foreach \d in {24,48,72}{
				\draw[color=cyan,line width=1pt] (4.5+\d,9)--(1+\d,9)--(1+\d,1)--(9+\d,1)--(9+\d,4.5);
				\draw[color=cyan,line width=1pt] (4.5+\d,13)--(1+\d,13)--(1+\d,21)--(9+\d,21)--(9+\d,17.5);
				\draw[color=cyan,line width=1pt] (13+\d,4.5)--(13+\d,1)--(21+\d,1)--(21+\d,9)--(17.5+\d,9);
				\draw[color=cyan,line width=1pt] (13+\d,17.5)--(13+\d,21)--(21+\d,21)--(21+\d,13)--(17.5+\d,13);
				\draw[color=red,line width=0.6pt] (9+\d,17.5)--(9+\d,17)--(13+\d,17)--(13+\d,17.5);
			}
			\draw[color=red,line width=0.6pt] (24,11)--(29,11)--(29,13)--(28.5,13);
			\draw[color=red,line width=0.6pt] (28.5,9)--(29,9)--(29,7)--(35,7)--(35,15)--(41,15)--(41,13)--(41.5,13);
			\draw[color=red,line width=0.6pt] (41.5,9)--(41,9)--(41,11)--(46,11);
			\draw[color=red,line width=0.6pt] (33,4.5)--(33,5)--(37,5)--(37,4.5);
			\draw[color=red,line width=0.6pt] (48,11)--(53,11)--(53,13)--(52.5,13);
			\draw[color=red,line width=0.6pt] (52.5,9)--(53,9)--(53,7)--(63,7)--(63,5)--(61,5)--(61,4.5);
			\draw[color=red,line width=0.6pt] (57,4.5)--(57,5)--(59,5)--(59,0);
			\draw[color=red,line width=0.6pt] (65.5,9)--(65,9)--(65,13)--(65.5,13);
			\draw[color=red,line width=0.6pt] (76.5,9)--(77,9)--(77,13)--(76.5,13);
			\draw[color=red,line width=0.6pt] (81,4.5)--(81,5)--(79,5)--(79,7)--(89,7)--(89,9)--(89.5,9);
			\draw[color=red,line width=0.6pt] (83,0)--(83,5)--(85,5)--(85,4.5);
			\draw[color=red,line width=0.6pt] (89.5,13)--(89,13)--(89,11)--(94,11);
			\foreach \d in {0,24,48,72}{
				\draw[line width=0.2pt] (\d,0)--(\d,22)--(22+\d,22)--(22+\d,0)--cycle;
				\draw[dashed,line width=0.8pt,color=orange] (\d,0)--(\d,22)--(22+\d,22)--(22+\d,0)--cycle;
			}
		\end{tikzpicture}
		\captionof{figure}{Balance Loop T-metacell and possible solutions}
	\end{figure}

	\subsection{Simple Loop}
	
	\textbf{Rule:} Draw a loop passing between adjacent unshaded cell centres that passes through all unshaded cells.
	
	This condition is trivial to check in $O(n^2)$ time. The gadget here can only be rotated and translated so the one-cell exits line up. Since all cells are visited, this ensures the gadget is visited. The genre's origin is unknown (and due to its simplicity, it may have been reinvented multiple times), though its satisfiability being NP-complete is a corollary of \cite{itai1982hamilton}. (Note that this gadget is the same as in Yajilin due to the similarity of the restrictions.)
	
	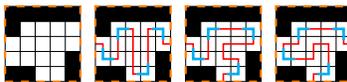
\begin{figure}[hbp]
		\centering
		\begin{tikzpicture}[scale=0.1]
			\foreach \d in {0,12,24,36} {
				\foreach \x in {2,4,6,8} \draw[line width=0.2pt] (\x+\d,0)--(\x+\d,10);
				\foreach \y in {2,4,6,8} \draw[line width=0.2pt] (\d,\y)--(10+\d,\y);
				\fill [color=black] (\d,6)--(\d,10)--(10+\d,10)--(10+\d,8)--(2+\d,8)--(2+\d,6)--cycle; 
				\fill [color=black] (\d,0)--(\d,2)--(4+\d,2)--(4+\d,0)--cycle; 
				\fill [color=black] (8+\d,0)--(8+\d,4)--(10+\d,4)--(10+\d,0)--cycle; 
			}
			\foreach \d in {12,24,36}{
				\draw[line width=1pt,color=cyan] (\d+1,4.5)--(\d+1,3)--(\d+2.5,3);
				\draw[line width=1pt,color=cyan] (\d+3,5.5)--(\d+3,7)--(\d+4.5,7);
				\draw[line width=1pt,color=cyan] (\d+5.5,1)--(\d+7,1)--(\d+7,2.5);
				\draw[line width=1pt,color=cyan] (\d+7.5,7)--(\d+9,7)--(\d+9,5.5);
			}
			\draw[line width=0.6pt,color=red] (12,5)--(13,5)--(13,4.5);
			\draw[line width=0.6pt,color=red] (14.5,3)--(15,3)--(15,5.5);
			\draw[line width=0.6pt,color=red] (16.5,7)--(17,7)--(17,1)--(17.5,1);
			\draw[line width=0.6pt,color=red] (19,2.5)--(19,7)--(19.5,7);
			\draw[line width=0.6pt,color=red] (21,5.5)--(21,5)--(22,5);
			\draw[line width=0.6pt,color=red] (24,5)--(25,5)--(25,4.5);
			\draw[line width=0.6pt,color=red] (26.5,3)--(27,3)--(27,5.5);
			\draw[line width=0.6pt,color=red] (28.5,7)--(31.5,7);
			\draw[line width=0.6pt,color=red] (33,5.5)--(33,5)--(29,5)--(29,3)--(31,3)--(31,2.5);
			\draw[line width=0.6pt,color=red] (29.5,1)--(29,1)--(29,0);
			\draw[line width=0.6pt,color=red] (41,0)--(41,1)--(41.5,1);
			\draw[line width=0.6pt,color=red] (43,2.5)--(43,5)--(41,5)--(41,3)--(38.5,3);
			\draw[line width=0.6pt,color=red] (37,4.5)--(37,5)--(39,5)--(39,5.5);
			\draw[line width=0.6pt,color=red] (40.5,7)--(43.5,7);
			\draw[line width=0.6pt,color=red] (45,5.5)--(45,5)--(46,5);
			\foreach \d in {0,12,24,36}{
				\draw[line width=0.2pt] (\d,0)--(\d,10)--(10+\d,10)--(10+\d,0)--cycle;
				\draw[dashed,line width=0.8pt,color=orange] (\d,0)--(\d,10)--(10+\d,10)--(10+\d,0)--cycle;
			}
		\end{tikzpicture}
		\captionof{figure}{Simple Loop T-metacell and possible solutions}
	\end{figure}
	
	\subsection{Detour}
	
	\textbf{Rules:} Draw a loop passing between adjacent cell centres passing through all cells. A number within a region indicates the number of times the loop turns within that region.
	
	We can check this in $O(n^2)$ by iterating over every cell going region-by-region. The gadget here can only be rotated and reflected over a side, ensuring that the free path ends line up. Consider pairs of 0 and 2 regions forming a $2\times3$ block. If there is a wall next to them, the lines must follow the stem of the L-shaped 0 region. Otherwise there is another 0-2 block (due to reflection) and notice if the lines go the other way, then the 2s force a small $2\times4$ loop to occur. Thus they must always follow the stem of the L. Then the 3 regions ensure the rest of the gadget is walled in, and the last 0 cell that the top entrance cannot be used. The genre was first published by Guowen Zhang at \cite{lmiwapc2ib}, and satisfiability was proved NP-complete in \cite{tang2020np}.
	
	\begin{figure}[hbp]
		\centering
		\begin{tikzpicture}[scale=0.1,font=\sffamily]
			\foreach \d in {0,16,32,48}{
				\foreach \x in {2,4,...,12} \draw[line width=0.2pt] (\x+\d,0)--(\x+\d,14);
				\foreach \y in {2,4,...,12} \draw[line width=0.2pt] (\d,\y)--(14+\d,\y);
				\draw[line width=0.8pt] (\d,12)--(10+\d,12)--(10+\d,10);
				\draw[line width=0.8pt] (2+\d,0)--(2+\d,10)--(12+\d,10);
				\draw[line width=0.8pt] (\d,6)--(4+\d,6)--(4+\d,12);
				\draw[line width=0.8pt] (4+\d,8)--(6+\d,8)--(6+\d,14);
				\draw[line width=0.8pt] (2+\d,4)--(8+\d,4)--(8+\d,0);
				\draw[line width=0.8pt] (4+\d,4)--(4+\d,2)--(14+\d,2);
				\draw[line width=0.8pt] (10+\d,2)--(10+\d,8)--(14+\d,8);
				\draw[line width=0.8pt] (10+\d,4)--(12+\d,4)--(12+\d,14);
				\draw (0.4+\d,5.5) node {\scalebox{0.2}{3}};
				\draw (0.4+\d,11.5) node {\scalebox{0.2}{0}};
				\draw (0.4+\d,13.5) node {\scalebox{0.2}{3}};
				\draw (2.4+\d,3.5) node {\scalebox{0.2}{0}};
				\draw (2.4+\d,9.5) node {\scalebox{0.2}{2}};
				\draw (4.4+\d,3.5) node {\scalebox{0.2}{2}};
				\draw (4.4+\d,9.5) node {\scalebox{0.2}{0}};
				\draw (6.4+\d,11.5) node {\scalebox{0.2}{2}};
				\draw (6.4+\d,13.5) node {\scalebox{0.2}{0}};
				\draw (8.4+\d,1.5) node {\scalebox{0.2}{3}};
				\draw (10.4+\d,3.5) node {\scalebox{0.2}{0}};
				\draw (10.4+\d,7.5) node {\scalebox{0.2}{2}};
				\draw (12.4+\d,13.5) node {\scalebox{0.2}{3}};
			}
			\foreach \d in {16,32,48}{
				\draw[color=cyan,line width=1pt] (1+\d,5.5)--(1+\d,13)--(3+\d,13)--(3+\d,9)--(7+\d,9)--(7+\d,11)
				--(5+\d,11)--(5+\d,13)--(13+\d,13)--(13+\d,11)--(9+\d,11)--(9+\d,9.5);
				\draw[color=cyan,line width=1pt] (3+\d,5.5)--(3+\d,7)--(4.5+\d,7);
				\draw[color=cyan,line width=1pt] (5+\d,4.5)--(5+\d,3)--(1+\d,3)--(1+\d,1)--(8.5+\d,1);
				\draw[color=cyan,line width=1pt] (7+\d,4.5)--(7+\d,3)--(8.5+\d,3);
				\draw[color=cyan,line width=1pt] (9.5+\d,5)--(11+\d,5)--(11+\d,1)--(13+\d,1)--(13+\d,8.5);
				\draw[color=cyan,line width=1pt] (9.5+\d,7)--(11+\d,7)--(11+\d,8.5);
			}
			\draw[color=red,line width=0.6pt] (16,5)--(17,5)--(17,5.5);
			\draw[color=red,line width=0.6pt] (19,5.5)--(19,5)--(21,5)--(21,4.5);
			\draw[color=red,line width=0.6pt] (23,4.5)--(23,5)--(25.5,5);
			\draw[color=red,line width=0.6pt] (24.5,1)--(25,1)--(25,3)--(24.5,3);
			\draw[color=red,line width=0.6pt] (20.5,7)--(25.5,7);
			\draw[color=red,line width=0.6pt] (25,9.5)--(25,9)--(27,9)--(27,8.5);
			\draw[color=red,line width=0.6pt] (29,8.5)--(29,9)--(30,9);
			\draw[color=red,line width=0.6pt] (32,5)--(33,5)--(33,5.5);
			\draw[color=red,line width=0.6pt] (35,5.5)--(35,5)--(37,5)--(37,4.5);
			\draw[color=red,line width=0.6pt] (36.5,7)--(39,7)--(39,4.5);
			\draw[color=red,line width=0.6pt] (40.5,1)--(41,1)--(41,0);
			\draw[color=red,line width=0.6pt] (40.5,3)--(41,3)--(41,5)--(41.5,5);
			\draw[color=red,line width=0.6pt] (41.5,7)--(41,7)--(41,9.5);
			\draw[color=red,line width=0.6pt] (43,8.5)--(43,9)--(45,9)--(45,8.5);
			\draw[color=red,line width=0.6pt] (49,5.5)--(49,5)--(51,5)--(51,5.5);
			\draw[color=red,line width=0.6pt] (52.5,7)--(53,7)--(53,4.5);
			\draw[color=red,line width=0.6pt] (55,4.5)--(55,7)--(57.5,7);
			\draw[color=red,line width=0.6pt] (56.5,1)--(57,1)--(57,0);
			\draw[color=red,line width=0.6pt] (56.5,3)--(57,3)--(57,5)--(57.5,5);
			\draw[color=red,line width=0.6pt] (57,9.5)--(57,9)--(59,9)--(59,8.5);
			\draw[color=red,line width=0.6pt] (61,8.5)--(61,9)--(62,9);
			\foreach \d in {0,16,32,48}{
				\draw[line width=0.8pt] (\d,0)--(\d,14)--(14+\d,14)--(14+\d,0)--cycle;
				\draw[dashed,line width=0.8pt,color=orange] (\d,0)--(\d,14)--(14+\d,14)--(14+\d,0)--cycle;
			}
		\end{tikzpicture}
		\captionof{figure}{Detour T-metacell and possible solutions}
	\end{figure}
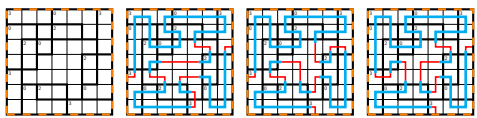

	\subsection{Haisu}
	
	\textbf{Rules:} Draw a path passing between adjacent cell centres from the S to the G, passing through all cells. The loop may only pass over a number $n$ on its $n^\text{th}$ visit to the region the number is in.
	
	We can check this in $O(n^2)$ time by simply following the path. The gadget here can only be rotated and translated to ensure only one free path end lines up between adjacent cells. Additionally, an S and a G must be placed in row 2, columns 2/3 in one of the gadgets (assuming the current orientation) to allow a path to be drawn. The clues in the outer regions force the entire length of the region to be traversed in one entry, thus creating the walls of the gadget. The genre was first published by William Hu at \cite{hu2017haisu}, and satisfiability was proved NP-complete in \cite{tang2020np}.
	
	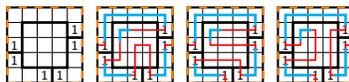
\begin{figure}[hbp]
		\centering
		\begin{tikzpicture}[scale=0.1,font=\sffamily]
			\foreach \d in {0,12,24,36}{
				\foreach \x in {2,4,6,8} \draw[line width=0.2pt] (\x+\d,0)--(\x+\d,10);
				\foreach \y in {2,4,6,8} \draw[line width=0.2pt] (\d,\y)--(10+\d,\y);
				\draw[line width=1pt] (\d+2,2)--(\d+2,8)--(\d+8,8)--(\d+8,2)--cycle;
				\draw[line width=1pt] (\d,4)--(\d+2,4);
				\draw[line width=1pt] (\d+8,6)--(\d+10,6);
				\draw[line width=1pt] (\d+6,0)--(\d+6,2);
				\draw (1+\d,3) node {\tiny 1};
				\draw (1+\d,5) node {\tiny 1};
				\draw (5+\d,1) node {\tiny 1};
				\draw (7+\d,1) node {\tiny 1};
				\draw (9+\d,5) node {\tiny 1};
				\draw (9+\d,7) node {\tiny 1};
			}
			\foreach \d in {12,24,36}{
				\draw[line width=1pt,color=cyan] (1+\d,5.5)--(1+\d,9)--(9+\d,9)--(9+\d,7.5);
				\draw[line width=1pt,color=cyan] (1+\d,2.5)--(1+\d,1)--(4.5+\d,1);
				\draw[line width=1pt,color=cyan] (7.5+\d,1)--(9+\d,1)--(9+\d,4.5);
				\draw[line width=1pt,color=cyan] (\d+3,5.5)--(\d+3,7)--(\d+4.5,7);
			}
			\draw[line width=0.6pt,color=red] (12,5)--(13,5)--(13,5.5);
			\draw[line width=0.6pt,color=red] (13,2.5)--(13,3)--(15,3)--(15,5.5);
			\draw[line width=0.6pt,color=red] (16.5,7)--(21,7)--(21,7.5);
			\draw[line width=0.6pt,color=red] (16.5,1)--(17,1)--(17,5)--(19,5)--(19,1)--(19.5,1);
			\draw[line width=0.6pt,color=red] (21,4.5)--(21,5)--(22,5);
			\draw[line width=0.6pt,color=red] (24,5)--(25,5)--(25,5.5);
			\draw[line width=0.6pt,color=red] (25,2.5)--(25,3)--(31,3)--(31,1)--(31.5,1);
			\draw[line width=0.6pt,color=red] (28.5,1)--(29,1)--(29,0);
			\draw[line width=0.6pt,color=red] (33,4.5)--(33,5)--(27,5)--(27,5.5);
			\draw[line width=0.6pt,color=red] (28.5,7)--(33,7)--(33,7.5);
			\draw[line width=0.6pt,color=red] (37,2.5)--(37,3)--(41,3)--(41,7)--(40.5,7);
			\draw[line width=0.6pt,color=red] (39,5.5)--(39,5)--(37,5)--(37,5.5);
			\draw[line width=0.6pt,color=red] (40.5,1)--(41,1)--(41,0);
			\draw[line width=0.6pt,color=red] (43.5,1)--(43,1)--(43,7)--(45,7)--(45,7.5);
			\draw[line width=0.6pt,color=red] (45,4.5)--(45,5)--(46,5);
			\foreach \d in {0,12,24,36}{
				\draw[line width=0.8pt] (\d,0)--(\d,10)--(10+\d,10)--(10+\d,0)--cycle;
				\draw[dashed,line width=0.8pt,color=orange] (\d,0)--(\d,10)--(10+\d,10)--(10+\d,0)--cycle;
			}
		\end{tikzpicture}
		\captionof{figure}{Haisu T-metacell and possible solutions}
	\end{figure}

	\clearpage

	\subsection{Reflect Link}
	
	\textbf{Rules:} Draw a loop passing between adjacent cell centres. The loop must path through all triangle cells (mirrors), `reflecting' off the mirror. The mirror might have a number in it, which indicates the total number of cells passed through by the segments emanating from the triangle. Additionally, the loop must cross itself at, and only at, cells marked with a plus.
	
	The gadget here can be reflected, rotated and translated as necessary. The external mirrors ensure the gadgets have the necessary walls. (We note that if we impose the restriction that each mirror must have a number, we can just place the numbers 6, 3, 3 and 3 in reading order without affecting the solutions.) The genre was first published by Nikoli in Puzzle Communication Nikoli \#106. No NP-completeness results could be found.
	
	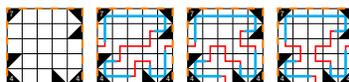
\begin{figure}[hbp]
		\centering
		\begin{tikzpicture}[scale=0.1,font=\sffamily]
			\foreach \d in {0,12,24,36}{
				\foreach \x in {2,4,6,8} \draw[line width=0.2pt] (\x+\d,0)--(\x+\d,10);
				\foreach \y in {2,4,6,8} \draw[line width=0.2pt] (\d,\y)--(10+\d,\y);
			}
			\foreach \d in {12,24,36}{
				\draw[color=cyan,line width=1pt] (1+\d,5.5)--(1+\d,9)--(9+\d,9)--(9+\d,7)--(7.5+\d,7);
				\draw[color=cyan,line width=1pt] (2.5+\d,3)--(1+\d,3)--(1+\d,1)--(4.5+\d,1);
				\draw[color=cyan,line width=1pt] (7+\d,2.5)--(7+\d,1)--(9+\d,1)--(9+\d,4.5);
			}
			\draw[color=red,line width=0.6pt] (12,5)--(13,5)--(13,5.5);
			\draw[color=red,line width=0.6pt] (14.5,3)--(15,3)--(15,5)--(19,5)--(19,7)--(19.5,7);
			\draw[color=red,line width=0.6pt] (16.5,1)--(17,1)--(17,3)--(19,3)--(19,2.5);
			\draw[color=red,line width=0.6pt] (21,4.5)--(21,5)--(22,5);
			\draw[color=red,line width=0.6pt] (24,5)--(25,5)--(25,5.5);
			\draw[color=red,line width=0.6pt] (26.5,3)--(27,3)--(27,5)--(29,5)--(29,3)--(31,3)--(31,2.5);
			\draw[color=red,line width=0.6pt] (28.5,1)--(29,1)--(29,0);
			\draw[color=red,line width=0.6pt] (31.5,7)--(31,7)--(31,5)--(33,5)--(33,4.5);
			\draw[color=red,line width=0.6pt] (37,5.5)--(37,5)--(39,5)--(39,3)--(38.5,3);
			\draw[color=red,line width=0.6pt] (40.5,1)--(41,1)--(41,0);
			\draw[color=red,line width=0.6pt] (43,2.5)--(43,3)--(41,3)--(41,5)--(43,5)--(43,7)--(43.5,7);
			\draw[color=red,line width=0.6pt] (45,4.5)--(45,5)--(46,5);
			\foreach \d in {0,12,24,36}{
				\fill[color=black] (\d,0)--(\d,2)--(2+\d,0)--cycle;
				\fill[color=black] (\d,2)--(\d,4)--(2+\d,4)--cycle;
				\fill[color=black] (\d,8)--(\d,10)--(2+\d,10)--cycle;
				\fill[color=black] (6+\d,0)--(6+\d,2)--(8+\d,0)--cycle;
				\fill[color=black] (8+\d,0)--(10+\d,2)--(10+\d,0)--cycle;
				\fill[color=black] (8+\d,6)--(10+\d,6)--(10+\d,8)--cycle;
				\fill[color=black] (8+\d,10)--(10+\d,10)--(10+\d,8)--cycle;
				\draw[color=white] (0.6+\d,0.7) node {\scalebox{0.3}{4}};
				\draw[color=white] (0.6+\d,9.3) node {\scalebox{0.3}{7}};
				\draw[color=white] (9.4+\d,0.7) node {\scalebox{0.3}{4}};
				\draw[line width=0.2pt] (\d,0)--(\d,10)--(10+\d,10)--(10+\d,0)--cycle;
				\draw[dashed,line width=0.8pt,color=orange] (\d,0)--(\d,10)--(10+\d,10)--(10+\d,0)--cycle;
			}
		\end{tikzpicture}
		\captionof{figure}{Reflect Link T-metacell and possible solutions}
	\end{figure}
	
	\subsection{Pipelink, Loop Special and Pipelink Returns}
	
	\textbf{Pipelink rules:} Draw a loop passing between adjacent cell centres passing through all cells. The loop must follow exactly the provided path on cells with shown path segments. The loop may cross itself at any empty cell or clued + cell (though it may not overlap itself except for at a crossing point).
	
	We can trivially check these conditions in $O(n^2)$ time. The gadget here can be rotated, reflect and translated as necessary. The given clue cells form the gadget's walls. The genre was first published by Nikoli in Puzzle Communication Nikoli \#45, and satisfiability was proved NP-complete in \cite{uejima2017complexity}.
	
	\textbf{Additional rules:} Loop Special allows for multiple loops each of which must contain at least one circled cell and all circled cells with a particular number, and also prevents lines crossing on a circled cell. Pipelink Returns adds the rule that the loop may only cross itself on a circled cell or a clued + cell, and the loop must pass straight through all circled cells.
	
	Both Loop Special and Pipelink Returns can also employ the same gadget, and all of these extra conditions can be checked in $O(n^2)$ time. For Loop Special we can replace all corners with a circled 1 to ensure there is exactly one loop. Loop Special and Pipelink Returns were first published by Nikoli in Puzzle Communication Nikoli \#57 and \#110 respectively. The proof from \cite{uejima2017complexity} carries over to Loop Special by the same addition, but not to Pipelink Returns as there are unclued crossing cells in the reduction which aren't in fixed locations. No other NP-completeness results could be found for Pipelink Returns.
	
	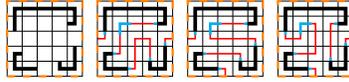
\begin{figure}
		\centering
		\begin{tikzpicture}[scale=0.1]
			\foreach \d in {12,24,36}{
				\draw[line width=1pt,color=cyan] (\d+1,6)--(\d+1,5.5);
				\draw[line width=1pt,color=cyan] (\d+8,7)--(\d+7.5,7);
				\draw[line width=1pt,color=cyan] (\d+2,3)--(\d+2.5,3);
				\draw[line width=1pt,color=cyan] (\d+4,1)--(\d+4.5,1);
				\draw[line width=1pt,color=cyan] (\d+7,2)--(\d+7,2.5);
				\draw[line width=1pt,color=cyan] (\d+9,4)--(\d+9,4.5);
			}
			\foreach \d in {0,12,24,36}{
				\foreach \x in {2,4,6,8} \draw[line width=0.2pt] (\x+\d,0)--(\x+\d,10);
				\foreach \y in {2,4,6,8} \draw[line width=0.2pt] (\d,\y)--(10+\d,\y);
				\draw[line width=1.4pt] (\d+1,6)--(\d+1,9)--(\d+9,9)--(\d+9,7)--(\d+8,7);
				\draw[line width=1.4pt] (\d+2,3)--(\d+1,3)--(\d+1,1)--(\d+4,1);
				\draw[line width=1.4pt] (\d+7,2)--(\d+7,1)--(\d+9,1)--(\d+9,4);
			}
			\foreach \d in {12,24,36}\draw[line width=1pt,color=cyan] (\d+3,5.5)--(\d+3,7)--(\d+4.5,7);
			\draw[color=red,line width=0.6pt] (12,5)--(13,5)--(13,5.5);
			\draw[color=red,line width=0.6pt] (14.5,3)--(15,3)--(15,5.5);
			\draw[color=red,line width=0.6pt] (16.5,7)--(19.5,7);
			\draw[color=red,line width=0.6pt] (16.5,1)--(17,1)--(17,5)--(19,5)--(19,2.5);
			\draw[color=red,line width=0.6pt] (21,4.5)--(21,5)--(22,5);
			\draw[color=red,line width=0.6pt] (24,5)--(25,5)--(25,5.5);
			\draw[color=red,line width=0.6pt] (26.5,3)--(31,3)--(31,2.5);
			\draw[color=red,line width=0.6pt] (28.5,1)--(29,1)--(29,0);
			\draw[color=red,line width=0.6pt] (27,5.5)--(27,5)--(33,5)--(33,4.5);
			\draw[color=red,line width=0.6pt] (28.5,7)--(31.5,7);
			\draw[color=red,line width=0.6pt] (37,5.5)--(37,5)--(39,5)--(39,5.5);
			\draw[color=red,line width=0.6pt] (40.5,7)--(41,7)--(41,3)--(38.5,3);
			\draw[color=red,line width=0.6pt] (40.5,1)--(41,1)--(41,0);
			\draw[color=red,line width=0.6pt] (43,2.5)--(43,7)--(43.5,7);
			\draw[color=red,line width=0.6pt] (45,4.5)--(45,5)--(46,5);
			\foreach \d in {0,12,24,36}{
				\draw[line width=0.2pt] (\d,0)--(\d,10)--(10+\d,10)--(10+\d,0)--cycle;
				\draw[dashed,line width=0.8pt,color=orange] (\d,0)--(\d,10)--(10+\d,10)--(10+\d,0)--cycle;
			}
		\end{tikzpicture}
		\captionof{figure}{Pipelink (and variants) T-metacell and possible solutions}
	\end{figure}
	
	\subsection{Icebarn}
	
	\textbf{Rules:} Draw a path passing between adjacent cell centres from the IN to the OUT. The path must pass over all arrows in the direction indicated. Coloured cells indicate `ice': any segment of path passing over ice must go straight, however the path may also cross itself only on ice cells. Connected ice cells are grouped into `icebarns', each of which must be passed through at least once.
	
	We can check all these conditions by iterating along in the path and along the arrows and icebarns, all in $O(n^2)$ time. The gadget here can only be rotated and reflected over a side, ensuring the exits of adjacent gadgets align. The arrows provide a wall to the gadget, though notice the ice cells provide `gaps' in the wall. In addition, instead of one of the gadgets in the corner of the grid we instead need an IN and OUT pointing into the grid from the sides. Note that there are six provided solutions, corresponding to each pair of exits and the direction of the loop between them. The genre was first published by Nikoli in Puzzle Communication Nikoli \#108. No NP-completeness results could be found.
	
	\begin{figure}[hbp]
		\centering
		\begin{tikzpicture}[scale=0.1]
			\foreach \d in {0,14,28,...,84}{
				\foreach \x in {2,4,6,8,10} \draw[line width=0.2pt] (\x+\d,0)--(\x+\d,12); 
				\foreach \y in {2,4,6,8,10} \draw[line width=0.2pt] (\d,\y)--(12+\d,\y); 
				\foreach \x in {0,6} \draw[line width=0.8pt,fill=cyan!30!white] (\x+\d,6)--(\x+\d+2,6)--(\x+\d+2,8)--(\x+\d,8)--cycle;
				\draw[line width=0.8pt,fill=cyan!30!white] (10+\d,6)--(12+\d,6)--(12+\d,4)--(10+\d,4)--cycle;
			}
			\draw[color=red,line width=0.6pt] (14,7)--(17,7)--(17,2.5);
			\draw[color=red,line width=0.6pt] (21,2.5)--(21,3)--(19,3)--(19,7)--(23,7)--(23,5)--(26,5);
			\draw[color=red,line width=0.6pt] (28,7)--(31,7)--(31,9)--(35,9)--(35,2.5);
			\draw[color=red,line width=0.6pt] (31,2.5)--(31,3)--(33,3)--(33,7)--(37,7)--(37,5)--(40,5);
			\draw[color=red,line width=0.6pt] (42,7)--(45,7)--(45,2.5);
			\draw[color=red,line width=0.6pt] (47,0)--(47,7)--(51,7)--(51,3)--(49,3)--(49,2.5);
			\draw[color=red,line width=0.6pt] (56,7)--(65,7)--(65,3)--(63,3)--(63,2.5);
			\draw[color=red,line width=0.6pt] (59,2.5)--(59,3)--(61,3)--(61,0);
			\draw[color=red,line width=0.6pt] (75,0)--(75,5)--(77,5)--(77,9)--(73,9)--(73,2.5);
			\draw[color=red,line width=0.6pt] (77,2.5)--(77,3)--(79,3)--(79,5)--(82,5);
			\draw[color=red,line width=0.6pt] (89,0)--(89,3)--(91,3)--(91,2.5);
			\draw[color=red,line width=0.6pt] (87,2.5)--(87,9)--(91,9)--(91,5)--(96,5);
			\foreach \d in {14,28,...,84} \draw[color=cyan,line width=1pt] (\d+3,2.5)--(\d+3,1)--(\d+1,1)--(\d+1,11)--(\d+11,11)--(\d+11,1)--(\d+7,1)--(\d+7,2.5);
			\foreach \d in {0,14,28,...,84}{
				\foreach \y in {2,4,6,8,10}{
					\draw (\d+1,\y-0.5)--(\d+1,\y+0.25);
					\fill (\d+0.75,\y+0.25)--(\d+1.25,\y+0.25)--(\d+1,\y+0.5)--cycle;
				}
				\foreach \y in {2,4,6,8,10}{
					\draw (\d+11,\y+0.5)--(\d+11,\y-0.25);
					\fill (\d+10.75,\y-0.25)--(\d+11.25,\y-0.25)--(\d+11,\y-0.5)--cycle;
				}
				\foreach \x in {2,4,6,8,10}{
					\draw (\d+\x-0.5,11)--(\d+\x+0.25,11);
					\fill (\d+\x+0.25,10.75)--(\d+\x+0.25,11.25)--(\d+\x+0.5,11)--cycle;
				}
				\foreach \x in {2,8,10}{
					\draw (\d+\x+0.5,1)--(\d+\x-0.25,1);
					\fill (\d+\x-0.25,0.75)--(\d+\x-0.25,1.25)--(\d+\x-0.5,1)--cycle;
				}
				\draw (3+\d,1.75)--(3+\d,2.5);
				\fill (2.75+\d,1.75)--(3.25+\d,1.75)--(3+\d,1.5)--cycle;
				\draw (7+\d,1.5)--(7+\d,2.25);
				\fill (6.75+\d,2.25)--(7.25+\d,2.25)--(7+\d,2.5)--cycle;
				\draw[line width=0.2pt] (\d,0)--(\d,12)--(12+\d,12)--(12+\d,0)--cycle; 
				\draw[line width=0.8pt] (\d,6)--(\d,8);
				\draw[line width=0.8pt] (\d+12,4)--(\d+12,6);
				\draw[dashed,line width=0.8pt,color=orange] (\d,0)--(\d,12)--(12+\d,12)--(12+\d,0)--cycle; 
			}
		\end{tikzpicture}
		\captionof{figure}{Icebarn T-metacell and possible solutions}
	\end{figure}
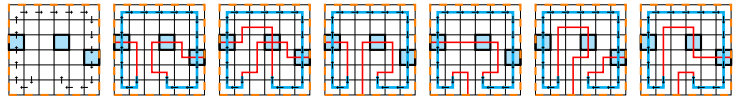
	
	\subsection{Barns}
	
	\textbf{Rules:} Draw a loop passing between adjacent cell centres visiting every cell. The loop may not cross any bolded edges. Coloured cells indicate `ice': any segment of path passing over ice must go straight, however the path may also cross itself only on ice cells.
	
	We can check all these conditions in $O(n^2)$ time by iterating over the cells and edges. The gadget here can be rotated, reflected and translated, and it trivially works. The genre was first published by Nikoli in Puzzle Communication Nikoli \#114. No NP-completeness results could be found.
	
	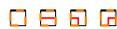
\begin{figure}[hbp]
		\centering
		\begin{tikzpicture}[scale=0.1]
			\foreach \d in {0,4,8,12}\draw[line width=0.8pt] (\d,2)--(\d+2,2);
			\draw[line width=0.6pt,color=red] (4,1)--(6,1);
			\draw[line width=0.6pt,color=red] (8,1)--(9,1)--(9,0);
			\draw[line width=0.6pt,color=red] (13,0)--(13,1)--(14,1);
			\foreach \d in {0,4,8,12}{
				\draw[line width=0.2pt] (\d,0)--(\d,2)--(2+\d,2)--(2+\d,0)--cycle; 
				\draw[dashed,line width=0.8pt,color=orange] (\d,0)--(\d,2)--(2+\d,2)--(2+\d,0)--cycle; 
			}
		\end{tikzpicture}
		\captionof{figure}{Barns T-metacell and possible solutions}
	\end{figure}
	
	\section{Limitations of the T-metacell gadget}
	
	The twenty-five genres listed above are amenable to a T-metacell gadget, but out of the remaining seven genres in the loop row of \cite{puzzlink}, we were only able to find T-metacell-inspired gadgets for four of them. The sections that follow use the same format as above, though we also discuss why a simple application of a T-metacell gadget is infeasible (regardless of whether a gadget is shown).
	
	\subsection{Angle Loop}
	
	\textbf{Rules:} Draw a loop via straight lines between shapes, visiting all the shapes. At every triangle, square or pentagon respectively, the loop must make an acute, right or obtuse angle. The loop may not cross itself.
	
	We can check this condition in $O(n^4)$ time by iterating around the loop in $O(n^2)$ time to check the angles and then $O(n^4)$ time to ensure no segments intersect. The issue with constructing a T-metacell gadget is that the lines can be at any angle and join arbitrarily distant points, so it is hard to make an enclosed gadget of a fixed size since arbitrarily distant points might be used.
	
	To fix this, in the gadget overleaf, each of the outer four rows/columns on each side (indicated by wavy lines) are allowed to vary in position among $2c^3+1$ different adjacent rows/columns, where $c\le27n^2$ is the total number of clues. Every square lies on one of those columns, and we would like to vary them so that each square can only be the centre of an angle among two points one of which is in the same row and other of which is in the same column as that square. Thus for any of the squares $Q$ in that row/column, noting that for every pair of other shapes $P,R$ not in the same row or column as $Q$ (clearly less than $c^2$ pairs), the circle with diameter $PR$ intersects $Q$'s column/row in at most 2 places, so each such $Q$ eliminates at most $2c^2$ potential row/column positions. Since there are at most $c$ shapes in the row/column, if we have $2c^3+1$ positions it guarantees we can choose a valid positioning of the row/column so as not to form an unintentional right angle. Then, iterating over all these ambiguous rows and columns we create a larger grid (with only polynomial blowup, so this can still be created in polynomial time) which satisfies the conditions we need.
	
	The gadget can be rotated, reflected and translated before the pertubation above is applied. To show the gadget works, notice that, by the above, in each squiggly row/column we can assume all the squares have exactly one segment passing along the row/column. Then iterating along the column from the side of the grid (taking advantage of the tiling pattern) we can show that all the cyan edges must exist. The genre was first published with only obtuse-angle clues by Maho Yokota (although no source could be found); the current version was published by the same author in Toketa?\ \#2. No NP-completeness results could be found.
	
	\begin{figure}
		\centering
		\begin{tikzpicture}[scale=0.1]
			\foreach \dx in {0,40} \foreach \dy in {0,40}{
				\foreach \x in {9,11,...,29} \draw[line width=0.2pt,color=gray] (\x+\dx,\dy)--(\x+\dx,\dy+38);
				\foreach \y in {9,11,...,29} \draw[line width=0.2pt,color=gray] (\dx,\y+\dy)--(\dx+38,\y+\dy);
				\foreach \x in {1,3,5,7,31,33,35,37} \draw[decorate, decoration={snake,amplitude=0.4,segment length=6},line width=0.2pt,color=gray] (\x+\dx,\dy)--(\x+\dx,\dy+38);
				\foreach \y in {1,3,5,7,31,33,35,37} \draw[decorate, decoration={snake,amplitude=0.4,segment length=6},line width=0.2pt,color=gray] (\dx,\y+\dy)--(\dx+38,\y+\dy);
			}
			\draw[color=cyan,line width=1pt] (1,19)--(1,3)--(19,3);
			\draw[color=cyan,line width=1pt] (5,19)--(5,7)--(19,7);
			\draw[color=cyan,line width=1pt] (19,1)--(35,1)--(35,19);
			\draw[color=cyan,line width=1pt] (19,5)--(31,5)--(31,19);
			\draw[color=cyan,line width=1pt] (3,19)--(3,35)--(37,35)--(37,19);
			\draw[color=cyan,line width=1pt] (7,19)--(7,31)--(33,31)--(33,19);
			\draw[color=cyan,line width=1pt] (41,19)--(41,3)--(59,3);
			\draw[color=cyan,line width=1pt] (45,19)--(45,7)--(59,7);
			\draw[color=cyan,line width=1pt] (59,1)--(75,1)--(75,19);
			\draw[color=cyan,line width=1pt] (59,5)--(71,5)--(71,19);
			\draw[color=cyan,line width=1pt] (43,19)--(43,35)--(77,35)--(77,19);
			\draw[color=cyan,line width=1pt] (47,19)--(47,31)--(73,31)--(73,19);
			\draw[color=cyan,line width=1pt] (41,59)--(41,43)--(59,43);
			\draw[color=cyan,line width=1pt] (45,59)--(45,47)--(59,47);
			\draw[color=cyan,line width=1pt] (59,41)--(75,41)--(75,59);
			\draw[color=cyan,line width=1pt] (59,45)--(71,45)--(71,59);
			\draw[color=cyan,line width=1pt] (43,59)--(43,75)--(77,75)--(77,59);
			\draw[color=cyan,line width=1pt] (47,59)--(47,71)--(73,71)--(73,59);
			\draw[color=red,line width=0.6pt] (40,59)--(41,59);
			\draw[color=red,line width=0.6pt] (43,59)--(45,59);
			\draw[color=red,line width=0.6pt] (59,41)--(59,43);
			\draw[color=red,line width=0.6pt] (59,45)--(59,47);
			\draw[color=red,line width=0.6pt] (73,59)--(75,59);
			\draw[color=red,line width=0.6pt] (77,59)--(78,59);
			\draw[rounded corners=.1mm,color=red,line width=0.6pt] (47,59)--(55,59)--(53,63)--(59,53)--(59,61)--(61,51)--(69,63)--(67,59)--(71,59);
			\draw[color=red,line width=0.6pt] (0,19)--(1,19);
			\draw[color=red,line width=0.6pt] (3,19)--(5,19);
			\draw[color=red,line width=0.6pt] (19,0)--(19,1);
			\draw[color=red,line width=0.6pt] (19,3)--(19,5);
			\draw[color=red,line width=0.6pt] (31,19)--(33,19);
			\draw[color=red,line width=0.6pt] (35,19)--(37,19);
			\draw[rounded corners=.1mm,color=red,line width=0.6pt] (7,19)--(15,19)--(13,23)--(29,23)--(19,21)--(27,19)--(21,11)--(19,13)--(19,7);
			\draw[color=red,line width=0.6pt] (41,19)--(43,19);
			\draw[color=red,line width=0.6pt] (45,19)--(47,19);
			\draw[color=red,line width=0.6pt] (59,0)--(59,1);
			\draw[color=red,line width=0.6pt] (59,3)--(59,5);
			\draw[color=red,line width=0.6pt] (73,19)--(75,19);
			\draw[color=red,line width=0.6pt] (77,19)--(78,19);
			\draw[rounded corners=.1mm,color=red,line width=0.6pt] (59,7)--(59,13)--(61,11)--(55,19)--(59,21)--(53,23)--(69,23)--(67,19)--(71,19);
			\foreach \dx in {0,40} \foreach \dy in {0,40}{
				\foreach \x in {1,3,5,7,31,33,35,37} \rightang{\x+\dx}{19+\dy};
				\foreach \y in {1,3,5,7} \rightang{19+\dx}{\y+\dy};
				\rightang{1+\dx}{3+\dy};
				\rightang{5+\dx}{7+\dy};
				\rightang{31+\dx}{5+\dy};
				\rightang{35+\dx}{1+\dy};
				\rightang{3+\dx}{35+\dy};
				\rightang{7+\dx}{31+\dy};
				\rightang{33+\dx}{31+\dy};
				\rightang{37+\dx}{35+\dy};
				\acuteang{19+\dx}{13+\dy};
				\acuteang{21+\dx}{11+\dy};
				\acuteang{19+\dx}{21+\dy};
				\acuteang{15+\dx}{19+\dy};
				\acuteang{13+\dx}{23+\dy};
				\acuteang{27+\dx}{19+\dy};
				\acuteang{29+\dx}{23+\dy};
				\draw[line width=0.8pt,color=white] (\dx,\dy)--(\dx,\dy+38)--(\dx+38,\dy+38)--(\dx+38,\dy)--cycle;
				\draw[dashed,line width=0.8pt,color=orange] (\dx,\dy)--(\dx,\dy+38)--(\dx+38,\dy+38)--(\dx+38,\dy)--cycle;
			}
		\end{tikzpicture}
		\captionof{figure}{Angle Loop variable-alignment T-metacell and possible solutions}
	\end{figure}
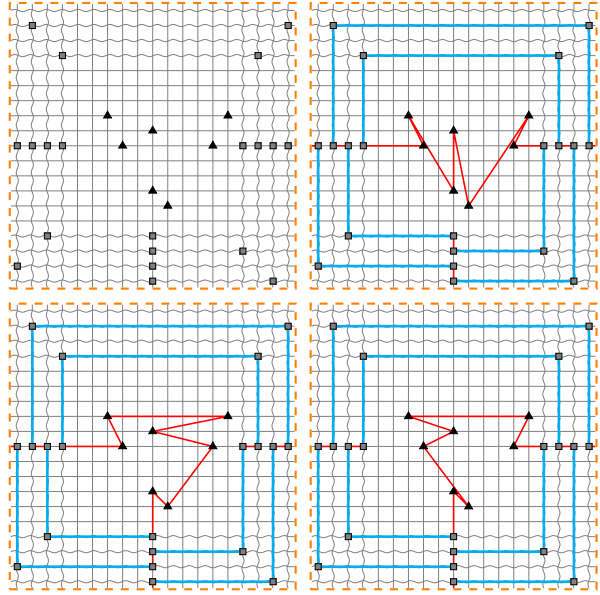
	
	\subsection{Kouchoku}
	
	\textbf{Rules:} Draw a loop via straight lines between dots and circles, visiting all the dots and circles. Circles with the same marking (here, letters) must be passed through all in one go with no dots or differently-marked circles between, and differently-marked circles may not be directly connected. Additionally, the loop may cross itself at any point which is not (the centre of) a dot or circle, as long as it intersects itself at a right-angle.
	
	We can check these conditions in $O(n^4)$ time: $O(n^2)$ for checking all the groups of circles are passed through in one go by iterating around the loop, and $O(n^4)$ time to check that if two segments cross they do so at right angles and not at their endpoints. The problem with this genre is that it is hard to make enclosed gadget because `gadget walls' may be crossable from points far away.
	
	To get around this problem, we ensure that the number of dots and the number of groups of circled cells are the same so that we must alternate between dots and groups of circled cells due to the conditions in the rules. Then, if we both ensure that the walls of the gadgets are horizontal or vertical, and that no dots are in the same column or row as a circled cell, we cannot cross the walls because it would require joining two dots or two circled cells with different markings (we construct such that in different gadgets, no circled cells have the same markings). To simplify, rows and columns alternate between circle rows/columns (full lines) and dot rows/columns (dotted lines). Unfortunately, this means that adjacent gadgets must join via a dot-circle connection. But then to ensure that the gadgets are always solvable with the given dots and circles in them, all entrances to the gadget must be all dots or all circles, and we must have at least one gadget of each type. Thus, for this construction we require two gadgets: one with dots on the openings (containing one more dot than circle groups) and one with circles on the openings (containing one more circle group than dots). Note that by parity an odd-by-odd Barred Simple Loop puzzle is trivially impossible, so we can assume that there is an even dimension of gadgets, and thus if we colour the board in a checkerboard pattern, we have the same number of squares of each colour. Then, if we place the two types of gadgets in this checkerboard fashion we have the same number of dot-gadgets and circle-gadgets, which ensures we have the same number of dots and circle groups. (This reduction is similar to the trick that we employed for Moon or Sun, except that our gadgets are more different than a simple inversion of clues.)
	
	The gadget can only be rotated and translated, so that the exits are at a diagonal from each other. To show both gadgets work, notice that there is only one way to connect each of the pairs from D to I (or L, for the circle gadget) inclusive (recalling that we will give different marks to different letters in different gadgets). Then observe that in each of A, B and C there is a unique way to connect the vertices without crossing D through I at a non-right angle, providing the vertical/horizontal walls of the gadget. Finally, the openings in the walls are positioned such that from the next gadget there is a line of sight with the midpoint of the orange-boundary between them to exactly one dot/circle (for the dot/circle gadget respectively), and no points beyond the walls of the immediately adjacent gadget have a line of sight with any of the points in the gadget. These internal and external lines of sight are illustrated by dashed brown lines in the first of the following figures. The genre was first published by Nikoli in Puzzle Communication Nikoli \#133. No NP-completeness results could be found.
	
	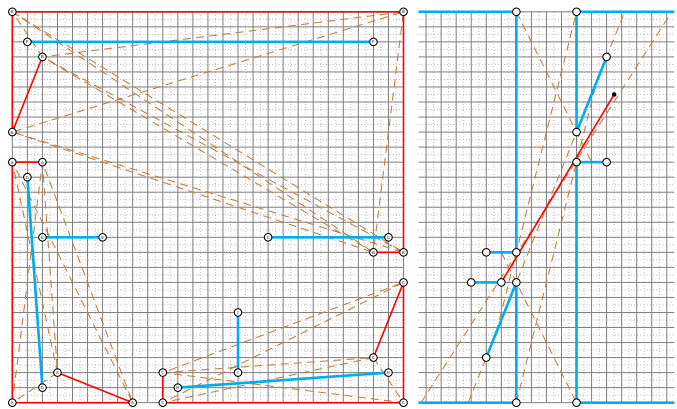
\begin{figure}[hbp]
		\centering
		\begin{tikzpicture}[scale=0.1,font=\sffamily]
			\foreach \x in {0,2,...,52} \draw[line width=0.2pt,color=gray] (\x,0)--(\x,52);
			\foreach \y in {0,2,...,52} \draw[line width=0.2pt,color=gray] (0,\y)--(52,\y);
			\foreach \x in {1,3,...,51} \draw[line width=0.2pt,color=gray,densely dotted] (\x,0)--(\x,52);
			\foreach \y in {1,3,...,51} \draw[line width=0.2pt,color=gray,densely dotted] (0,\y)--(52,\y);
			\draw[line width=0.4pt,color=brown,densely dashed] (6,4)--(0,0)--(4,32)--(16,0)--(0,32)--(6,4)--(4,32);
			\draw[line width=0.4pt,color=brown,densely dashed] (48,6)--(52,0)--(20,4)--(52,16)--(20,0)--(48,6)--(20,4);
			\draw[line width=0.4pt,color=brown,densely dashed] (52,20)--(4,46)--(0,52)--(52,20)--(0,36)--(52,52)--(48,20)--(4,46)--(52,52);
			\draw[line width=0.4pt,color=brown,densely dashed] (0,52)--(48,20)--(0,36);
			\draw[line width=1pt,color=cyan] (4,2)--(2,30);
			\draw[line width=1pt,color=cyan] (4,22)--(12,22);
			\draw[line width=1pt,color=cyan] (2,48)--(48,48);
			\draw[line width=1pt,color=cyan] (50,22)--(34,22);
			\draw[line width=1pt,color=cyan] (50,4)--(22,2);
			\draw[line width=1pt,color=cyan] (30,4)--(30,12);
			\draw[line width=0.6pt,color=red] (6,4)--(16,0)--(0,0)--(0,32)--(4,32);
			\draw[line width=0.6pt,color=red] (4,46)--(0,36)--(0,52)--(52,52)--(52,20)--(48,20);
			\draw[line width=0.6pt,color=red] (48,6)--(52,16)--(52,0)--(20,0)--(20,4);
			\letcirc{6}{4}{A};
			\letcirc{16}{0}{A};
			\letcirc{0}{0}{A};
			\letcirc{0}{32}{A};
			\letcirc{4}{32}{A};
			\letcirc{4}{46}{B};
			\letcirc{0}{36}{B};
			\letcirc{0}{52}{B};
			\letcirc{52}{52}{B};
			\letcirc{52}{20}{B};
			\letcirc{48}{20}{B};
			\letcirc{48}{6}{C};
			\letcirc{52}{16}{C};
			\letcirc{52}{0}{C};
			\letcirc{20}{0}{C};
			\letcirc{20}{4}{C};
			\letcirc{4}{2}{D};
			\letcirc{2}{30}{D};
			\letcirc{4}{22}{E};
			\letcirc{12}{22}{E};
			\letcirc{2}{48}{F};
			\letcirc{48}{48}{F};
			\letcirc{50}{22}{G};
			\letcirc{34}{22}{G};
			\letcirc{50}{4}{H};
			\letcirc{22}{2}{H};
			\letcirc{30}{4}{I};
			\letcirc{30}{12}{I};
			
			\foreach \x in {55,57,...,87} \draw[line width=0.2pt,color=gray] (\x,0)--(\x,52);
			\foreach \y in {0,2,...,52} \draw[line width=0.2pt,color=gray] (54,\y)--(88,\y);
			\foreach \x in {56,58,...,86} \draw[line width=0.2pt,color=gray,densely dotted] (\x,0)--(\x,52);
			\foreach \y in {1,3,...,51} \draw[line width=0.2pt,color=gray,densely dotted] (54,\y)--(88,\y);
			\draw[line width=0.4pt,color=brown,densely dashed] (60.6,0)--(81.4,52);
			\draw[line width=0.4pt,color=brown,densely dashed] (54+1/3,0)--(87+2/3,52);
			\draw[line width=0.4pt,color=brown,densely dashed] (65,20)--(75,0);
			\draw[line width=0.4pt,color=brown,densely dashed] (64+1/3,9+1/3)--(75,52);
			\draw[line width=0.4pt,color=brown,densely dashed] (67,52)--(77,32);
			\draw[line width=0.4pt,color=brown,densely dashed] (67,0)--(77+2/3,42+2/3);
			\draw[line width=1pt,color=cyan] (63,6)--(67,16);
			\draw[line width=1pt,color=cyan] (67,16)--(67,0)--(54,0);
			\draw[line width=1pt,color=cyan] (63,20)--(67,20)--(67,52)--(54,52);
			\draw[line width=1pt,color=cyan] (88,0)--(75,0)--(75,32)--(79,32);
			\draw[line width=1pt,color=cyan] (79,46)--(75,36);
			\draw[line width=1pt,color=cyan] (75,36)--(75,52)--(88,52);
			\draw[line width=1pt,color=cyan] (61,16)--(65,16);
			\draw[line width=0.6pt,color=red] (65,16)--(80,41);
			\letcirc{63}{6}{};
			\letcirc{67}{16}{};
			\letcirc{67}{0}{};
			\letcirc{63}{20}{};
			\letcirc{67}{20}{};
			\letcirc{67}{52}{};
			\letcirc{75}{0}{};
			\letcirc{75}{32}{};
			\letcirc{79}{32}{};
			\letcirc{79}{46}{};
			\letcirc{75}{36}{};
			\letcirc{75}{52}{};
			\letcirc{61}{16}{};
			\letcirc{65}{16}{};
			\fill[color=black] (80,41) circle (0.3);
		\end{tikzpicture}
		\captionof{figure}{Relevant lines of sight for both gadgets}
	\end{figure}

	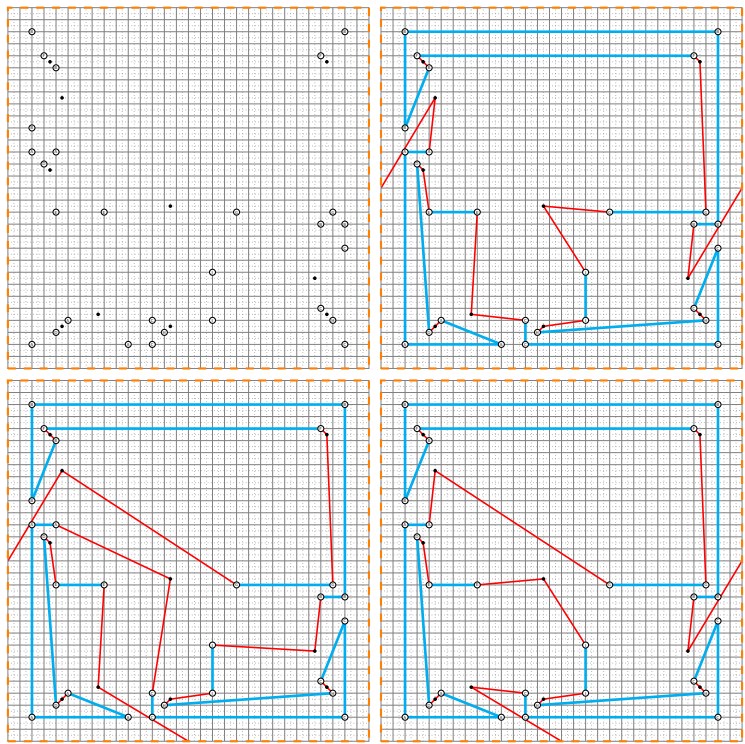
\begin{figure}
		\centering
		\begin{tikzpicture}[scale=0.08,font=\sffamily]
			\foreach \dx in {0,62} \foreach \dy in {0,62}{
				\foreach \x in {2,4,...,58} \draw[line width=0.2pt,color=gray] (\dx+\x,\dy)--(\dx+\x,\dy+60);
				\foreach \y in {2,4,...,58} \draw[line width=0.2pt,color=gray] (\dx,\dy+\y)--(\dx+60,\dy+\y);
				\foreach \x in {1,3,...,59} \draw[line width=0.2pt,color=gray,densely dotted] (\dx+\x,\dy)--(\dx+\x,\dy+60);
				\foreach \y in {1,3,...,59} \draw[line width=0.2pt,color=gray,densely dotted] (\dx,\dy+\y)--(\dx+60,\dy+\y);
			}
			\draw[line width=1pt,color=cyan] (4+4,2+4)--(4+2,30+4);
			\draw[line width=1pt,color=cyan] (4+4,22+4)--(4+12,22+4);
			\draw[line width=1pt,color=cyan] (4+2,48+4)--(4+48,48+4);
			\draw[line width=1pt,color=cyan] (4+50,22+4)--(4+34,22+4);
			\draw[line width=1pt,color=cyan] (4+50,4+4)--(4+22,2+4);
			\draw[line width=1pt,color=cyan] (4+30,4+4)--(4+30,12+4);
			\draw[line width=1pt,color=cyan] (4+6,4+4)--(4+16,0+4);
			\draw[line width=1pt,color=cyan] (4+16,0+4)--(4+0,0+4)--(4+0,32+4)--(4+4,32+4);
			\draw[line width=1pt,color=cyan] (4+4,46+4)--(4+0,36+4);
			\draw[line width=1pt,color=cyan] (4+0,36+4)--(4+0,52+4)--(4+52,52+4)--(4+52,20+4)--(4+48,20+4);
			\draw[line width=1pt,color=cyan] (4+48,6+4)--(4+52,16+4);
			\draw[line width=1pt,color=cyan] (4+52,16+4)--(4+52,0+4)--(4+20,0+4)--(4+20,4+4);
			\draw[line width=0.6pt,color=red] (4+4,2+4)--(4+6,4+4);
			\draw[line width=0.6pt,color=red] (4+2,30+4)--(4+3,29+4)--(4+4,22+4);
			\draw[line width=0.6pt,color=red] (4+2,48+4)--(4+4,46+4);
			\draw[line width=0.6pt,color=red] (4+48,48+4)--(4+49,47+4)--(4+50,22+4);
			\draw[line width=0.6pt,color=red] (4+48,6+4)--(4+50,4+4);
			\draw[line width=0.6pt,color=red] (4+22,2+4)--(4+23,3+4)--(4+30,4+4);
			\draw[line width=1pt,color=cyan] (62+4+4,2+4)--(62+4+2,30+4);
			\draw[line width=1pt,color=cyan] (62+4+4,22+4)--(62+4+12,22+4);
			\draw[line width=1pt,color=cyan] (62+4+2,48+4)--(62+4+48,48+4);
			\draw[line width=1pt,color=cyan] (62+4+50,22+4)--(62+4+34,22+4);
			\draw[line width=1pt,color=cyan] (62+4+50,4+4)--(62+4+22,2+4);
			\draw[line width=1pt,color=cyan] (62+4+30,4+4)--(62+4+30,12+4);
			\draw[line width=1pt,color=cyan] (62+4+6,4+4)--(62+4+16,0+4);
			\draw[line width=1pt,color=cyan] (62+4+16,0+4)--(62+4+0,0+4)--(62+4+0,32+4)--(62+4+4,32+4);
			\draw[line width=1pt,color=cyan] (62+4+4,46+4)--(62+4+0,36+4);
			\draw[line width=1pt,color=cyan] (62+4+0,36+4)--(62+4+0,52+4)--(62+4+52,52+4)--(62+4+52,20+4)--(62+4+48,20+4);
			\draw[line width=1pt,color=cyan] (62+4+48,6+4)--(62+4+52,16+4);
			\draw[line width=1pt,color=cyan] (62+4+52,16+4)--(62+4+52,0+4)--(62+4+20,0+4)--(62+4+20,4+4);
			\draw[line width=0.6pt,color=red] (62+4+4,2+4)--(62+4+6,4+4);
			\draw[line width=0.6pt,color=red] (62+4+2,30+4)--(62+4+3,29+4)--(62+4+4,22+4);
			\draw[line width=0.6pt,color=red] (62+4+2,48+4)--(62+4+4,46+4);
			\draw[line width=0.6pt,color=red] (62+4+48,48+4)--(62+4+49,47+4)--(62+4+50,22+4);
			\draw[line width=0.6pt,color=red] (62+4+48,6+4)--(62+4+50,4+4);
			\draw[line width=0.6pt,color=red] (62+4+22,2+4)--(62+4+23,3+4)--(62+4+30,4+4);
			\draw[line width=1pt,color=cyan] (62+4+4,2+4+62)--(62+4+2,30+4+62);
			\draw[line width=1pt,color=cyan] (62+4+4,22+4+62)--(62+4+12,22+4+62);
			\draw[line width=1pt,color=cyan] (62+4+2,48+4+62)--(62+4+48,48+4+62);
			\draw[line width=1pt,color=cyan] (62+4+50,22+4+62)--(62+4+34,22+4+62);
			\draw[line width=1pt,color=cyan] (62+4+50,4+4+62)--(62+4+22,2+4+62);
			\draw[line width=1pt,color=cyan] (62+4+30,4+4+62)--(62+4+30,12+4+62);
			\draw[line width=1pt,color=cyan] (62+4+6,4+4+62)--(62+4+16,0+4+62);
			\draw[line width=1pt,color=cyan] (62+4+16,0+4+62)--(62+4+0,0+4+62)--(62+4+0,32+4+62)--(62+4+4,32+4+62);
			\draw[line width=1pt,color=cyan] (62+4+4,46+4+62)--(62+4+0,36+4+62);
			\draw[line width=1pt,color=cyan] (62+4+0,36+4+62)--(62+4+0,52+4+62)--(62+4+52,52+4+62)--(62+4+52,20+4+62)--(62+4+48,20+4+62);
			\draw[line width=1pt,color=cyan] (62+4+48,6+4+62)--(62+4+52,16+4+62);
			\draw[line width=1pt,color=cyan] (62+4+52,16+4+62)--(62+4+52,0+4+62)--(62+4+20,0+4+62)--(62+4+20,4+4+62);
			\draw[line width=0.6pt,color=red] (62+4+4,2+4+62)--(62+4+6,4+4+62);
			\draw[line width=0.6pt,color=red] (62+4+2,30+4+62)--(62+4+3,29+4+62)--(62+4+4,22+4+62);
			\draw[line width=0.6pt,color=red] (62+4+2,48+4+62)--(62+4+4,46+4+62);
			\draw[line width=0.6pt,color=red] (62+4+48,48+4+62)--(62+4+49,47+4+62)--(62+4+50,22+4+62);
			\draw[line width=0.6pt,color=red] (62+4+48,6+4+62)--(62+4+50,4+4+62);
			\draw[line width=0.6pt,color=red] (62+4+22,2+4+62)--(62+4+23,3+4+62)--(62+4+30,4+4+62);
			\draw[line width=0.6pt,color=red] (62,92)--(71,107);
			\draw[line width=0.6pt,color=red] (71,107)--(70,98);
			\draw[line width=0.6pt,color=red] (78,88)--(77,71)--(86,70);
			\draw[line width=0.6pt,color=red] (96,78)--(89,89)--(100,88);
			\draw[line width=0.6pt,color=red] (114,86)--(113,77);
			\draw[line width=0.6pt,color=red] (113,77)--(122,92);
			\draw[line width=0.6pt,color=red] (92,0)--(77,9);
			\draw[line width=0.6pt,color=red] (77,9)--(86,8);
			\draw[line width=0.6pt,color=red] (78,26)--(89,27)--(96,16);
			\draw[line width=0.6pt,color=red] (70,36)--(71,45)--(100,26);
			\draw[line width=0.6pt,color=red] (114,24)--(113,15);
			\draw[line width=0.6pt,color=red] (113,15)--(122,30);
			\draw[line width=0.6pt,color=red] (0,30)--(9,45)--(38,26);
			\draw[line width=0.6pt,color=red] (8,36)--(27,27)--(24,8);
			\draw[line width=0.6pt,color=red] (16,26)--(15,9)--(30,0);
			\draw[line width=0.6pt,color=red] (34,16)--(51,15)--(52,24);
			\foreach \dx in {0,62} \foreach \dy in {0,62}{
				\letcirc{\dx+4+6}{\dy+4+4}{A};
				\letcirc{\dx+4+16}{\dy+4+0}{A};
				\letcirc{\dx+4+0}{\dy+4+0}{A};
				\letcirc{\dx+4+0}{\dy+4+32}{A};
				\letcirc{\dx+4+4}{\dy+4+32}{A};
				\letcirc{\dx+4+4}{\dy+4+46}{B};
				\letcirc{\dx+4+0}{\dy+4+36}{B};
				\letcirc{\dx+4+0}{\dy+4+52}{B};
				\letcirc{\dx+4+52}{\dy+4+52}{B};
				\letcirc{\dx+4+52}{\dy+4+20}{B};
				\letcirc{\dx+4+48}{\dy+4+20}{B};
				\letcirc{\dx+4+48}{\dy+4+6}{C};
				\letcirc{\dx+4+52}{\dy+4+16}{C};
				\letcirc{\dx+4+52}{\dy+4+0}{C};
				\letcirc{\dx+4+20}{\dy+4+0}{C};
				\letcirc{\dx+4+20}{\dy+4+4}{C};
				\letcirc{\dx+4+4}{\dy+4+2}{D};
				\letcirc{\dx+4+2}{\dy+4+30}{D};
				\letcirc{\dx+4+4}{\dy+4+22}{E};
				\letcirc{\dx+4+12}{\dy+4+22}{E};
				\letcirc{\dx+4+2}{\dy+4+48}{F};
				\letcirc{\dx+4+48}{\dy+4+48}{F};
				\letcirc{\dx+4+50}{\dy+4+22}{G};
				\letcirc{\dx+4+34}{\dy+4+22}{G};
				\letcirc{\dx+4+50}{\dy+4+4}{H};
				\letcirc{\dx+4+22}{\dy+4+2}{H};
				\letcirc{\dx+4+30}{\dy+4+4}{I};
				\letcirc{\dx+4+30}{\dy+4+12}{I};
				\fill[color=black] (\dx+4+5,\dy+4+3) circle (0.3);
				\fill[color=black] (\dx+4+3,\dy+4+29) circle (0.3);
				\fill[color=black] (\dx+4+3,\dy+4+47) circle (0.3);
				\fill[color=black] (\dx+4+23,\dy+4+3) circle (0.3);
				\fill[color=black] (\dx+4+49,\dy+4+5) circle (0.3);
				\fill[color=black] (\dx+4+49,\dy+4+47) circle (0.3);
				\fill[color=black] (\dx+4+23,\dy+4+23) circle (0.3);
				\fill[color=black] (\dx+4+11,\dy+4+5) circle (0.3);
				\fill[color=black] (\dx+4+5,\dy+4+41) circle (0.3);
				\fill[color=black] (\dx+4+47,\dy+4+11) circle (0.3);
				\fill[color=white] (\dx-.5,\dy-.5)--(\dx-.5,\dy+60)--(\dx,\dy+60)--(\dx,\dy)--(\dx+60,\dy)--(\dx+60,\dy+60)--(\dx+60.5,\dy+60)--(\dx+60.5,\dy-.5)--cycle;
				\draw[line width=0.2pt,color=gray] (\dx,\dy)--(\dx,\dy+60)--(60+\dx,60+\dy)--(60+\dx,\dy)--cycle;
				\draw[dashed,line width=0.8pt,color=orange] (\dx,\dy)--(\dx,\dy+60)--(60+\dx,60+\dy)--(60+\dx,\dy)--cycle;
			}
		\end{tikzpicture}
		\captionof{figure}{Kouchoku dot T-metacell gadget and possible solutions}
	\end{figure}

	\begin{figure}
		\centering
		\begin{tikzpicture}[scale=0.08,font=\sffamily]
			\foreach \dx in {0,62} \foreach \dy in {0,62}{
				\foreach \x in {2,4,...,58} \draw[line width=0.2pt,color=gray] (\dx+\x,\dy)--(\dx+\x,\dy+60);
				\foreach \y in {2,4,...,58} \draw[line width=0.2pt,color=gray] (\dx,\dy+\y)--(\dx+60,\dy+\y);
				\foreach \x in {1,3,...,59} \draw[line width=0.2pt,color=gray,densely dotted] (\dx+\x,\dy)--(\dx+\x,\dy+60);
				\foreach \y in {1,3,...,59} \draw[line width=0.2pt,color=gray,densely dotted] (\dx,\dy+\y)--(\dx+60,\dy+\y);
			}
			\draw[line width=1pt,color=cyan] (4+4,2+4)--(4+2,30+4);
			\draw[line width=1pt,color=cyan] (4+4,22+4)--(4+12,22+4);
			\draw[line width=1pt,color=cyan] (4+2,48+4)--(4+48,48+4);
			\draw[line width=1pt,color=cyan] (4+50,22+4)--(4+34,22+4);
			\draw[line width=1pt,color=cyan] (4+50,4+4)--(4+22,2+4);
			\draw[line width=1pt,color=cyan] (4+30,4+4)--(4+30,12+4);
			\draw[line width=1pt,color=cyan] (4+6,4+4)--(4+16,0+4);
			\draw[line width=1pt,color=cyan] (4+16,0+4)--(4+0,0+4)--(4+0,32+4)--(4+4,32+4);
			\draw[line width=1pt,color=cyan] (4+4,46+4)--(4+0,36+4);
			\draw[line width=1pt,color=cyan] (4+0,36+4)--(4+0,52+4)--(4+52,52+4)--(4+52,20+4)--(4+48,20+4);
			\draw[line width=1pt,color=cyan] (4+48,6+4)--(4+52,16+4);
			\draw[line width=1pt,color=cyan] (4+52,16+4)--(4+52,0+4)--(4+20,0+4)--(4+20,4+4);
			\draw[line width=0.6pt,color=red] (4+4,2+4)--(4+6,4+4);
			\draw[line width=0.6pt,color=red] (4+2,30+4)--(4+3,29+4)--(4+4,22+4);
			\draw[line width=0.6pt,color=red] (4+2,48+4)--(4+4,46+4);
			\draw[line width=0.6pt,color=red] (4+48,48+4)--(4+49,47+4)--(4+50,22+4);
			\draw[line width=0.6pt,color=red] (4+48,6+4)--(4+50,4+4);
			\draw[line width=0.6pt,color=red] (4+22,2+4)--(4+23,3+4)--(4+30,4+4);
			\draw[line width=1pt,color=cyan] (62+4+4,2+4)--(62+4+2,30+4);
			\draw[line width=1pt,color=cyan] (62+4+4,22+4)--(62+4+12,22+4);
			\draw[line width=1pt,color=cyan] (62+4+2,48+4)--(62+4+48,48+4);
			\draw[line width=1pt,color=cyan] (62+4+50,22+4)--(62+4+34,22+4);
			\draw[line width=1pt,color=cyan] (62+4+50,4+4)--(62+4+22,2+4);
			\draw[line width=1pt,color=cyan] (62+4+30,4+4)--(62+4+30,12+4);
			\draw[line width=1pt,color=cyan] (62+4+6,4+4)--(62+4+16,0+4);
			\draw[line width=1pt,color=cyan] (62+4+16,0+4)--(62+4+0,0+4)--(62+4+0,32+4)--(62+4+4,32+4);
			\draw[line width=1pt,color=cyan] (62+4+4,46+4)--(62+4+0,36+4);
			\draw[line width=1pt,color=cyan] (62+4+0,36+4)--(62+4+0,52+4)--(62+4+52,52+4)--(62+4+52,20+4)--(62+4+48,20+4);
			\draw[line width=1pt,color=cyan] (62+4+48,6+4)--(62+4+52,16+4);
			\draw[line width=1pt,color=cyan] (62+4+52,16+4)--(62+4+52,0+4)--(62+4+20,0+4)--(62+4+20,4+4);
			\draw[line width=0.6pt,color=red] (62+4+4,2+4)--(62+4+6,4+4);
			\draw[line width=0.6pt,color=red] (62+4+2,30+4)--(62+4+3,29+4)--(62+4+4,22+4);
			\draw[line width=0.6pt,color=red] (62+4+2,48+4)--(62+4+4,46+4);
			\draw[line width=0.6pt,color=red] (62+4+48,48+4)--(62+4+49,47+4)--(62+4+50,22+4);
			\draw[line width=0.6pt,color=red] (62+4+48,6+4)--(62+4+50,4+4);
			\draw[line width=0.6pt,color=red] (62+4+22,2+4)--(62+4+23,3+4)--(62+4+30,4+4);
			\draw[line width=1pt,color=cyan] (62+4+4,2+4+62)--(62+4+2,30+4+62);
			\draw[line width=1pt,color=cyan] (62+4+4,22+4+62)--(62+4+12,22+4+62);
			\draw[line width=1pt,color=cyan] (62+4+2,48+4+62)--(62+4+48,48+4+62);
			\draw[line width=1pt,color=cyan] (62+4+50,22+4+62)--(62+4+34,22+4+62);
			\draw[line width=1pt,color=cyan] (62+4+50,4+4+62)--(62+4+22,2+4+62);
			\draw[line width=1pt,color=cyan] (62+4+30,4+4+62)--(62+4+30,12+4+62);
			\draw[line width=1pt,color=cyan] (62+4+6,4+4+62)--(62+4+16,0+4+62);
			\draw[line width=1pt,color=cyan] (62+4+16,0+4+62)--(62+4+0,0+4+62)--(62+4+0,32+4+62)--(62+4+4,32+4+62);
			\draw[line width=1pt,color=cyan] (62+4+4,46+4+62)--(62+4+0,36+4+62);
			\draw[line width=1pt,color=cyan] (62+4+0,36+4+62)--(62+4+0,52+4+62)--(62+4+52,52+4+62)--(62+4+52,20+4+62)--(62+4+48,20+4+62);
			\draw[line width=1pt,color=cyan] (62+4+48,6+4+62)--(62+4+52,16+4+62);
			\draw[line width=1pt,color=cyan] (62+4+52,16+4+62)--(62+4+52,0+4+62)--(62+4+20,0+4+62)--(62+4+20,4+4+62);
			\draw[line width=0.6pt,color=red] (62+4+4,2+4+62)--(62+4+6,4+4+62);
			\draw[line width=0.6pt,color=red] (62+4+2,30+4+62)--(62+4+3,29+4+62)--(62+4+4,22+4+62);
			\draw[line width=0.6pt,color=red] (62+4+2,48+4+62)--(62+4+4,46+4+62);
			\draw[line width=0.6pt,color=red] (62+4+48,48+4+62)--(62+4+49,47+4+62)--(62+4+50,22+4+62);
			\draw[line width=0.6pt,color=red] (62+4+48,6+4+62)--(62+4+50,4+4+62);
			\draw[line width=0.6pt,color=red] (62+4+22,2+4+62)--(62+4+23,3+4+62)--(62+4+30,4+4+62);
			\draw[line width=1pt,color=cyan] (4+2,36+4)--(4+6,36+4);
			\draw[line width=1pt,color=cyan] (62+4+2,36+4)--(62+4+6,36+4);
			\draw[line width=1pt,color=cyan] (62+4+2,36+4+62)--(62+4+6,36+4+62);
			\draw[line width=1pt,color=cyan] (4+16,2+4)--(4+16,6+4);
			\draw[line width=1pt,color=cyan] (62+4+16,2+4)--(62+4+16,6+4);
			\draw[line width=1pt,color=cyan] (62+4+16,2+4+62)--(62+4+16,6+4+62);
			\draw[line width=1pt,color=cyan] (4+46,16+4)--(4+50,16+4);
			\draw[line width=1pt,color=cyan] (62+4+46,16+4)--(62+4+50,16+4);
			\draw[line width=1pt,color=cyan] (62+4+46,16+4+62)--(62+4+50,16+4+62);
			\draw[line width=0.6pt,color=red] (62,92)--(68,102);
			\draw[line width=0.6pt,color=red] (72,102)--(73,99)--(70,98);
			\draw[line width=0.6pt,color=red] (78,88)--(89,87)--(82,72);
			\draw[line width=0.6pt,color=red] (82,68)--(85,73)--(86,70);
			\draw[line width=0.6pt,color=red] (96,78)--(91,89)--(100,88);
			\draw[line width=0.6pt,color=red] (114,86)--(111,85)--(112,82);
			\draw[line width=0.6pt,color=red] (116,82)--(122,92);
			\draw[line width=0.6pt,color=red] (92,0)--(82,6);
			\draw[line width=0.6pt,color=red] (82,10)--(85,11)--(86,8);
			\draw[line width=0.6pt,color=red] (96,16)--(89,25)--(78,26);
			\draw[line width=0.6pt,color=red] (70,36)--(73,37);
			\draw[line width=0.6pt,color=red] (73,37)--(68,40);
			\draw[line width=0.6pt,color=red] (72,40)--(91,27)--(100,26);
			\draw[line width=0.6pt,color=red] (114,24)--(111,23)--(112,20);
			\draw[line width=0.6pt,color=red] (116,20)--(122,30);
			\draw[line width=0.6pt,color=red] (0,30)--(6,40);
			\draw[line width=0.6pt,color=red] (10,40)--(11,37)--(38,26);
			\draw[line width=0.6pt,color=red] (52,24)--(49,23);
			\draw[line width=0.6pt,color=red] (49,23)--(54,20);
			\draw[line width=0.6pt,color=red] (50,20)--(29,27);
			\draw[line width=0.6pt,color=red] (29,27)--(34,16);
			\draw[line width=0.6pt,color=red] (24,8)--(27,25)--(8,36);
			\draw[line width=0.6pt,color=red] (16,26)--(23,11)--(20,10);
			\draw[line width=0.6pt,color=red] (20,6)--(30,0);
			\foreach \dx in {0,62} \foreach \dy in {0,62}{
				\letcirc{\dx+4+6}{\dy+4+4}{A};
				\letcirc{\dx+4+16}{\dy+4+0}{A};
				\letcirc{\dx+4+16}{\dy+4+2}{K};
				\letcirc{\dx+4+16}{\dy+4+6}{K};
				\letcirc{\dx+4+0}{\dy+4+0}{A};
				\letcirc{\dx+4+0}{\dy+4+32}{A};
				\letcirc{\dx+4+4}{\dy+4+32}{A};
				\letcirc{\dx+4+4}{\dy+4+46}{B};
				\letcirc{\dx+4+0}{\dy+4+36}{B};
				\letcirc{\dx+4+2}{\dy+4+36}{L};
				\letcirc{\dx+4+6}{\dy+4+36}{L};
				\letcirc{\dx+4+0}{\dy+4+52}{B};
				\letcirc{\dx+4+52}{\dy+4+52}{B};
				\letcirc{\dx+4+52}{\dy+4+20}{B};
				\letcirc{\dx+4+48}{\dy+4+20}{B};
				\letcirc{\dx+4+48}{\dy+4+6}{C};
				\letcirc{\dx+4+52}{\dy+4+16}{C};
				\letcirc{\dx+4+46}{\dy+4+16}{J};
				\letcirc{\dx+4+50}{\dy+4+16}{J};
				\letcirc{\dx+4+52}{\dy+4+0}{C};
				\letcirc{\dx+4+20}{\dy+4+0}{C};
				\letcirc{\dx+4+20}{\dy+4+4}{C};
				\letcirc{\dx+4+4}{\dy+4+2}{D};
				\letcirc{\dx+4+2}{\dy+4+30}{D};
				\letcirc{\dx+4+4}{\dy+4+22}{E};
				\letcirc{\dx+4+12}{\dy+4+22}{E};
				\letcirc{\dx+4+2}{\dy+4+48}{F};
				\letcirc{\dx+4+48}{\dy+4+48}{F};
				\letcirc{\dx+4+50}{\dy+4+22}{G};
				\letcirc{\dx+4+34}{\dy+4+22}{G};
				\letcirc{\dx+4+50}{\dy+4+4}{H};
				\letcirc{\dx+4+22}{\dy+4+2}{H};
				\letcirc{\dx+4+30}{\dy+4+4}{I};
				\letcirc{\dx+4+30}{\dy+4+12}{I};
				\fill[color=black] (\dx+4+5,\dy+4+3) circle (0.3);
				\fill[color=black] (\dx+4+3,\dy+4+29) circle (0.3);
				\fill[color=black] (\dx+4+3,\dy+4+47) circle (0.3);
				\fill[color=black] (\dx+4+23,\dy+4+3) circle (0.3);
				\fill[color=black] (\dx+4+49,\dy+4+5) circle (0.3);
				\fill[color=black] (\dx+4+49,\dy+4+47) circle (0.3);
				\fill[color=black] (\dx+4+25,\dy+4+23) circle (0.3);
				\fill[color=black] (\dx+4+23,\dy+4+21) circle (0.3);
				\fill[color=black] (\dx+4+19,\dy+4+7) circle (0.3);
				\fill[color=black] (\dx+4+7,\dy+4+33) circle (0.3);
				\fill[color=black] (\dx+4+45,\dy+4+19) circle (0.3);
				\fill[color=white] (\dx-.5,\dy-.5)--(\dx-.5,\dy+60)--(\dx,\dy+60)--(\dx,\dy)--(\dx+60,\dy)--(\dx+60,\dy+60)--(\dx+60.5,\dy+60)--(\dx+60.5,\dy-.5)--cycle;
				\draw[line width=0.2pt,color=gray] (\dx,\dy)--(\dx,\dy+60)--(60+\dx,60+\dy)--(60+\dx,\dy)--cycle;
				\draw[dashed,line width=0.8pt,color=orange] (\dx,\dy)--(\dx,\dy+60)--(60+\dx,60+\dy)--(60+\dx,\dy)--cycle;
			}
		\end{tikzpicture}
		\captionof{figure}{Kouchoku circle T-metacell gadget and possible solutions}
	\end{figure}
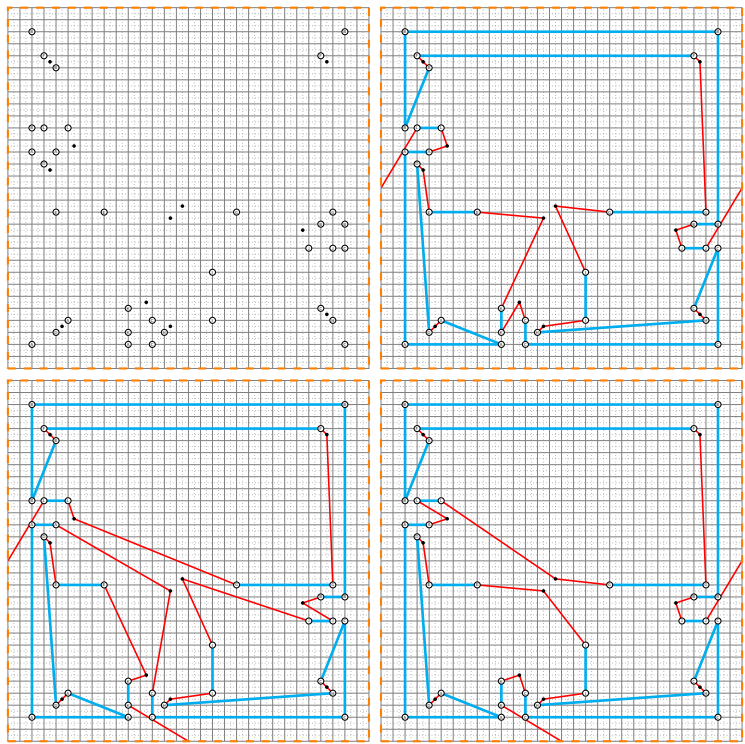
	
	\clearpage
	
	\subsection{Icelom and Icelom2}
	
	\textbf{Rules:} Draw a path passing between adjacent cell centres from the IN to the OUT. The path must pass over all numbers in increasing order. Coloured cells indicate `ice': the path must pass straight over ice cells, though it is allowed to cross itself only on ice cells (though it may not overlap itself except for at a crossing point). Connected ice cells are grouped into `icebarns'. In Icelom, every non-ice square must be passed through. In Icelom2, every icebarn must be passed through.
	
	All these conditions can be checked in $O(n^2)$ time by iterating along the path and then iterating among the cells/icebarns. The problem with constructing a T-metacell for this gadget is that the number condition is global (so we can't isolate these clues to a particular gadget.) Apart from that, we only have the constraints on which cells are passed through and the behaviour of the ice cells, which aren't particularly constraining and in particular don't provide an easy way to make barriers between gadgets.
	
	To get around this problem, we allow the alignment of T-metacell to vary within a column or row of T-metacells. In particular, in the following diagrams, the arrows align between adjacent T-metacells, and these gadgets are flowing in a `sea' of ice which forms an entirely-ice column around all gadgets and between each column/row of gadgets (which ensures this `sea' is connected). This ice sea allows two gadgets to be reachable from each other only if their arrows are on their respective sides, forming `walls'. Additionally, we note that since each gadget within a column/row of gadgets takes up at most a constant number of grid columns/rows, the number of grid columns/rows per column/row of gadgets is $O(n)$. Thus, the entire grid size is at most $O(n^2)\times O(n^2)$, which is still polynomial in the input grid size. (It is not entirely trivial to construct a puzzle with the virtual arrows aligning, however if we place the relative positions of the gadgets within the gadget column/rows before fixing the number of grid column/rows that these gadget column/rows take up, it is not too hard.)
	
	The gadget can be rotated and translated (before alignment). The Icelom gadget has to be passed through since it contains an empty cell, and the Icelom2 gadget has to be passed through since it contains an isolated icebarn. Finally, one corner gadget needs to be replaced with an IN/OUT on the edge of the grid pointing towards the arrows of the next cells so as to satsify the path constraint (noting that a corner gadget must be visited and contain a turn). Icelom and Icelom2 were first published by Nikoli in Puzzle Communication Nikoli \#128 and \#129 respectively. No NP-completeness results could be found for either genre.
	
	\begin{figure}[hbp]
		\centering
		\begin{tikzpicture}[scale=0.1]
			\foreach \d in {0,8,16,24}{
				\fill[color=cyan!30!white] (\d,0)--(\d,6)--(\d+6,6)--(\d+6,0)--cycle;
				\draw[line width=0.8pt,fill=white] (\d+2,2)--(\d+2,4)--(\d+4,4)--(\d+4,2)--cycle;
			}
			\draw[line width=0.6pt,color=red] (8,3)--(14,3);
			\draw[line width=0.6pt,color=red] (16,3)--(19,3)--(19,0);
			\draw[line width=0.6pt,color=red] (27,0)--(27,3)--(30,3);
			\foreach \d in {0,8,16,24}{
				\draw[line width=0.8pt,color=orange] (\d+0.5,4.5)--(\d+1,5.5)--(\d+1.5,4.5);
				\draw[line width=0.8pt,color=orange] (\d+1.5,2.5)--(\d+0.5,3)--(\d+1.5,3.5);
				\draw[line width=0.8pt,color=orange] (\d+2.5,1.5)--(\d+3,0.5)--(\d+3.5,1.5);
				\draw[line width=0.8pt,color=orange] (\d+4.5,2.5)--(\d+5.5,3)--(\d+4.5,3.5);
				\foreach \x in {2,4} \draw[line width=0.2pt] (\x+\d,0)--(\x+\d,2);
				\foreach \x in {2,4} \draw[line width=0.2pt] (\x+\d,4)--(\x+\d,6);
				\foreach \y in {2,4} \draw[line width=0.2pt] (\d,\y)--(\d+2,\y);
				\foreach \y in {2,4} \draw[line width=0.2pt] (\d+4,\y)--(\d+6,\y);
				\draw[line width=0.2pt] (\d,0)--(\d,6)--(\d+6,6)--(\d+6,0)--cycle;
				\draw[line width=0.8pt,color=orange,dashed] (\d,0)--(\d,6)--(\d+6,6)--(\d+6,0)--cycle;
			}
		\end{tikzpicture}
		\captionof{figure}{Icelom variable-position T-metacell and possible solutions}
	\end{figure}
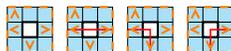
	
	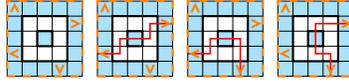
\begin{figure}
		\centering
		\begin{tikzpicture}[scale=0.1]
			\foreach \d in {0,12,24,36}{
				\fill[color=cyan!30!white] (\d,0)--(\d,10)--(\d+10,10)--(\d+10,0)--cycle;
				\draw[line width=0.8pt,fill=white] (\d+2,2)--(\d+2,8)--(\d+8,8)--(\d+8,2)--cycle;
				\draw[line width=0.2pt] (\d+2,4)--(\d+8,4);
				\draw[line width=0.2pt] (\d+2,6)--(\d+8,6);
				\draw[line width=0.2pt] (\d+4,2)--(\d+4,8);
				\draw[line width=0.2pt] (\d+6,2)--(\d+6,8);
				\draw[line width=0.8pt,fill=cyan!30!white] (\d+4,4)--(\d+4,6)--(\d+6,6)--(\d+6,4)--cycle;
			}
			\draw[line width=0.6pt,color=red] (12,3)--(15,3)--(15,5)--(19,5)--(19,7)--(22,7);
			\draw[line width=0.6pt,color=red] (24,3)--(27,3)--(27,5)--(31,5)--(31,0);
			\draw[line width=0.6pt,color=red] (43,0)--(43,3)--(41,3)--(41,7)--(46,7);
			\foreach \d in {0,12,24,36}{
				\draw[line width=0.8pt,color=orange] (\d+0.5,8.5)--(\d+1,9.5)--(\d+1.5,8.5);
				\draw[line width=0.8pt,color=orange] (\d+1.5,2.5)--(\d+0.5,3)--(\d+1.5,3.5);
				\draw[line width=0.8pt,color=orange] (\d+6.5,1.5)--(\d+7,0.5)--(\d+7.5,1.5);
				\draw[line width=0.8pt,color=orange] (\d+8.5,6.5)--(\d+9.5,7)--(\d+8.5,7.5);
				\foreach \x in {2,4,6,8} \draw[line width=0.2pt] (\x+\d,0)--(\x+\d,2);
				\foreach \x in {2,4,6,8} \draw[line width=0.2pt] (\x+\d,8)--(\x+\d,10);
				\foreach \y in {2,4,6,8} \draw[line width=0.2pt] (\d,\y)--(\d+2,\y);
				\foreach \y in {2,4,6,8} \draw[line width=0.2pt] (\d+8,\y)--(\d+10,\y);
				\draw[line width=0.2pt] (\d,0)--(\d,10)--(\d+10,10)--(\d+10,0)--cycle;
				\draw[line width=0.8pt,color=orange,dashed] (\d,0)--(\d,10)--(\d+10,10)--(\d+10,0)--cycle;
			}
		\end{tikzpicture}
		\captionof{figure}{Icelom2 variable-position T-metacell and possible solutions}
	\end{figure}
	
	\subsection{Ring-ring and Nagenawa}
	
	\textbf{Rules:} Draw some number of (potentially crossing) rectangles with vertices on cell centres, such that no two rectangles share any part of their boundaries, except for crossing points. In Ring-ring, all unshaded cells must be used by some rectangle and no shaded cells may be used by any rectangle. In Nagenawa, numbers in some regions indicate the number of cells used by any rectangle.
	
	Both of these conditions can be checked in $O(n^2)$ time by iterating through each cell, so satisfiability is at least in NP. The issue with constructing a T-metacell gadget for both these genres is that the rectangles are purely local structures in the sense that there is no connectivity requirement between different rectangles (unlike Scrin), and being rectangles they do not have enough freedom to form a single loop or path visiting various points in an arbitrary order. As a result, the T-metacell construction does not work for these particular genres. Ring-ring and Nagenawa were first published by Nikoli in Puzzle Communication Nikoli \#135 and \#123 respectively. No NP-completeness results could be found for either genre.
	
	\subsection{Regional Yajilin}
	
	\textbf{Rules:} Draw a single loop passing between adjacent cell centres. Any cell which the loop does not pass through must be shaded, and no two shaded cells may be adjacent. A region's number indicates the number of shaded cells within it.
	
	We can check both these conditions in $O(n^2)$ time by iterating through the cells in the grid to check no unshaded cells are adjacent, then by iterating through the cells in each region and counting the unshaded cells, so satisfiability is at least in NP. The issue with constructing a T-metacell gadget for this genre is that the only features restricting the path of the loop are the shaded cells, but since no shaded cells can be adjacent it is hard to make a horizontal wall. (A diagonal chain of shaded cells could be used to make a diagonal wall, but it is harder to fill in the corners where walls meet, and diagonal squares don't neatly tile the grid so there would be large corner sections which need filling.) It might yet be possible to construct a T-metacell for this genre, but this genre seems significantly more difficult to construct for than other similar genres. Regional Yajilin was first published by Naoki Inaba at \cite{inaba2005bbox}. No NP-completeness results could be found.
	
	\section{Closing Remarks}
	
	Firstly, some of these puzzles were first published in the magazines Puzzle Communication Nikoli (パズル通信ニコリ) and Toketa?\ (トケタ？), and while they have not been directly referenced inline, their websites can be found at \cite{puzzcomm} and \cite{toketa} respectively.
		
	The main open question from this paper is whether satisfaction of a Regional Yajilin puzzle is NP-complete. Given the similarity of the genre to Simple Loop -- they are identical on a fully-clued grid except Regional Yajilin has the additional constraint of no adjacent shaded cells -- the answer is almost certainly yes, but this restriction may prevent a simple metacell gadget from resolving this question. Another related question is whether Angle Loop, Icelom and Icelom2 have a reduction from Barred Simple Loop with a constant factor of dilation. The existence of such a reduction would produce more comparable results in terms of the exact time complexities required to solve each problem.
	
	Another potential direction for investigation is the computational complexity of verifying whether certain loop puzzle genres have a unique solution. However, this would require a different complexity class, since while checking whether a puzzle has any possible solution is a problem characteristic of NP, checking whether a puzzle has at most one solution is a problem characteristic of co-NP. It would also be interesting to investigate if other types of logic puzzle genres can be proven NP-complete to satisfy via a different gadget-based reduction.
	
	I would like to express my sincere gratitude to Ivan Adrian Koswara for their assistance with proofreading and suggesting improvements in my paper, and William Hu for verifying all of the example puzzles and rules.
	
	An additional NP-completeness result can be found in Appendix A, a full list of genres covered by this paper sorted by known results can be found in Appendix B, and example puzzles for each genre can be found in Appendix C.
	
	\printbibliography[heading=bibnumbered]
	
	\appendix
	
	\section*{Appendices}
	
	\section{Extra NP-Completeness result}
	
	After this paper had been written but before it was published, Tapa-Like Loop was added to the loop category on the puzz.link repository. We provide a proof of its satisfiability being NP-complete below in the same style as before.
	
	\subsection{Tapa-Like Loop}
	
	\textbf{Rules:} Draw a loop passing between adjacent cell centres, not passing through any clue cells. For each clue cell, consider the eight touching cells in the $3\times3$ region centred on that cell. Consider only the portions of the loop in those cells. Then, the numbers in the cell represent the lengths of the different segments of loop in that region, in arbitrary order. However, a single 0 instead indicates that the loop does not pass in any of the eight touching cells.
	
	We can check this condition in $O(n^2)$ time by iterating over the clue cells. The gadget here can be rotated, reflected and translated as necessary. The 3 clue ensures each gadget is visited, and the 0 clues leave only three entrances to the gadget. The genre was first published by Takeya Saikachi at \cite{saikachi2012tapalikeloop}. No NP-completeness results could be found.
	
	\begin{figure}[hbp]
		\centering
		\begin{tikzpicture}[scale=0.1,font=\sffamily]
			\foreach \d in {0,16,32,48} {
				\foreach \x in {2,4,...,12} \draw[line width=0.2pt] (\x+\d,0)--(\x+\d,14);
				\foreach \y in {2,4,...,12} \draw[line width=0.2pt] (\d,\y)--(14+\d,\y);
				\foreach \x in {1,3,11,13} \foreach \y in {1,13} \draw (\x+\d,\y) node {\tiny 0};
				\foreach \x in {1,13} \foreach \y in {3,11} \draw (\x+\d,\y) node {\tiny 0};
				\foreach \x in {5,7,9} \draw (\x+\d,13) node {\tiny 0};
				\draw (7+\d,9) node {\tiny 3}; 
			}
			\foreach \d in {16,32,48} \draw [line width=1pt,color=cyan] (\d+5.5,7)--(\d+8.5,7);
			\draw[color=red,line width=0.6pt] (16,7)--(21.5,7);
			\draw[color=red,line width=0.6pt] (24.5,7)--(30,7);
			\draw[color=red,line width=0.6pt] (32,7)--(37.5,7);
			\draw[color=red,line width=0.6pt] (40.5,7)--(41,7)--(41,5)--(39,5)--(39,0);
			\draw[color=red,line width=0.6pt] (55,0)--(55,5)--(53,5)--(53,7)--(53.5,7);
			\draw[color=red,line width=0.6pt] (56.5,7)--(62,7);
			\foreach \d in {0,16,32,48} {
				\draw[line width=0.2pt] (\d,0)--(\d,14)--(14+\d,14)--(14+\d,0)--cycle;
				\draw[dashed,line width=0.8pt,color=orange] (\d,0)--(\d,14)--(14+\d,14)--(14+\d,0)--cycle;
			}
		\end{tikzpicture}
		\captionof{figure}{Tapa-Like Loop T-metacell and possible solutions}
	\end{figure}
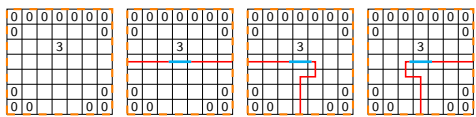
	
	\newpage
	
	\section{List of genres addressed in this paper}
	
	Genres whose satisfiability was proven NP-complete for the first time (as far as could be found):
	\begin{itemize}[noitemsep,topsep=0pt]
		\item Onsen-meguri (温泉めぐり)
		\item Mejilink (メジリンク)
		\item Double Back
		\item Scrin (スクリン)
		\item Geradeweg (グラーデヴェグ)
		\item Castle Wall
		\item Maxi Loop (ＭＡＸライン)
		\item Mid-loop (ミッドループ)
		\item Balance Loop
		\item Reflect Loop (リフレクトリンク)
		\item Pipelink Returns (帰ってきたパイプリンク)
		\item Icebarn (アイスバーン)
		\item Barns (バーンズ)
		\item Angle Loop (鋭直鈍ループ)
		\item Kouchoku (交差は直角に限る)
		\item Icelom (アイスローム)
		\item Icelom2 (アイスローム２)
		\item Tapa-Like Loop
	\end{itemize}
	
	Genres whose satisfiability was proven NP-complete previously but also proven NP-complete in this paper:
	\begin{itemize}[noitemsep,topsep=0pt]
		\item Slitherlink (スリザーリンク)
		\item Masyu (ましゅ)
		\item Yajilin (ヤジリン)
		\item Slalom (スラローム')
		\item Nagareru Loop (流れるループ)
		\item Moon or Sun (月か太陽)
		\item Country Road (カントリーロード)
		\item Simple Loop
		\item Detour
		\item Haisu
		\item Pipelink (パイプリンク)
		\item Loop Special (環状線スペシャル)
	\end{itemize}
	
	Genres whose satisfiability lies within NP but are not currently known to be NP-hard (as far as could be found):
	\begin{itemize}[noitemsep,topsep=0pt]
		\item Ring-ring (リングリング)
		\item Nagenawa (なげなわ)
		\item Regional Yajilin (ヘヤジリン or ブロックボックス)
	\end{itemize}

	\section{Example Puzzles}
	
	Puzzle rules have been restated here for convenience. Many example puzzles are copied or derived from example puzzles in the \href{https://github.com/robx/pzprjs}{puzz.link github repository}, whose software is open-source under the \href{https://github.com/robx/pzprjs/blob/main/LICENSE.txt}{MIT License}. 
	
	\subsection{Barred Simple Loop}
	
	\textbf{Rules:} Draw a single loop passing between adjacent cell centres passing through all cells. The loop may not pass through any marked edges.
	
	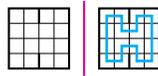
\begin{figure}[hbp]
		\centering
		\begin{tikzpicture}[scale=0.1]
			\foreach \d in {0,12} {
				\foreach \x in {2,4,6} \draw[line width=0.2pt] (\x+\d,1)--(\x+\d,9);
				\foreach \y in {3,5,7} \draw[line width=0.2pt] (\d,\y)--(\d+8,\y);
				\draw[line width=0.8pt] (\d,1)--(\d,9)--(\d+8,9)--(\d+8,1)--cycle;
				\draw[line width=0.8pt] (\d+4,1)--(\d+4,3);
				\draw[line width=0.8pt] (\d+4,7)--(\d+4,9);
			}
			\draw[color=cyan,line width=1pt] (13,2)--(13,8)--(15,8)--(15,6)--(17,6)--(17,8)--(19,8)--(19,2)--(17,2)--(17,4)--(15,4)--(15,2)--cycle;
			\draw[line width=0.8pt,color=magenta] (10,0)--(10,10);
		\end{tikzpicture}
		\captionof{figure}{Barred Simple Loop example puzzle and solution}
	\end{figure}
	
	\subsection{Cubic Barred Simple Loop}
	
	\textbf{Rules:} Draw a single loop passing between adjacent cell centres passing through all cells. The loop may not pass through any marked edges. Additionally, the solver is guaranteed that all cells not on the border of the grid are adjacent to at least one edge.
	
	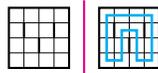
\begin{figure}[hbp]
		\centering
		\begin{tikzpicture}[scale=0.1]
			\foreach \d in {0,12} {
				\foreach \x in {2,4,6} \draw[line width=0.2pt] (\x+\d,1)--(\x+\d,9);
				\foreach \y in {3,5,7} \draw[line width=0.2pt] (\d,\y)--(\d+8,\y);
				\draw[line width=0.8pt] (\d,1)--(\d,9)--(\d+8,9)--(\d+8,1)--cycle;
				\draw[line width=0.8pt] (\d+4,3)--(\d+4,5);
				\draw[line width=0.8pt] (\d+2,5)--(\d+2,7);
				\draw[line width=0.8pt] (\d+6,5)--(\d+6,7);
			}
			\draw[color=cyan,line width=1pt] (13,2)--(13,8)--(19,8)--(19,2)--(17,2)--(17,6)--(15,6)--(15,2)--cycle;
			\draw[line width=0.8pt,color=magenta] (10,0)--(10,10);
		\end{tikzpicture}
		\captionof{figure}{Cubic Barred Simple Loop example puzzle and solution}
	\end{figure}
	
	\subsection{Slitherlink}
	
	\textbf{Rules:} Draw a single loop passing between pairs of adjacent dots. Numbers within the grid indicate how many of the edges of the square formed by the nearest 4 dots are used by the loop. 
	
	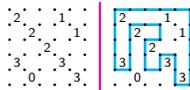
\begin{figure}[hbp]
		\centering
		\begin{tikzpicture}[scale=0.1,font=\sffamily]
			\draw[color=cyan,line width=1pt] (14,3)--(14,11)--(24,11)--(24,1)--(22,1)--(22,5)--(20,5)--(20,3)--(18,3)--(18,7)--(20,7)--(20,9)--(16,9)--(16,3)--cycle;
			\foreach \d in {0,14} {
				\foreach \x in {0,2,...,10} \foreach \y in {1,3,...,11} \draw[fill=black] (\x+\d,\y) circle (3pt);
				\draw (1+\d,4) node {\tiny 3};
				\draw (1+\d,10) node {\tiny 2};
				\draw (3+\d,2) node {\tiny 0};
				\draw (3+\d,8) node {\tiny 2};
				\draw (5+\d,6) node {\tiny 2};
				\draw (7+\d,4) node {\tiny 3};
				\draw (7+\d,10) node {\tiny 1};
				\draw (9+\d,2) node {\tiny 3};
				\draw (9+\d,8) node {\tiny 1};
			}
			\draw[line width=0.8pt,color=magenta] (12,0)--(12,12);
		\end{tikzpicture}
		\captionof{figure}{Slitherlink example puzzle and solution}
	\end{figure}
	
	\subsection{Masyu}
	
	\textbf{Rules:} Draw a single loop through all pearls and passing between adjacent cell centres. The loop must turn on each black pearl but go straight on the cells immediately before and after, and it must go straight through each white pearl and turn on a cell immediately before and/or after. 
	
	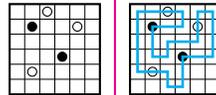
\begin{figure}[hbp]
		\centering
		\begin{tikzpicture}[scale=0.1]
			\foreach \d in {0,16} {
				\foreach \x in {2,4,6,8,10} \draw[line width=0.2pt] (\x+\d,1)--(\x+\d,13);
				\foreach \y in {3,5,7,9,11} \draw[line width=0.2pt] (\d,\y)--(\d+12,\y);
				\draw[line width=0.8pt] (\d,1)--(\d,13)--(\d+12,13)--(\d+12,1)--cycle;
				\draw (3+\d,4) circle (0.6);
				\draw[fill=black] (3+\d,10) circle (0.6);
				\draw (5+\d,12) circle (0.6);
				\draw[fill=black] (7+\d,6) circle (0.6);
				\draw (9+\d,10) circle (0.6);
			}
			\draw[color=cyan,line width=1pt] (17,4)--(17,12)--(23,12)--(23,10)--(19,10)--(19,6)--(21,6)--(21,8)--(25,8)--(25,12)--(27,12)--(27,6)--(23,6)--(23,2)--(21,2)--(21,4)--cycle;
			\draw[line width=0.8pt,color=magenta] (14,0)--(14,14);
		\end{tikzpicture}
		\captionof{figure}{Masyu example puzzle and solution}
	\end{figure}
	
	\subsection{Yajilin}
	
	\textbf{Rules:} Draw a single loop passing between adjacent cell centres. Some squares are marked in grey; these are not part of loop. Any other square not passed through by the loop must be shaded, and no two shaded cells may be adjacent. Grey cells may contain a number $n$ and an arrow pointing orthogonally: this means that there are $n$ shaded cells in the direction of the arrow.
	
	\begin{figure}[hbp]
		\centering
		\begin{tikzpicture}[scale=0.1,font=\sffamily]
			\fill[color=cyan] (16,5)--(16,7)--(18,7)--(18,5);
			\fill[color=cyan] (16,11)--(16,13)--(18,13)--(18,11);
			\fill[color=cyan] (20,9)--(20,7)--(22,7)--(22,9);
			\fill[color=cyan] (22,5)--(22,7)--(24,7)--(24,5);
			\fill[color=cyan] (26,9)--(26,11)--(28,11)--(28,9);
			\foreach \d in {0,16} {
				\foreach \x in {2,4,6,8,10} \draw[line width=0.2pt] (\x+\d,1)--(\x+\d,13);
				\foreach \y in {3,5,7,9,11} \draw[line width=0.2pt] (\d,\y)--(\d+12,\y);
				\draw[line width=0.8pt,fill=gray!40!white] (\d+2,13)--(\d+2,11)--(\d+4,11)--(\d+4,13);
				\draw[line width=0.8pt,fill=gray!40!white] (\d+6,5)--(\d+6,7)--(\d+4,7)--(\d+4,5)--cycle;
				\draw[line width=0.8pt,fill=gray!40!white] (\d+8,1)--(\d+8,3)--(\d+6,3)--(\d+6,1);
				\draw[line width=0.8pt,fill=gray!40!white] (\d+10,5)--(\d+10,7)--(\d+8,7)--(\d+8,5)--cycle;
				\draw[line width=0.8pt,fill=gray!40!white] (\d+10,13)--(\d+10,11)--(\d+12,11)--(\d+12,13);
				\draw[line width=0.8pt] (\d,1)--(\d,13)--(\d+12,13)--(\d+12,1)--cycle;
				\draw (\d+3,12) node {\scalebox{0.4}{0$\downarrow$}};
				\draw (\d+5,6.2) node {\scalebox{0.4}{1}};
				\draw (\d+5,5.5) node {\scalebox{0.4}{$\leftarrow$}};
				\draw (\d+11,12.2) node {\scalebox{0.4}{1}};
				\draw (\d+11,11.5) node {\scalebox{0.4}{$\leftarrow$}};
			}
			\draw[color=cyan,line width=1pt] (17,2)--(17,4)--(19,4)--(19,8)--(17,8)--(17,10)--(21,10)--(21,12)--(25,12)--(25,10)--(23,10)--(23,8)--(27,8)--(27,2)--(25,2)--(25,4)--(21,4)--(21,2)--cycle;
			\draw[line width=0.8pt,color=magenta] (14,0)--(14,14);
		\end{tikzpicture}
		\captionof{figure}{Yajilin example puzzle and solution}
	\end{figure}
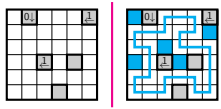
	
	\subsection{Slalom}
	
	\textbf{Rules:} Draw a single loop passing between adjacent unshaded cell centres which visits each gate (dotted line) once. When passing through a gate, the loop must go straight perpendicular to the gate. Furthermore, the loop must be assigned a direction such that, starting at the circle containing the number $n$ (the number of gates), any gate with an arrow and a number $k$ pointing to it is the $k^\text{th}$ gate the loop passes through.
	
	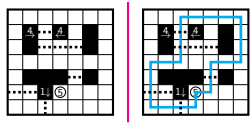
\begin{figure}[hbp]
		\centering
		\begin{tikzpicture}[scale=0.1,font=\sffamily]
			\foreach \d in {0,18} {
				\foreach \x in {2,4,...,12} \draw[line width=0.2pt] (\x+\d,1)--(\x+\d,15);
				\foreach \y in {3,5,...,13} \draw[line width=0.2pt] (\d,\y)--(\d+14,\y);
				\fill (\d+2,7)--(\d+2,5)--(\d+4,5)--(\d+4,3)--(\d+6,3)--(\d+6,5)--(\d+8,5)--(\d+8,7);
				\fill (\d+2,9)--(\d+2,13)--(\d+4,13)--(\d+4,9);
				\fill (\d+6,11)--(\d+6,13)--(\d+8,13)--(\d+8,11);
				\fill (\d+10,5)--(\d+10,7)--(\d+12,7)--(\d+12,5);
				\fill (\d+10,9)--(\d+10,13)--(\d+12,13)--(\d+12,9);
				\draw[line width=0.8pt] (\d,1)--(\d,15)--(\d+14,15)--(\d+14,1)--cycle;
				\draw[color=white] (\d+3,12.2) node {\scalebox{0.4}{4}};
				\draw[color=white] (\d+3,11.5) node {\scalebox{0.4}{$\rightarrow$}};
				\draw[color=white] (\d+7,12.2) node {\scalebox{0.4}{4}};
				\draw[color=white] (\d+7,11.5) node {\scalebox{0.4}{$\leftarrow$}};
				\draw[color=white] (\d+5,4) node {\scalebox{0.4}{1$\downarrow$}};
				\draw[fill=white] (\d+7,4) circle (0.7);
				\draw (\d+7,4) node {\scalebox{.4}{5}};
			}
			\draw[color=cyan,line width=1pt] (19,2)--(19,8)--(23,8)--(23,14)--(31,14)--(31,8)--(27,8)--(27,4)--(25,4)--(25,2)--cycle;
			\draw[line width=0.8pt,color=magenta] (16,0)--(16,16);
			\foreach \d in {0,18}{
				\draw[densely dotted, line width=1pt] (\d,4)--(\d+4,4);
				\draw[densely dotted, line width=1pt] (\d+5,1)--(\d+5,3);
				\draw[densely dotted, line width=1pt] (\d+8,6)--(\d+10,6);
				\draw[densely dotted, line width=1pt] (\d+4,10)--(\d+10,10);
				\draw[densely dotted, line width=1pt] (\d+4,12)--(\d+6,12);
			}
		\end{tikzpicture}
		\captionof{figure}{Slalom example puzzle and solution}
	\end{figure}
	
	\subsection{Nagareru Loop}
	
	\textbf{Rules:} Draw a single directed loop passing between adjacent unshaded cell centres. It must pass straight through every black arrow in the direction of the arrow, and if it passes in front of a white arrow (with no shaded cells between), it must turn away from the arrow and travel that way for at least one unit.
	
	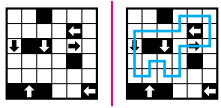
\begin{figure}[hbp]
		\centering
		\begin{tikzpicture}[scale=0.1]
			\foreach \d in {0,16} {
				\foreach \x in {2,4,6,8,10} \draw[line width=0.2pt] (\x+\d,1)--(\x+\d,13);
				\foreach \y in {3,5,7,9,11} \draw[line width=0.2pt] (\d,\y)--(\d+12,\y);
				\fill (\d,1)--(\d,3)--(\d+6,3)--(\d+6,1);
				\fill (\d+10,1)--(\d+10,3)--(\d+12,3)--(\d+12,1);
				\fill (\d+2,7)--(\d+2,9)--(\d+6,9)--(\d+6,7);
				\fill (\d+4,11)--(\d+4,13)--(\d+6,13)--(\d+6,11);
				\fill (\d+10,11)--(\d+10,9)--(\d+8,9)--(\d+8,11);
				\fill (\d+10,5)--(\d+10,7)--(\d+8,7)--(\d+8,5);
				\draw[line width=0.8pt] (\d,1)--(\d,13)--(\d+12,13)--(\d+12,1)--cycle;
				\fill (0.7+\d,8.8)--(0.7+\d,7.9)--(0.3+\d,7.9)--(1+\d,7.2)--(1.7+\d,7.9)--(1.3+\d,7.9)--(1.3+\d,8.8)--cycle;
				\fill (8.2+\d,7.7)--(9.1+\d,7.7)--(9.1+\d,7.3)--(9.8+\d,8)--(9.1+\d,8.7)--(9.1+\d,8.3)--(8.2+\d,8.3)--cycle;
				\fill[color=white] (4.7+\d,8.8)--(4.7+\d,7.9)--(4.3+\d,7.9)--(5+\d,7.2)--(5.7+\d,7.9)--(5.3+\d,7.9)--(5.3+\d,8.8)--cycle;
				\fill[color=white] (2.7+\d,1.2)--(2.7+\d,2.1)--(2.3+\d,2.1)--(3+\d,2.8)--(3.7+\d,2.1)--(3.3+\d,2.1)--(3.3+\d,1.2)--cycle;
				\fill[color=white] (9.8+\d,9.7)--(8.9+\d,9.7)--(8.9+\d,9.3)--(8.2+\d,10)--(8.9+\d,10.7)--(8.9+\d,10.3)--(9.8+\d,10.3)--cycle;
				\fill[color=white] (11.8+\d,1.7)--(10.9+\d,1.7)--(10.9+\d,1.3)--(10.2+\d,2)--(10.9+\d,2.7)--(10.9+\d,2.3)--(11.8+\d,2.3)--cycle;
			}
			\draw[color=cyan,line width=1pt] (17,4)--(17,10)--(23,10)--(23,12)--(27,12)--(27,8)--(23,8)--(23,4)--(21,4)--(21,6)--(19,6)--(19,4)--cycle;
			\draw[line width=0.8pt,color=magenta] (14,0)--(14,14);
		\end{tikzpicture}
		\captionof{figure}{Nagareru Loop example puzzle and solution}
	\end{figure}
	
	\subsection{Moon or Sun}
	
	\textbf{Rules:} Draw a single loop passing between adjacent cell centres. It must visit each region exactly once, either passing through all the moon cells and none of the sun cells, or vice versa. Furthermore, each time it crosses a region boundary it must alternate from visiting all suns to visiting all moons, or vice versa.
	
	\begin{figure}[hbp]
		\centering
		\begin{tikzpicture}[scale=0.1]
			\foreach \d in {0,16} {
				\foreach \x in {2,4,6,8,10} \draw[line width=0.2pt] (\x+\d,1)--(\x+\d,13);
				\foreach \y in {3,5,7,9,11} \draw[line width=0.2pt] (\d,\y)--(\d+12,\y);
				\draw[line width=0.8pt] (\d,1)--(\d,13)--(\d+12,13)--(\d+12,1)--cycle;
				\draw[line width=0.8pt] (\d,5)--(\d+8,5)--(\d+8,1);
				\draw[line width=0.8pt] (\d,9)--(\d+4,9)--(\d+4,1);
				\draw[line width=0.8pt] (\d+6,5)--(\d+6,13);
				\draw[line width=0.8pt] (\d+6,7)--(\d+12,7);
				\draw (\d+1,12) circle (0.6);
				\draw (\d+3,6) circle (0.6);
				\draw (\d+5,2) circle (0.6);
				\draw (\d+5,8) circle (0.6);
				\draw (\d+5,10) circle (0.6);
				\draw (\d+7,8) circle (0.6);
				\draw (\d+11,12) circle (0.6);
				\fill (1+\d+.7/2^.5,4+.7/2^.5) arc (45:-135:.7) arc (-90:0:.7*2^.5);
				\fill (3+\d+.7/2^.5,8+.7/2^.5) arc (45:-135:.7) arc (-90:0:.7*2^.5);
				\fill (3+\d+.7/2^.5,10+.7/2^.5) arc (45:-135:.7) arc (-90:0:.7*2^.5);
				\fill (3+\d+.7/2^.5,12+.7/2^.5) arc (45:-135:.7) arc (-90:0:.7*2^.5);
				\fill (7+\d+.7/2^.5,4+.7/2^.5) arc (45:-135:.7) arc (-90:0:.7*2^.5);
				\fill (9+\d+.7/2^.5,10+.7/2^.5) arc (45:-135:.7) arc (-90:0:.7*2^.5);
				\fill (11+\d+.7/2^.5,2+.7/2^.5) arc (45:-135:.7) arc (-90:0:.7*2^.5);
				\fill (11+\d+.7/2^.5,6+.7/2^.5) arc (45:-135:.7) arc (-90:0:.7*2^.5);
			}
			\draw[color=cyan,line width=1pt] (17,2)--(17,4)--(19,4)--(19,6)--(17,6)--(17,10)--(19,10)--(19,12)--(27,12)--(27,8)--(23,8)--(23,6)--(27,6)--(27,2)--cycle;
			\draw[line width=0.8pt,color=magenta] (14,0)--(14,14);
		\end{tikzpicture}
		\captionof{figure}{Moon or Sun example puzzle and solution}
	\end{figure}
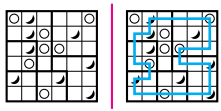
	
	\subsection{Country Road}
	
	\textbf{Rules:} Draw a single loop passing between adjacent cell centres and visiting each region once. A number within a region indicates the number of cells in that region through which the loop passes. Out of any pair of adjacent cells in different regions, at least one must be used by the loop.
	
	\begin{figure}[hbp]
		\centering
		\begin{tikzpicture}[scale=0.1,font=\sffamily]
			\foreach \d in {0,14} {
				\foreach \x in {2,4,6,8} \draw[line width=0.2pt] (\x+\d,1)--(\x+\d,11);
				\foreach \y in {3,5,7,9} \draw[line width=0.2pt] (\d,\y)--(\d+10,\y);
				\draw[line width=0.8pt] (\d,1)--(\d,11)--(\d+10,11)--(\d+10,1)--cycle;
				\draw[line width=0.8pt] (\d,7)--(\d+8,7);
				\draw[line width=0.8pt] (\d+2,1)--(\d+2,7);
				\draw[line width=0.8pt] (\d+4,7)--(\d+4,11);
				\draw[line width=0.8pt] (\d+2,5)--(\d+10,5);
				\draw[line width=0.8pt] (\d+6,1)--(\d+6,5);
				\draw[line width=0.8pt] (\d+8,5)--(\d+8,11);
				\draw (0.4+\d,6.5) node {\scalebox{.2}{1}};
				\draw (0.4+\d,10.5) node {\scalebox{.2}{2}};
				\draw (2.4+\d,6.5) node {\scalebox{.2}{2}};
			}
			\draw[color=cyan,line width=1pt] (15,6)--(15,8)--(19,8)--(19,10)--(21,10)--(21,8)--(23,8)--(23,4)--(21,4)--(21,2)--(17,2)--(17,4)--(19,4)--(19,6)--cycle;
			\draw[line width=0.8pt,color=magenta] (12,0)--(12,12);
		\end{tikzpicture}
		\captionof{figure}{Country Road example puzzle and solution}
	\end{figure}
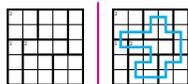
	
	\subsection{Onsen-Meguri}
	
	\textbf{Rules:} Draw non-overlapping loops passing between adjacent cell centres. At least one loop passes through each region, but each loop visits a region at most once. Each loop must pass over exactly one circled cell, and all such cells are passed over. For each loop the number of cells the loop passes through in each region must be fixed, and if there is a number on the circled cell, equal to that.
	
	\begin{figure}[hbp]
		\centering
		\begin{tikzpicture}[scale=0.1,font=\sffamily]
			\foreach \d in {0,16} {
				\foreach \x in {2,4,6,8,10} \draw[line width=0.2pt] (\x+\d,1)--(\x+\d,13);
				\foreach \y in {3,5,7,9,11} \draw[line width=0.2pt] (\d,\y)--(\d+12,\y);
				\draw[line width=0.8pt] (\d,1)--(\d,13)--(\d+12,13)--(\d+12,1)--cycle;
				\draw[line width=0.8pt] (\d,5)--(\d+4,5)--(\d+4,13);
				\draw[line width=0.8pt] (\d,7)--(\d+4,7);
				\draw[line width=0.8pt] (\d,11)--(\d+4,11);
				\draw[line width=0.8pt] (\d+4,9)--(\d+12,9);
				\draw (\d+1,12) circle (0.7);
				\draw (\d+3,2) circle (0.7);
				\draw (\d+3,2) node {\scalebox{.4}{4}};
				\draw (\d+5,2) circle (0.7);
				\draw (\d+5,2) node {\scalebox{.4}{4}};
				\draw (\d+11,10) circle (0.7);
				\draw (\d+11,10) node {\scalebox{.4}{3}};
			}
			\draw[color=cyan,line width=1pt] (19,2)--(19,4)--(17,4)--(17,2)--cycle;
			\draw[color=cyan,line width=1pt] (21,2)--(21,4)--(23,4)--(23,2)--cycle;
			\draw[color=cyan,line width=1pt] (17,6)--(17,12)--(21,12)--(21,6)--cycle;
			\draw[color=cyan,line width=1pt] (23,8)--(23,10)--(27,10)--(27,8)--cycle;
			\draw[line width=0.8pt,color=magenta] (14,0)--(14,14);
		\end{tikzpicture}
		\captionof{figure}{Onsen-Meguri example puzzle and solution}
	\end{figure}
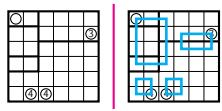
	
	\subsection{Mejilink}
	
	\textbf{Rule:} Draw a loop along region boundaries so that the unused perimeter of each region is equal to that region's area. 
	
	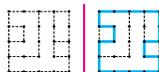
\begin{figure}[hbp]
		\centering
		\begin{tikzpicture}[scale=0.1]
			\foreach \d in {0,12} {
				\draw[dash pattern=on 1pt off 0.5pt,line width=0.2pt] (\d,1)--(\d,9)--(\d+8,9)--(\d+8,1)--cycle;
				\draw[dash pattern=on 1pt off 0.5pt,line width=0.2pt] (\d,5)--(\d+2,5)--(\d+2,7)--(\d,7);
				\draw[dash pattern=on 1pt off 0.5pt,line width=0.2pt] (\d,3)--(\d+4,3)--(\d+4,9);
				\draw[dash pattern=on 1pt off 0.5pt,line width=0.2pt] (\d+6,9)--(\d+6,3)--(\d+8,3);
				\draw[dash pattern=on 1pt off 0.5pt,line width=0.2pt] (\d+6,5)--(\d+8,5);
			}
			\draw[color=cyan,line width=1pt] (12,1)--(12,5)--(14,5)--(14,7)--(12,7)--(12,9)--(20,9)--(20,5)--(18,5)--(18,3)--(20,3)--(20,1)--cycle;
			\draw[line width=0.8pt,color=magenta] (10,0)--(10,10);
			\foreach \d in {0,12} \foreach \x in {0,2,4,6,8} \foreach \y in {1,3,5,7,9} \draw[fill=black] (\x+\d,\y) circle (3pt);
		\end{tikzpicture}
		\captionof{figure}{Mejilink example puzzle and solution}
	\end{figure}
	
	\subsection{Double Back}
	
	\textbf{Rule:} Draw a loop passing between adjacent cell centres visiting every cell once and visiting each region exactly twice.
	
	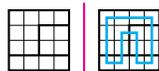
\begin{figure}[hbp]
		\centering
		\begin{tikzpicture}[scale=0.1]
			\foreach \d in {0,12} {
				\foreach \x in {2,4,6} \draw[line width=0.2pt] (\x+\d,1)--(\x+\d,9);
				\foreach \y in {3,5,7} \draw[line width=0.2pt] (\d,\y)--(\d+8,\y);
				\draw[line width=0.8pt] (\d,1)--(\d,9)--(\d+8,9)--(\d+8,1)--cycle;
				\draw[line width=0.8pt] (\d,3)--(\d+8,3);
				\draw[line width=0.8pt] (\d+4,3)--(\d+4,7)--(\d+8,7);
			}
			\draw[color=cyan,line width=1pt] (13,2)--(13,8)--(19,8)--(19,2)--(17,2)--(17,6)--(15,6)--(15,2)--cycle;
			\draw[line width=0.8pt,color=magenta] (10,0)--(10,10);
		\end{tikzpicture}
		\captionof{figure}{Double Back example puzzle and solution}
	\end{figure}
	
	\newpage
	
	\subsection{Scrin}
	
	\textbf{Rules:} Draw at least five non-overlapping grid-aligned rectangles that do not share any cell edge. Considering rectangles touching at a corner to be adjacent, the rectangles should form a single loop. All circled cells must be in a rectangle, and if they contain a number that must be the area of the rectangle.
	
	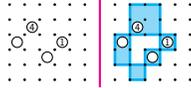
\begin{figure}[hbp]
		\centering
		\begin{tikzpicture}[scale=0.1,font=\sffamily]
			\draw[color=cyan,fill=cyan!40!white,line width=1pt] (14,3)--(14,7)--(16,7)--(16,3)--cycle;
			\draw[color=cyan,fill=cyan!40!white,line width=1pt] (16,7)--(16,11)--(20,11)--(20,7)--cycle;
			\draw[color=cyan,fill=cyan!40!white,line width=1pt]
			(16,1)--(16,3)--(18,3)--(18,1)--cycle;
			\draw[color=cyan,fill=cyan!40!white,line width=1pt] (18,3)--(18,5)--(20,5)--(20,3)--cycle;
			\draw[color=cyan,fill=cyan!40!white,line width=1pt] (20,5)--(20,7)--(22,7)--(22,5)--cycle;
			\foreach \d in {0,14} {
				\foreach \x in {0,2,...,10} \foreach \y in {1,3,...,11} \draw[fill=black] (\x+\d,\y) circle (3pt);
				\draw[fill=white] (\d+1,6) circle (0.7);
				\draw[fill=white] (\d+3,8) circle (0.7);
				\draw (\d+3,8) node {\scalebox{.4}{4}};
				\draw[fill=white] (\d+5,4) circle (0.7);
				\draw[fill=white] (\d+7,6) circle (0.7);
				\draw (\d+7,6) node {\scalebox{.4}{1}};
			}
			\draw[line width=0.8pt,color=magenta] (12,0)--(12,12);
		\end{tikzpicture}
		\captionof{figure}{Scrin example puzzle and solution}
	\end{figure}
	
	\subsection{Geradeweg}
	
	\textbf{Rules:} Draw a loop passing between adjacent cell centres that visits every circled cell. Divide the loop into segments where it turns. Any numbers on circled cells indicate the lengths of all segments passing through that cell. At a circled corner, both segments must have equal length.
	
	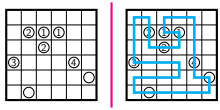
\begin{figure}[hbp]
		\centering
		\begin{tikzpicture}[scale=0.1,font=\sffamily]
			\foreach \d in {0,16} {
				\foreach \x in {2,4,6,8,10} \draw[line width=0.2pt] (\x+\d,1)--(\x+\d,13);
				\foreach \y in {3,5,7,9,11} \draw[line width=0.2pt] (\d,\y)--(\d+12,\y);
				\draw[line width=0.8pt] (\d,1)--(\d,13)--(\d+12,13)--(\d+12,1)--cycle;
				\draw[fill=white] (\d+1,6) circle (0.7);
				\draw (\d+1,6) node {\scalebox{.4}{3}};
				\draw[fill=white] (\d+3,2) circle (0.7);
				\draw[fill=white] (\d+3,10) circle (0.7);
				\draw (\d+3,10) node {\scalebox{.4}{2}};
				\draw[fill=white] (\d+5,8) circle (0.7);
				\draw (\d+5,8) node {\scalebox{.4}{2}};
				\draw[fill=white] (\d+5,10) circle (0.7);
				\draw (\d+5,10) node {\scalebox{.4}{1}};
				\draw[fill=white] (\d+7,10) circle (0.7);
				\draw (\d+7,10) node {\scalebox{.4}{1}};
				\draw[fill=white] (\d+9,6) circle (0.7);
				\draw (\d+9,6) node {\scalebox{.4}{4}};
				\draw[fill=white] (\d+11,4) circle (0.7);
			}
			\draw[color=cyan,line width=1pt] (17,2)--(17,4)--(23,4)--(23,6)--(17,6)--(17,12)--(19,12)--(19,8)--(23,8)--(23,10)--(21,10)--(21,12)--(25,12)--(25,4)--(27,4)--(27,2)--cycle;
			\draw[line width=0.8pt,color=magenta] (14,0)--(14,14);
		\end{tikzpicture}
		\captionof{figure}{Geradeweg example puzzle and solution}
	\end{figure}
	
	\subsection{Castle Wall}
	
	\textbf{Rules:} Outlined blocks of one or more cells are called walls. Draw a loop passing between adjacent non-wall cell centres. White walls and black walls can only be inside/outside the loop respectively. Walls may contain a number $n$ and an arrow pointing orthogonally: this means that the loop segments in the direction of the arrow have a total length of $n$.
	
	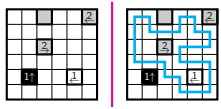
\begin{figure}[hbp]
		\centering
		\begin{tikzpicture}[scale=0.1,font=\sffamily]
			\foreach \d in {0,16} {
				\foreach \x in {2,4,6,8,10} \draw[line width=0.2pt] (\x+\d,1)--(\x+\d,13);
				\foreach \y in {3,5,7,9,11} \draw[line width=0.2pt] (\d,\y)--(\d+12,\y);
				\draw[line width=0.8pt,fill=gray!40!white] (\d+4,13)--(\d+4,11)--(\d+6,11)--(\d+6,13);
				\draw[line width=0.8pt,fill=gray!40!white] (\d+10,13)--(\d+10,11)--(\d+12,11)--(\d+12,13);
				\draw[line width=0.8pt,fill=gray!40!white] (\d+4,7)--(\d+4,9)--(\d+6,9)--(\d+6,7)--cycle;
				\draw[line width=0.8pt,fill=black] (\d+4,3)--(\d+4,5)--(\d+2,5)--(\d+2,3)--cycle;
				\draw[line width=0.8pt] (\d+10,3)--(\d+10,5)--(\d+8,5)--(\d+8,3)--cycle;
				\draw[line width=0.8pt] (\d,1)--(\d,13)--(\d+12,13)--(\d+12,1)--cycle;
				\draw (\d+5,8.2) node {\scalebox{0.4}{2}};
				\draw (\d+5,7.5) node {\scalebox{0.4}{$\rightarrow$}};
				\draw (\d+11,12.2) node {\scalebox{0.4}{2}};
				\draw (\d+11,11.5) node {\scalebox{0.4}{$\leftarrow$}};
				\draw (\d+9,4.2) node {\scalebox{0.4}{1}};
				\draw (\d+9,3.5) node {\scalebox{0.4}{$\leftarrow$}};
				\draw[color=white] (\d+3,4) node {\scalebox{0.4}{1$\uparrow$}};
			}
			\draw[color=cyan,line width=1pt] (17,6)--(17,12)--(19,12)--(19,10)--(23,10)--(23,12)--(25,12)--(25,10)--(27,10)--(27,8)--(23,8)--(23,6)--(27,6)--(27,2)--(23,2)--(23,4)--(21,4)--(21,6)--cycle;
			\draw[line width=0.8pt,color=magenta] (14,0)--(14,14);
		\end{tikzpicture}
		\captionof{figure}{Castle Wall example puzzle and solution}
	\end{figure}
	
	\subsection{Maxi Loop}
	
	\textbf{Rules:} Draw a loop passing between adjacent cell centres that visits every cell. A number within the region indicates the maximum number of cells the loop visits within the region on any one entrance to the region, across all such visits.
	
	\begin{figure}[hbp]
		\centering
		\begin{tikzpicture}[scale=0.1,font=\sffamily]
			\foreach \d in {0,16} {
				\foreach \x in {2,4,6,8,10} \draw[line width=0.2pt] (\x+\d,1)--(\x+\d,13);
				\foreach \y in {3,5,7,9,11} \draw[line width=0.2pt] (\d,\y)--(\d+12,\y);
				\draw[line width=0.8pt] (\d,1)--(\d,13)--(\d+12,13)--(\d+12,1)--cycle;
				\draw[line width=0.8pt] (\d,9)--(\d+4,9)--(\d+4,11)--(\d+8,11)--(\d+8,13);
				\draw[line width=0.8pt] (\d+4,1)--(\d+4,3)--(\d+2,3)--(\d+2,5)--(\d+8,5)--(\d+8,3)--(\d+10,3)--(\d+10,9)--(\d+6,9)--(\d+6,7)--(\d+4,7)--(\d+4,5);
				\draw[line width=0.8pt]
				(\d+6,1)--(\d+6,5);
				\draw (0.4+\d,12.5) node {\scalebox{.2}{6}};
				\draw (2.4+\d,4.5) node {\scalebox{.2}{1}};
				\draw (4.4+\d,6.5) node {\scalebox{.2}{2}};
			}
			\draw[color=cyan,line width=1pt] (17,2)--(17,4)--(19,4)--(19,6)--(17,6)--(17,8)--(21,8)--(21,4)--(23,4)--(23,10)--(17,10)--(17,12)--(27,12)--(27,10)--(25,10)--(25,8)--(27,8)--(27,6)--(25,6)--(25,4)--(27,4)--(27,2)--cycle;
			\draw[line width=0.8pt,color=magenta] (14,0)--(14,14);
		\end{tikzpicture}
		\captionof{figure}{Maxi Loop example puzzle and solution}
	\end{figure}
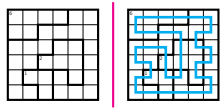
	
	\subsection{Mid-loop}
	
	\textbf{Rules:} Draw a loop passing between adjacent cell centres passing over all the black dots. Black dots indicate the midpoints of their corresponding loop segments (and cannot be on corners), though not all possible dots are necessarily given.
	
	\begin{figure}[hbp]
		\centering
		\begin{tikzpicture}[scale=0.1]
			\foreach \d in {0,14} {
				\foreach \x in {2,4,6,8} \draw[line width=0.2pt] (\x+\d,1)--(\x+\d,11);
				\foreach \y in {3,5,7,9} \draw[line width=0.2pt] (\d,\y)--(\d+10,\y);
				\draw[line width=0.8pt] (\d,1)--(\d,11)--(\d+10,11)--(\d+10,1)--cycle;
			}
			\draw[color=cyan,line width=1pt] (15,2)--(15,4)--(19,4)--(19,8)--(17,8)--(17,6)--(15,6)--(15,10)--(21,10)--(21,8)--(23,8)--(23,6)--(21,6)--(21,2)--cycle;
			\draw[line width=0.8pt,color=magenta] (12,0)--(12,12);
			\foreach \d in {0,14} {
				\draw[fill=black] (\d+1,8) circle (7pt);
				\draw[fill=black] (\d+4,2) circle (7pt);
				\draw[fill=black] (\d+4,8) circle (7pt);
				\draw[fill=black] (\d+4,10) circle (7pt);
				\draw[fill=black] (\d+7,4) circle (7pt);
			}
		\end{tikzpicture}
		\captionof{figure}{Mid-Loop example puzzle and solution}
	\end{figure}
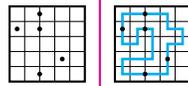
	
	\subsection{Balance Loop}
	
	\textbf{Rules:} Draw a loop passing between adjacent cell centres passing through all the pearls. Divide the loop into segments where it turns. A white pearl indicates the corner of two equal-length segments or the midpoint of a segment, and a black pearl indicates the corner of two unequal-length segments or a non-midpoint cell in some segment's interior. Additionally, numbers are given on some pearls, this represents the length of the segment passing straight through the pearl or the total length of the two segments touching the pearl.
	
	\begin{figure}[hbp]
		\centering
		\begin{tikzpicture}[scale=0.1,font=\sffamily]
			\foreach \d in {0,14} {
				\foreach \x in {2,4,6,8} \draw[line width=0.2pt] (\x+\d,1)--(\x+\d,11);
				\foreach \y in {3,5,7,9} \draw[line width=0.2pt] (\d,\y)--(\d+10,\y);
				\draw[line width=0.8pt] (\d,1)--(\d,11)--(\d+10,11)--(\d+10,1)--cycle;
				\draw[fill=black] (\d+3,10) circle (0.7);
				\draw[fill=white] (\d+5,6) circle (0.7);
				\draw[fill=white] (\d+7,2) circle (0.7);
				\draw (\d+7,2) node {\scalebox{.4}{4}};
				\draw[fill=black] (\d+9,8) circle (0.7);
				\draw[color=white] (\d+9,8) node {\scalebox{.4}{4}};
			}
			\draw[color=cyan,line width=1pt] (15,4)--(15,10)--(23,10)--(23,8)--(17,8)--(17,6)--(21,6)--(21,2)--(17,2)--(17,4)--cycle;
			\draw[line width=0.8pt,color=magenta] (12,0)--(12,12);
		\end{tikzpicture}
		\captionof{figure}{Balance Loop example puzzle and solution}
	\end{figure}
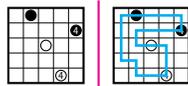
	
	\subsection{Simple Loop}
	
	\textbf{Rule:} Draw a loop passing between adjacent unshaded cell centres that passes through all unshaded cells.
	
	\begin{figure}[hbp]
		\centering
		\begin{tikzpicture}[scale=0.1]
			\foreach \d in {0,14} {
				\foreach \x in {2,4,6,8} \draw[line width=0.2pt] (\x+\d,1)--(\x+\d,11);
				\foreach \y in {3,5,7,9} \draw[line width=0.2pt] (\d,\y)--(\d+10,\y);
				\draw[line width=0.8pt] (\d,1)--(\d,11)--(\d+10,11)--(\d+10,1)--cycle;
				\fill (\d,1)--(\d,3)--(\d+2,3)--(\d+2,1);
				\fill (\d+4,7)--(\d+4,11)--(\d+6,11)--(\d+6,7)--cycle;
				\fill (\d+6,1)--(\d+6,3)--(\d+10,3)--(\d+10,1)--cycle;
			}
			\draw[color=cyan,line width=1pt] (15,4)--(15,10)--(17,10)--(17,6)--(21,6)--(21,10)--(23,10)--(23,4)--(19,4)--(19,2)--(17,2)--(17,4)--cycle;
			\draw[line width=0.8pt,color=magenta] (12,0)--(12,12);
		\end{tikzpicture}
		\captionof{figure}{Simple Loop example puzzle and solution}
	\end{figure}
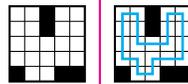
	
	\subsection{Detour}
	
	\textbf{Rules:} Draw a loop passing between adjacent cell centres passing through all cells. A number within a region indicates the number of times the loop turns within that region.
	
	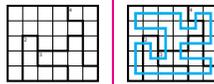
\begin{figure}[hbp]
		\centering
		\begin{tikzpicture}[scale=0.1,font=\sffamily]
			\foreach \d in {0,16} {
				\foreach \x in {2,4,6,8,10} \draw[line width=0.2pt] (\x+\d,1)--(\x+\d,11);
				\foreach \y in {3,5,7,9} \draw[line width=0.2pt] (\d,\y)--(\d+12,\y);
				\draw[line width=0.8pt] (\d,1)--(\d,11)--(\d+12,11)--(\d+12,1)--cycle;
				\draw[line width=0.8pt] (\d+2,1)--(\d+2,7)--(\d+4,7)--(\d+4,3)--(\d+8,3)--(\d+8,1);
				\draw[line width=0.8pt] (\d+4,5)--(\d+8,5)--(\d+8,7)--(\d+10,7);
				\draw[line width=0.8pt] (\d+8,11)--(\d+8,9)--(\d+10,9)--(\d+10,3)--(\d+12,3);
				\draw (2.4+\d,6.5) node {\scalebox{.2}{2}};
				\draw (4.4+\d,4.5) node {\scalebox{.2}{3}};
				\draw (8.4+\d,10.5) node {\scalebox{.2}{4}};
			}
			\draw[color=cyan,line width=1pt] (17,2)--(17,6)--(19,6)--(19,4)--(21,4)--(21,8)--(17,8)--(17,10)--(23,10)--(23,8)--(25,8)--(25,10)--(27,10)--(27,6)--(23,6)--(23,4)--(27,4)--(27,2)--cycle;
			\draw[line width=0.8pt,color=magenta] (14,0)--(14,12);
		\end{tikzpicture}
		\captionof{figure}{Detour example puzzle and solution}
	\end{figure}
	
	\subsection{Haisu}
	
	\textbf{Rules:} Draw a path passing between adjacent cell centres from the S to the G, passing through all cells. The loop may only pass over a number $n$ on its $n^\text{th}$ visit to the region the number is in.
	
	\begin{figure}[hbp]
		\centering
		\begin{tikzpicture}[scale=0.1,font=\sffamily]
			\foreach \d in {0,14} {
				\foreach \x in {2,4,6,8} \draw[line width=0.2pt] (\x+\d,1)--(\x+\d,11);
				\foreach \y in {3,5,7,9} \draw[line width=0.2pt] (\d,\y)--(\d+10,\y);
				\draw[line width=0.8pt] (\d,1)--(\d,11)--(\d+10,11)--(\d+10,1)--cycle;
				\draw[line width=0.8pt] (\d,3)--(\d+2,3)--(\d+2,1);
				\draw[line width=0.8pt] (\d+4,1)--(\d+4,3)--(\d+8,3)--(\d+8,7)--(\d+6,7)--(\d+6,11);
				\draw[line width=0.8pt] (\d+8,11)--(\d+8,9)--(\d+10,9);
				\draw (1+\d,2) node {\tiny S};
				\draw (1+\d,6) node {\tiny 1};
				\draw (3+\d,10) node {\tiny 2};
				\draw (5+\d,6) node {\tiny 2};
				\draw (7+\d,2) node {\tiny 1};
				\draw (9+\d,6) node {\tiny 3};
				\draw (9+\d,10) node {\tiny G};
			}
			\draw[color=cyan,line width=1pt] (15,2)--(15,6)--(17,6)--(17,2)--(23,2)--(23,4)--(19,4)--(19,8)--(15,8)--(15,10)--(21,10)--(21,6)--(23,6)--(23,10);
			\draw[line width=0.8pt,color=magenta] (12,0)--(12,12);
		\end{tikzpicture}
		\captionof{figure}{Haisu example puzzle and solution}
	\end{figure}
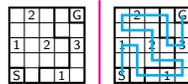
	
	\newpage
	
	\subsection{Reflect Link}
	
	\textbf{Rules:} Draw a loop passing between adjacent cell centres. The loop must path through all triangle cells (mirrors), `reflecting' off the mirror. The mirror might have a number in it, which indicates the total number of cells passed through by the segments emanating from the triangle. Additionally, the loop must cross itself at, and only at, cells marked with a plus.
	
	\begin{figure}[hbp]
		\centering
		\begin{tikzpicture}[scale=0.1,font=\sffamily]
			\foreach \d in {0,14} {
				\foreach \x in {2,4,6,8} \draw[line width=0.2pt] (\x+\d,1)--(\x+\d,11);
				\foreach \y in {3,5,7,9} \draw[line width=0.2pt] (\d,\y)--(\d+10,\y);
				\draw[line width=0.8pt] (\d,1)--(\d,11)--(\d+10,11)--(\d+10,1)--cycle;
			}
			\draw[color=cyan,line width=1pt] (15,8)--(15,10)--(17,10)--(17,6)--(19,6)--(19,2)--(23,2)--(23,8)--(21,8)--(21,10)--(19,10)--(19,8)--cycle;
			\draw[line width=0.2pt] (16,7)--(16,9)--(18,9)--(18,7)--cycle;
			\draw[line width=0.8pt,color=magenta] (12,0)--(12,12);
			\foreach \d in {0,14} {
				\fill (\d+2,5)--(\d+2,7)--(\d+4,5)--cycle;
				\fill (\d+4,1)--(\d+4,3)--(\d+6,1)--cycle;
				\fill (\d+4,11)--(\d+4,9)--(\d+6,11)--cycle;
				\fill (\d+10,1)--(\d+10,3)--(\d+8,1)--cycle;
				\draw[line width=1.4pt] (\d+3,7)--(\d+3,9);
				\draw[line width=1.4pt] (\d+2,8)--(\d+4,8);
				\draw[color=white] (2.6+\d,5.7) node {\scalebox{0.3}{4}};
				\draw[color=white] (9.4+\d,1.7) node {\scalebox{0.3}{6}};
			}
		\end{tikzpicture}
		\captionof{figure}{Reflect Link example puzzle and solution}
	\end{figure}
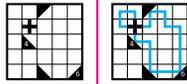
	
	\subsection{Pipelink}
	
	\textbf{Rules:} Draw a loop passing between adjacent cell centres passing through all cells. The loop must follow exactly the provided path on cells with shown path segments. The loop may cross itself at any empty cell or clued + cell (though it may not overlap itself except for at a crossing point).
	
	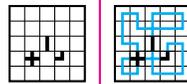
\begin{figure}[hbp]
		\centering
		\begin{tikzpicture}[scale=0.1]
			\foreach \d in {0,14} {
				\foreach \x in {2,4,6,8} \draw[line width=0.2pt] (\x+\d,1)--(\x+\d,11);
				\foreach \y in {3,5,7,9} \draw[line width=0.2pt] (\d,\y)--(\d+10,\y);
				\draw[line width=0.8pt] (\d,1)--(\d,11)--(\d+10,11)--(\d+10,1)--cycle;
			}
			\draw[color=cyan,line width=1pt] (15,2)--(15,4)--(21,4)--(21,6)--(23,6)--(23,2)--(19,2)--(19,10)--(23,10)--(23,8)--(17,8)--(17,10)--(15,10)--(15,6)--(17,6)--(17,2)--cycle;
			\draw[line width=0.2pt] (16,3)--(16,5)--(22,5)--(22,3)--(20,3)--(20,7)--(18,7)--(18,3)--cycle;
			\draw[line width=0.8pt,color=magenta] (12,0)--(12,12);
			\foreach \d in {0,14} {
				\draw[line width=1.4pt] (\d+3,3)--(\d+3,5);
				\draw[line width=1.4pt] (\d+2,4)--(\d+4,4);
				\draw[line width=1.4pt] (\d+5,5)--(\d+5,7);
				\draw[line width=1.4pt] (\d+6,4)--(\d+7,4)--(\d+7,5);
			}
		\end{tikzpicture}
		\captionof{figure}{Pipelink example puzzle and solution}
	\end{figure}
	
	\subsection{Loop Special}
	
	\textbf{Rules:} Draw some number of loops passing between adjacent cell centres that together pass through all cells. Each loop must pass through at least one circled cell, and all circled cells of the same number must be passed through by the same loop. The loop must follow exactly the provided path on cells with shown path segments. The loop may cross itself at any empty cell or clued + cell, but not on a circled cell (and it may not overlap itself except for at a crossing point).
	
	\begin{figure}[hbp]
		\centering
		\begin{tikzpicture}[scale=0.1,font=\sffamily]
			\foreach \d in {0,16} {
				\foreach \x in {2,4,6,8,10} \draw[line width=0.2pt] (\x+\d,1)--(\x+\d,13);
				\foreach \y in {3,5,7,9,11} \draw[line width=0.2pt] (\d,\y)--(\d+12,\y);
				\draw[line width=0.8pt] (\d,1)--(\d,13)--(\d+12,13)--(\d+12,1)--cycle;
				\draw[fill=white] (\d+3,4) circle (0.7);
				\draw (\d+3,4) node {\scalebox{.4}{1}};
				\draw[fill=white] (\d+5,12) circle (0.7);
				\draw (\d+5,12) node {\scalebox{.4}{1}};
				\draw[fill=white] (\d+7,4) circle (0.7);
				\draw (\d+7,4) node {\scalebox{.4}{1}};
				\draw[fill=white] (\d+7,10) circle (0.7);
				\draw (\d+7,10) node {\scalebox{.4}{1}};
				\draw[fill=white] (\d+11,12) circle (0.7);
				\draw (\d+11,12) node {\scalebox{.4}{2}};
			}
			\draw[color=cyan,line width=1pt] (17,2)--(17,4)--(21,4)--(21,10)--(17,10)--(17,12)--(23,12)--(23,8)--(17,8)--(17,6)--(25,6)--(25,8)--(27,8)--(27,2)--(25,2)--(25,4)--(23,4)--(23,2)--cycle;
			\draw[color=cyan,line width=1pt] (25,10)--(25,12)--(27,12)--(27,10)--cycle;
			\draw[line width=0.2pt] (18,5)--(18,9)--(22,9)--(22,7)--(16,7);
			\draw[line width=0.2pt] (22,1)--(22,3)--(24,3);
			\draw[line width=0.2pt] (24,7)--(26,7)--(26,9);
			\draw[line width=0.2pt] (28,3)--(26,3)--(26,5)--(28,5);
			\draw[line width=0.8pt,color=magenta] (14,0)--(14,14);
			\foreach \d in {0,16} {
				\draw[line width=1.4pt] (\d+1,7)--(\d+1,6)--(\d+2,6);
				\draw[line width=1.4pt] (\d+2,8)--(\d+6,8);
				\draw[line width=1.4pt] (\d+5,7)--(\d+5,9);
				\draw[line width=1.4pt] (\d+6,2)--(\d+7,2)--(\d+7,3);
				\draw[line width=1.4pt] (\d+9,7)--(\d+9,8)--(\d+10,8);
				\draw[line width=1.4pt] (\d+11,3)--(\d+11,5);
			}
		\end{tikzpicture}
		\captionof{figure}{Loop Special example puzzle and solution}
	\end{figure}
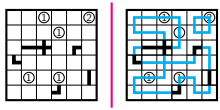
	
	\subsection{Pipelink Returns}
	
	\textbf{Rules:} Draw a loop passing between adjacent cell centres passing through all cells. The loop must follow exactly the provided path on cells with shown path segments. The loop may cross itself at any circled cell or clued + cell, but not on an empty cell (and it may not overlap itself except for at a crossing point). The loop must pass straight through all circled cells.
	
	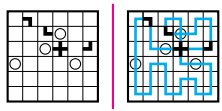
\begin{figure}[hbp]
		\centering
		\begin{tikzpicture}[scale=0.1]
			\foreach \d in {0,16} {
				\foreach \x in {2,4,6,8,10} \draw[line width=0.2pt] (\x+\d,1)--(\x+\d,13);
				\foreach \y in {3,5,7,9,11} \draw[line width=0.2pt] (\d,\y)--(\d+12,\y);
				\draw[line width=0.8pt] (\d,1)--(\d,13)--(\d+12,13)--(\d+12,1)--cycle;
				\draw[fill=white] (\d+1,6) circle (0.7);
				\draw[fill=white] (\d+5,8) circle (0.7);
				\draw[fill=white] (\d+7,10) circle (0.7);
				\draw[fill=white] (\d+9,6) circle (0.7);
			}
			\draw[color=cyan,line width=1pt] (17,2)--(17,12)--(19,12)--(19,8)--(27,8)--(27,12)--(25,12)--(25,10)--(21,10)--(21,12)--(23,12)--(23,6)--(27,6)--(27,2)--(25,2)--(25,4)--(23,4)--(23,2)--(21,2)--(21,6)--(19,6)--(19,2)--cycle;
			\draw[line width=0.2pt] (18,13)--(18,11)--(22,11)--(22,7)--(24,7)--(24,9)--(22,9);
			\draw[line width=0.2pt] (26,7)--(26,9)--(28,9);
			\draw[line width=0.8pt,color=magenta] (14,0)--(14,14);
			\foreach \d in {0,16} {
				\draw[line width=1.4pt] (\d+2,12)--(\d+3,12)--(\d+3,11);
				\draw[line width=1.4pt] (\d+5,11)--(\d+5,10)--(\d+6,10);
				\draw[line width=1.4pt] (\d+6,8)--(\d+8,8);
				\draw[line width=1.4pt] (\d+7,7)--(\d+7,9);
				\draw[line width=1.4pt] (\d+10,8)--(\d+11,8)--(\d+11,9);
			}
		\end{tikzpicture}
		\captionof{figure}{Pipelink Returns example puzzle and solution}
	\end{figure}
	
	\subsection{Icebarn}
	
	\textbf{Rules:} Draw a path passing between adjacent cell centres from the IN to the OUT. The path must pass over all arrows in the direction indicated. Coloured cells indicate `ice': any segment of path passing over ice must go straight, however the path may also cross itself only on ice cells. Connected ice cells are grouped into `icebarns', each of which must be passed through at least once.
	
	\begin{figure}[hbp]
		\centering
		\begin{tikzpicture}[scale=0.1,font=\sffamily]
			\foreach \d in {2,20} {
				\draw[line width=0.8pt,fill=cyan!30!white] (\d,3)--(\d+4,3)--(\d+4,5)--(\d+8,5)--(\d+8,7)--(\d+6,7)--(\d+6,11)--(\d+12,11)--(\d+12,3)--(\d+8,3)--(\d+8,1)--(\d,1);
				\draw[line width=0.8pt,fill=cyan!30!white] (\d+2,9)--(\d+2,11)--(\d+4,11)--(\d+4,9)--cycle;
				\foreach \x in {2,4,6,8,10} \draw[line width=0.2pt] (\x+\d,1)--(\x+\d,13);
				\foreach \y in {3,5,7,9,11} \draw[line width=0.2pt] (\d,\y)--(\d+12,\y);
				\draw[line width=0.8pt] (\d,1)--(\d,13)--(\d+12,13)--(\d+12,1)--cycle;
				\draw (\d-1.75,10) node {\scalebox{0.3}{OUT}};
				\draw (\d+3,14) node {\scalebox{0.3}{IN}};
			}
			\draw[color=cyan,line width=1pt] (23,12.75)--(23,8)--(21,8)--(21,6)--(27,6)--(27,12)--(25,12)--(25,10)--(19.75,10);
			\draw[line width=0.8pt,color=magenta] (16,0)--(16,15);
			\foreach \d in {2,20} {
				\draw (\d-0.25,10)--(\d+0.5,10);
				\fill (\d-0.25,9.75)--(\d-0.25,10.25)--(\d-0.5,10)--cycle;
				\draw (\d+2.25,6)--(\d+1.5,6);
				\fill (\d+2.25,5.75)--(\d+2.25,6.25)--(\d+2.5,6)--cycle;
				\draw (\d+3,13.5)--(\d+3,12.75);
				\fill (\d+2.75,12.75)--(\d+3.25,12.75)--(\d+3,12.5)--cycle;
			}
		\end{tikzpicture}
		\captionof{figure}{Icebarn example puzzle and solution}
	\end{figure}
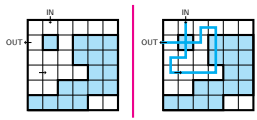
	
	\subsection{Barns}
	
	\textbf{Rules:} Draw a loop passing between adjacent cell centres visiting every cell. The loop may not cross any bolded edges. Coloured cells indicate `ice': any segment of path passing over ice must go straight, however the path may also cross itself only on ice cells.
	
	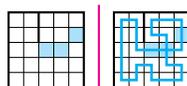
\begin{figure}[hbp]
		\centering
		\begin{tikzpicture}[scale=0.1]
			\foreach \d in {0,14} {
				\fill[color=cyan!30!white] (\d+4,5)--(\d+4,7)--(\d+10,7)--(\d+10,9)--(\d+8,9)--(\d+8,5)--cycle;
				\foreach \x in {2,4,6,8} \draw[line width=0.2pt] (\x+\d,1)--(\x+\d,11);
				\foreach \y in {3,5,7,9} \draw[line width=0.2pt] (\d,\y)--(\d+10,\y);
				\draw[line width=0.8pt] (\d,1)--(\d,11)--(\d+10,11)--(\d+10,1)--cycle;
				\draw[line width=0.8pt] (\d+4,7)--(\d+4,11);
			}
			\draw[color=cyan,line width=1pt] (15,2)--(15,10)--(17,10)--(17,6)--(23,6)--(23,10)--(19,10)--(19,8)--(21,8)--(21,4)--(23,4)--(23,2)--(19,2)--(19,4)--(17,4)--(17,2)--cycle;
			\draw[line width=0.8pt,color=magenta] (12,0)--(12,12);
		\end{tikzpicture}
		\captionof{figure}{Barns example puzzle and solution}
	\end{figure}
	
	\subsection{Angle Loop}
	
	\textbf{Rules:} Draw a loop via straight lines between shapes, visiting all the shapes. At every triangle, square or pentagon respectively, the loop must make an acute, right or obtuse angle. The loop may not cross itself.
	
	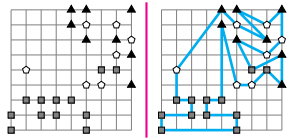
\begin{figure}[hbp]
		\centering
		\begin{tikzpicture}[scale=0.1]
			\foreach \d in {0,20} {
				\foreach \x in {0,2,...,16} \draw[color=gray,line width=0.2pt] (\d+\x,1)--(\d+\x,17);
				\foreach \y in {1,3,...,17} \draw[color=gray,line width=0.2pt] (\d,\y)--(\d+16,\y);
			}
			\draw[rounded corners=.1mm,color=cyan,line width=1pt] (20,1)--(20,3)--(24,3)--(24,5)--(22,5)--(22,9)--(28,17)--(30,15)--(34,13)--(30,17)--(34,15)--(36,17)--(36,13)--(34,11)--(30,13)--(32,9)--(36,11)--(36,7)--(34,9)--(32,7)--(30,7)--(28,15)--(28,5)--(26,5)--(26,3)--(30,3)--(30,1)--cycle;
			\draw[line width=0.8pt,color=magenta] (18,0)--(18,18);
			\foreach \d in {0,20} {
				\rightang{\d}{1};
				\rightang{\d}{3};
				\rightang{\d+2}{5};
				\rightang{\d+4}{3};
				\rightang{\d+4}{5};
				\rightang{\d+6}{3};
				\rightang{\d+6}{5};
				\rightang{\d+8}{5};
				\rightang{\d+10}{1};
				\rightang{\d+10}{3};
				\rightang{\d+12}{9};
				\rightang{\d+14}{9};
				\acuteang{\d+8}{15};
				\acuteang{\d+8}{17};
				\acuteang{\d+10}{13};
				\acuteang{\d+10}{17};
				\acuteang{\d+14}{13};
				\acuteang{\d+16}{7};
				\acuteang{\d+16}{11};
				\acuteang{\d+16}{17};
				\obtuseang{\d+2}{9};
				\obtuseang{\d+10}{7};
				\obtuseang{\d+10}{15};
				\obtuseang{\d+12}{7};
				\obtuseang{\d+14}{11};
				\obtuseang{\d+14}{15};
				\obtuseang{\d+16}{13};
			}
		\end{tikzpicture}
		\captionof{figure}{Angle Loop example puzzle and solution}
	\end{figure}
	
	\subsection{Kouchoku}
	
	\textbf{Rules:} Draw a loop via straight lines between dots and circles, visiting all the dots and circles. Circles with the same marking (here, letters) must be passed through all in one go with no dots or differently-marked circles between, and differently-marked circles may not be directly connected. Additionally, the loop may cross itself at any point which is not (the centre of) a dot or circle, as long as it intersects itself at a right-angle.
	
	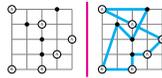
\begin{figure}[hbp]
		\centering
		\begin{tikzpicture}[scale=0.1,font=\sffamily]
			\foreach \d in {0,12} {
				\foreach \x in {0,2,...,8} \draw[color=gray,line width=0.2pt] (\d+\x,1)--(\d+\x,9);
				\foreach \y in {1,3,...,9} \draw[color=gray,line width=0.2pt] (\d,\y)--(\d+8,\y);
			}
			\draw[rounded corners=.1mm,color=cyan,line width=1pt] (12,1)--(14,7)--(16,5)--(18,9)--(12,9)--(20,5)--(18,3)--(16,3)--(16,1)--cycle;
			\draw[line width=0.8pt,color=magenta] (10,0)--(10,10);
			\foreach \d in {0,12} {
				\letcirc{\d+0}{1}{B};
				\letcirc{\d+4}{1}{B};
				\letcirc{\d+0}{9}{A};
				\letcirc{\d+4}{7}{A};
				\letcirc{\d+6}{3}{A};
				\letcirc{\d+8}{5}{A};
				\fill[color=black] (\d+2,7) circle (0.3);
				\fill[color=black] (\d+4,3) circle (0.3);
				\fill[color=black] (\d+4,5) circle (0.3);
				\fill[color=black] (\d+6,9) circle (0.3);
			}
		\end{tikzpicture}
		\captionof{figure}{Kouchoku example puzzle and solution}
	\end{figure}
	
	\subsection{Icelom}
	
	\textbf{Rules:} Draw a path passing between adjacent cell centres from the IN to the OUT. The path must pass over all numbers in increasing order. Coloured cells indicate `ice': the path must pass straight over ice cells, though it is allowed to cross itself only on ice cells (though it may not overlap itself except for at a crossing point). Every non-ice square must be passed through.
	
	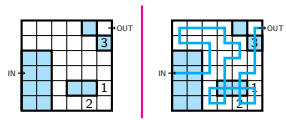
\begin{figure}[hbp]
		\centering
		\begin{tikzpicture}[scale=0.1,font=\sffamily]
			\foreach \d in {2,22} {
				\draw[line width=0.8pt,fill=cyan!30!white] (\d,1)--(\d,9)--(\d+4,9)--(\d+4,1);
				\draw[line width=0.8pt,fill=cyan!30!white] (\d+6,3)--(\d+6,5)--(\d+10,5)--(\d+10,3)--cycle;
				\draw[line width=0.8pt,fill=cyan!30!white] (\d+8,13)--(\d+8,11)--(\d+12,11)--(\d+12,9)--(\d+10,9)--(\d+10,13);
				\foreach \x in {2,4,6,8,10} \draw[line width=0.2pt] (\x+\d,1)--(\x+\d,13);
				\foreach \y in {3,5,7,9,11} \draw[line width=0.2pt] (\d,\y)--(\d+12,\y);
				\draw[line width=0.8pt] (\d,1)--(\d,13)--(\d+12,13)--(\d+12,1)--cycle;
				\draw (\d-1.3,6) node {\scalebox{0.3}{IN}};
				\draw (\d+13.75,12) node {\scalebox{0.3}{OUT}};
				\draw (\d+9,2) node {\tiny 2};
				\draw (\d+11,4) node {\tiny 1};
				\draw (\d+11,10) node {\tiny 3};
			}
			\draw[color=cyan,line width=1pt] (22.25,6)--(27,6)--(27,10)--(23,10)--(23,12)--(29,12)--(29,10)--(31,10)--(31,8)--(29,8)--(29,2)--(27,2)--(27,4)--(33,4)--(33,2)--(31,2)--(31,6)--(33,6)--(33,12)--(34.25,12);
			\draw[line width=0.8pt,color=magenta] (18,0)--(18,15);
			\foreach \d in {2,22} {
				\draw (\d-0.5,6)--(\d+0.25,6);
				\fill (\d+0.25,5.75)--(\d+0.25,6.25)--(\d+0.5,6)--cycle;
				\draw (\d+11.5,12)--(\d+12.25,12);
				\fill (\d+12.25,11.75)--(\d+12.25,12.25)--(\d+12.5,12)--cycle;
			}
		\end{tikzpicture}
		\captionof{figure}{Icelom example puzzle and solution}
	\end{figure}
	
	\subsection{Icelom2}
	
	\textbf{Rules:} Draw a path passing between adjacent cell centres from the IN to the OUT. The path must pass over all numbers in increasing order. Coloured cells indicate `ice': the path must pass straight over ice cells, though it is allowed to cross itself only on ice cells (though it may not overlap itself except for at a crossing point). Connected ice cells are grouped into `icebarns'. Every icebarn must be passed through.
	
	\begin{figure}[hbp]
		\centering
		\begin{tikzpicture}[scale=0.1,font=\sffamily]
			\foreach \d in {0,16} {
				\draw[line width=0.8pt,fill=cyan!30!white] (\d+2,2)--(\d+2,4)--(\d+4,4)--(\d+4,2);
				\draw[line width=0.8pt,fill=cyan!30!white] (\d+6,2)--(\d+6,6)--(\d+12,6)--(\d+12,2);
				\draw[line width=0.8pt,fill=cyan!30!white] (\d+8,8)--(\d+8,10)--(\d+10,10)--(\d+10,8)--cycle;
				\draw[line width=0.8pt,fill=cyan!30!white] (\d,8)--(\d+4,8)--(\d+4,6)--(\d+6,6)--(\d+6,14)--(\d+4,14)--(\d+4,10)--(\d+2,10)--(\d+2,12)--(\d,12);
				\foreach \x in {2,4,6,8,10} \draw[line width=0.2pt] (\x+\d,2)--(\x+\d,14);
				\foreach \y in {4,6,...,12} \draw[line width=0.2pt] (\d,\y)--(\d+12,\y);
				\draw[line width=0.8pt] (\d,2)--(\d,14)--(\d+12,14)--(\d+12,2)--cycle;
				\draw (\d+3,15) node {\scalebox{0.3}{IN}};
				\draw (\d+9,1) node {\scalebox{0.3}{OUT}};
				\draw (\d+7,13) node {\tiny 2};
				\draw (\d+11,13) node {\tiny 1};
			}
			\draw[color=cyan,line width=1pt] (19,13.75)--(19,13)--(17,13)--(17,3)--(21,3)--(21,5)--(19,5)--(19,7)--(23,7)--(23,9)--(27,9)--(27,13)--(23,13)--(23,11)--(25,11)--(25,1.75);
			\draw[line width=0.8pt,color=magenta] (14,0)--(14,16);
			\foreach \d in {0,16} {
				\draw (\d+3,13.75)--(\d+3,14.5);
				\fill (\d+2.75,13.75)--(\d+3.25,13.75)--(\d+3,13.5)--cycle;
				\draw (\d+9,1.75)--(\d+9,2.5);
				\fill (\d+8.75,1.75)--(\d+9.25,1.75)--(\d+9,1.5)--cycle;
			}
		\end{tikzpicture}
		\captionof{figure}{Icelom2 example puzzle and solution}
	\end{figure}
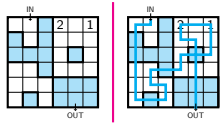
	
	\subsection{Ring-ring}
	
	\textbf{Rules:} Draw some number of (potentially crossing) rectangles with vertices on cell centres, such that no two rectangles share any part of their boundaries, except for crossing points. All unshaded cells must be used by some rectangle and no shaded cells may be used by any rectangle.
	
	\begin{figure}[hbp]
		\centering
		\begin{tikzpicture}[scale=0.1]
			\foreach \d in {0,16} {
				\foreach \x in {2,4,6,8,10} \draw[line width=0.2pt] (\x+\d,1)--(\x+\d,13);
				\foreach \y in {3,5,...,11} \draw[line width=0.2pt] (\d,\y)--(\d+12,\y);
				\draw[line width=0.8pt] (\d,1)--(\d,13)--(\d+12,13)--(\d+12,1)--cycle;
				\fill (\d,1)--(\d,3)--(\d+8,3)--(\d+8,1);
				\fill (\d,11)--(\d,13)--(\d+2,13)--(\d+2,11);
				\fill (\d+6,13)--(\d+6,11)--(\d+8,11)--(\d+8,5)--(\d+12,5)--(\d+12,13);
			}
			\draw[color=cyan,line width=1pt] (17,4)--(17,10)--(23,10)--(23,4)--cycle;
			\draw[color=cyan,line width=1pt] (19,6)--(19,12)--(21,12)--(21,6)--cycle;
			\draw[color=cyan,line width=1pt] (25,2)--(25,4)--(27,4)--(27,2)--cycle;
			\draw[line width=0.8pt,color=magenta] (14,0)--(14,14);
		\end{tikzpicture}
		\captionof{figure}{Ring-ring example puzzle and solution}
	\end{figure}
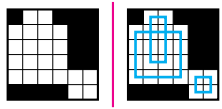
	
	\subsection{Nagenawa}
	
	\textbf{Rules:} Draw some number of (potentially crossing) rectangles with vertices on cell centres, such that no two rectangles share any part of their boundaries, except for crossing points. Numbers in some regions indicate the number of cells used by any rectangle.
	
	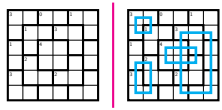
\begin{figure}[hbp]
		\centering
		\begin{tikzpicture}[scale=0.1,font=\sffamily]
			\foreach \d in {0,16} {
				\foreach \x in {2,4,6,8,10} \draw[line width=0.2pt] (\x+\d,1)--(\x+\d,13);
				\foreach \y in {3,5,...,11} \draw[line width=0.2pt] (\d,\y)--(\d+12,\y);
				\draw[line width=0.8pt] (\d,1)--(\d,13)--(\d+12,13)--(\d+12,1)--cycle;
				\draw[line width=0.8pt] (\d+2,3)--(\d+2,11)--(\d+10,11)--(\d+10,3)--cycle;
				\draw[line width=0.8pt] (\d+4,5)--(\d+4,9)--(\d+8,9)--(\d+8,5)--cycle;
				\draw[line width=0.8pt] (\d,5)--(\d+2,5);
				\draw[line width=0.8pt] (\d,9)--(\d+2,9);
				\draw[line width=0.8pt] (\d+10,5)--(\d+12,5);
				\draw[line width=0.8pt] (\d+10,9)--(\d+12,9);
				\draw[line width=0.8pt] (\d+2,7)--(\d+4,7);
				\draw[line width=0.8pt] (\d+8,7)--(\d+10,7);
				\draw[line width=0.8pt] (\d+4,1)--(\d+4,3);
				\draw[line width=0.8pt] (\d+8,1)--(\d+8,3);
				\draw[line width=0.8pt] (\d+6,3)--(\d+6,5);
				\draw[line width=0.8pt] (\d+6,9)--(\d+6,11);
				\draw[line width=0.8pt] (\d+4,11)--(\d+4,13);
				\draw[line width=0.8pt] (\d+8,11)--(\d+8,13);
				\draw (0.4+\d,4.5) node {\scalebox{.2}{3}};
				\draw (0.4+\d,8.5) node {\scalebox{.2}{1}};
				\draw (0.4+\d,12.5) node {\scalebox{.2}{3}};
				\draw (2.4+\d,6.5) node {\scalebox{.2}{2}};
				\draw (2.4+\d,10.5) node {\scalebox{.2}{1}};
				\draw (4.4+\d,8.5) node {\scalebox{.2}{4}};
				\draw (4.4+\d,12.5) node {\scalebox{.2}{0}};
				\draw (6.4+\d,4.5) node {\scalebox{.2}{2}};
				\draw (6.4+\d,10.5) node {\scalebox{.2}{3}};
				\draw (8.4+\d,12.5) node {\scalebox{.2}{1}};
			}
			\draw[color=cyan,line width=1pt] (17,2)--(17,6)--(19,6)--(19,2)--cycle;
			\draw[color=cyan,line width=1pt] (17,12)--(17,10)--(19,10)--(19,12)--cycle;
			\draw[color=cyan,line width=1pt] (21,6)--(21,8)--(25,8)--(25,6)--cycle;
			\draw[color=cyan,line width=1pt] (23,2)--(23,10)--(27,10)--(27,2)--cycle;
			\draw[line width=0.8pt,color=magenta] (14,0)--(14,14);
		\end{tikzpicture}
		\captionof{figure}{Nagenawa example puzzle and solution}
	\end{figure}
	
	\subsection{Regional Yajilin}
	
	\textbf{Rules:} Draw a single loop passing between adjacent cell centres. Any cell which the loop does not pass through must be shaded, and no two shaded cells may be adjacent. A region's number indicates the number of shaded cells within it.
	
	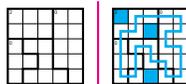
\begin{figure}[hbp]
		\centering
		\begin{tikzpicture}[scale=0.1,font=\sffamily]
			\fill[color=cyan] (14,9)--(14,11)--(16,11)--(16,9);
			\fill[color=cyan] (18,7)--(18,9)--(20,9)--(20,7);
			\fill[color=cyan] (18,1)--(18,3)--(20,3)--(20,1);
			\foreach \d in {0,14} {
				\foreach \x in {2,4,6,8} \draw[line width=0.2pt] (\x+\d,1)--(\x+\d,11);
				\foreach \y in {3,5,7,9} \draw[line width=0.2pt] (\d,\y)--(\d+10,\y);
				\draw[line width=0.8pt] (\d,1)--(\d,11)--(\d+10,11)--(\d+10,1)--cycle;
				\draw[line width=0.8pt] (\d,7)--(\d+6,7);
				\draw[line width=0.8pt] (\d+2,1)--(\d+2,3)--(\d+4,3)--(\d+4,5)--(\d+2,5)--(\d+2,7);
				\draw[line width=0.8pt] (\d+8,1)--(\d+8,3)--(\d+6,3)--(\d+6,11);
				\draw (0.4+\d,6.5) node {\scalebox{.2}{0}};
				\draw (0.4+\d,10.5) node {\scalebox{.2}{2}};
				\draw (6.4+\d,10.5) node {\scalebox{.2}{0}};
			}
			\draw[color=cyan,line width=1pt] (15,2)--(15,8)--(17,8)--(17,10)--(23,10)--(23,8)--(21,8)--(21,6)--(23,6)--(23,2)--(21,2)--(21,4)--(19,4)--(19,6)--(17,6)--(17,2)--cycle;
			\draw[line width=0.8pt,color=magenta] (12,0)--(12,12);
		\end{tikzpicture}
		\captionof{figure}{Regional Yajilin example puzzle and solution}
	\end{figure}
	
	\subsection{Tapa-Like Loop}
	
	\textbf{Rules:} Draw a loop passing between adjacent cell centres, not passing through any clue cells. For each clue cell, consider the eight touching cells in the $3\times3$ region centred on that cell. Consider only the portions of the loop in those cells. Then, the numbers in the cell represent the lengths of the different segments of loop in that region, in arbitrary order. However, a single 0 instead indicates that the loop does not pass in any of the eight touching cells.
	
	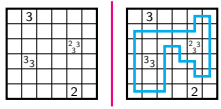
\begin{figure}[hbp]
		\centering
		\begin{tikzpicture}[scale=0.1,font=\sffamily]
			\foreach \d in {0,16} {
				\foreach \x in {2,4,6,8,10} \draw[line width=0.2pt] (\x+\d,1)--(\x+\d,13);
				\foreach \y in {3,5,...,11} \draw[line width=0.2pt] (\d,\y)--(\d+12,\y);
				\draw[line width=0.8pt] (\d,1)--(\d,13)--(\d+12,13)--(\d+12,1)--cycle;
				\draw (\d+3,12) node {\tiny 3};
				\draw (\d+9,2) node {\tiny 2};
				\draw (\d+2.6,6.3) node {\scalebox{0.4}{3}};
				\draw (\d+3.4,5.7) node {\scalebox{0.4}{3}};
				\draw (\d+8.5,8.4) node {\scalebox{0.3}{2}};
				\draw (\d+9.5,8.2) node {\scalebox{0.3}{3}};
				\draw (\d+8.95,7.5) node {\scalebox{0.3}{3}};
			}
			\draw[color=cyan,line width=1pt] (17,2)--(17,10)--(25,10)--(25,12)--(27,12)--(27,4)--(25,4)--(25,6)--(23,6)--(23,8)--(21,8)--(21,2)--cycle;
			\draw[line width=0.8pt,color=magenta] (14,0)--(14,14);
		\end{tikzpicture}
		\captionof{figure}{Tapa-Like Loop example puzzle and solution}
	\end{figure}

\end{document}